%% file: BryceHensonHonoursThesis.tex
\documentclass[twoside,onecolumn,11pt,a4paper]{book}
\pdfoutput=1
\usepackage{graphicx}  % most common package for figure handling
\usepackage[cc]{titlepic} %title page pic
\usepackage{sty/anuthesis}
\usepackage{gensymb} %gives a degree symbol
\usepackage{fancyhdr}
\usepackage{epstopdf}
\usepackage{float}
\usepackage{enumitem} %change the style of lists
\usepackage{pdfpages}  %includes pdf documents
\usepackage{subfig}
\usepackage{bm}%bold maths
\usepackage{etoolbox}
\usepackage{color,soul} %highlighting
\usepackage{amsmath}  % American Mathematical Society (AMS) standard symbols and math expressions
\usepackage{amssymb}
\usepackage{bm}
\usepackage{indentfirst} %adds a indent after title
\usepackage{verbatim} % useful for displaying computer code or similar verbatim
\usepackage{full page} % This package changes the otherwise very generous margins to 1 inch on all sides
\usepackage{textcomp}
\usepackage{verbatimbox}

\usepackage[backend=biber,
            citestyle=numeric-comp,
            bibstyle=phys,
            biblabel=brackets,
            eprint=false,
            doi=true,
            sorting=none,
            maxnames = 12,
            minbibnames = 12,
            mincitenames= 3,
            hyperref=true,
            ]{biblatex}
\DeclareFieldFormat[thesis]{title}{
    \href{\thefield{url}}{#1}%
}

\DeclareSourcemap{
  \maps[datatype=bibtex]{
    \map[overwrite]{
      \step[fieldsource=month, match=\regexp{\A(j|J)an(uary)?\Z}, replace=1]
      \step[fieldsource=month, match=\regexp{\A(f|F)eb(ruary)?\Z}, replace=2]
      \step[fieldsource=month, match=\regexp{\A(m|M)ar(ch)?\Z}, replace=3]
      \step[fieldsource=month, match=\regexp{\A(a|A)pr(il)?\Z}, replace=4]
      \step[fieldsource=month, match=\regexp{\A(m|M)ay\Z}, replace=5]
      \step[fieldsource=month, match=\regexp{\A(j|J)un(e)?\Z}, replace=6]
      \step[fieldsource=month, match=\regexp{\A(j|J)ul(y)?\Z}, replace=7]
      \step[fieldsource=month, match=\regexp{\A(a|A)ug(ust)?\Z}, replace=8]
      \step[fieldsource=month, match=\regexp{\A(s|S)ep(tember)?\Z}, replace=9]
      \step[fieldsource=month, match=\regexp{\A(o|O)ct(ober)?\Z}, replace=10]
      \step[fieldsource=month, match=\regexp{\A(n|N)ov(ember)?\Z}, replace=11]
      \step[fieldsource=month, match=\regexp{\A(d|D)ec(ember)?\Z}, replace=12]
    }
  }
}

\usepackage{doi}
\usepackage{hyperref} %hyperlinks on citations

\newcommand{\bra}[1]{\langle #1 |} %quantum sumbols
\newcommand{\ket}[1]{| #1 \rangle}
\newcommand{\expt}[1]{\langle #1 \rangle}

\newcommand{\TO}{2^{3\!}S_1 - 2^{3\!}P / 3^{3\!}P}%

\begin{document}

\pagenumbering{roman}  % first use Roman numerals for page numbers

\begin{titlepage}
\titlepic{\includegraphics[height=4cm]{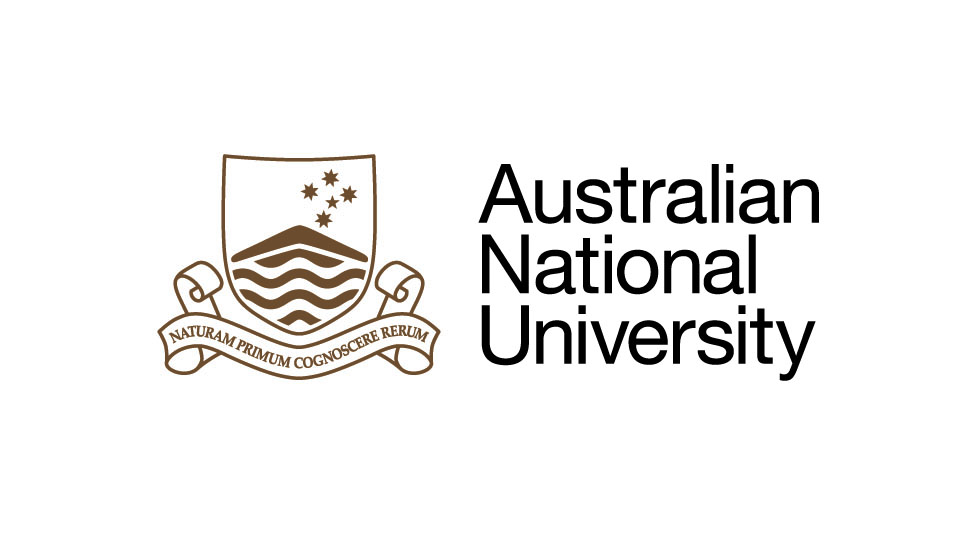}}
\title{\textbf{The First Measurement of the $\TO$ Tune-Out Wavelength in He*}\\[2cm]}
 \author{\textbf{Bryce Maxwell Henson}\\[6cm]
 \textbf{A thesis submitted for the degree of}\\
 \textbf{Bachelor of Science with Honours in Physics at} \\
 \textbf{The Australian National University}\\[1cm]}
 \date{\textbf{January-October, 2014}}
 
\maketitle

 \end{titlepage}
 \sloppy

{\let\cleardoublepage\relax \chapter*{Citation}}
\noindent
To cite this thesis using BibTeX format: \\
\begin{verbnobox}
@thesis{henson_hnrs_thesis_2014,
  author       = {Bryce Henson}, 
  title        = {The First Measurement of the 
  {$2^{3\!}S_1 - 2^{3\!}P / 3^{3\!}P$} Tune-Out Wavelength in {He*}},
  institution  = {Australian National University},
  url ={https://arxiv.org/abs/1708.08200},
  year         = 2014,
  type = {Honours Thesis}
}                                                        
\end{verbnobox}
\noindent
Alternatively for manual referencing:

\vspace{0.5cm}
\noindent
\fullcite{henson_hnrs_thesis_2014}

\chapter*{The First Measurement of the $\TO$ Tune-Out Wavelength in He*}
 \noindent
 \textbf{Bryce Maxwell Henson}\\
 \textit{\href{mailto:bryce.henson@live.com}{bryce.henson@live.com}}
 \\[0.4cm]
Atomic \& Molecular Physics Laboratories,\\
Research School of Physics \& Engineering,\\
College of Physical \& Mathematical Sciences,\\
Australian National University, Canberra, Australia\\[1cm]
 \textit{Supervisory committee:}
\begin{tabular}[t]{ l}
Dr Andrew Truscott  \\
Professor Kenneth Baldwin  \\
Dr Robert Dall  \\
\end{tabular}
 \\[1cm]
 \noindent
{\Large \textbf{Declaration}}\\
Except where acknowledged in the customary manner, the material 
presented in this thesis is, to the best of my knowledge, original and 
has not been submitted in whole or part for a degree in any 
university.

\vspace{10mm}  % vertical space

\hspace{80mm}\rule{70mm}{.15mm}\par   % horizontal space, line, start new line
\hspace{80mm} Bryce Maxwell Henson\par
\hspace{80mm} October, 2014

\newpage
{\let\cleardoublepage\relax \chapter*{Revisions}}
\section*{version 2 2017 }
This version has been modified from the original version, submitted in October 2014 as part of my Honours degree, to improve clarity and fix typographical errors. To see the version originally submitted please see v1 on \href{https://arxiv.org/abs/1708.08200}{arxiv}. If you are embarking on a related research topic please send me an \href{mailto:bryce.henson@live.com}{email}. I would like to thank Jacob Ross for help with a great deal of the corrections.

\section*{version 3 2022-01}
I have changed the notation of the tune-out from $2^{3}S_{1} \rightarrow 3^{3}P - 2^{3}P$ to the new notation  $\TO$ to match the use in our recent work [arXiv:2107.00149].

\section*{version 4 2022-01}
I have fixed a typo in \autoref{alingment} and added the citation page. The citations have been migrated to biblatex to give links for almost all references.

\chapter*{Abstract}
The workhorse of atomic physics, quantum electrodynamics, is one of the best tested theories in physics. However, recent discrepancies have shed doubt on its accuracy for complex atomic systems. To facilitate the development of the theory further we aim to measure transition dipole matrix elements of metastable helium (He*) (the ideal 3 body test-bed) to the highest accuracy thus far. We have undertaken a measurement of the `tune-out wavelength' which occurs when the contributions to the dynamic polarizability from all atomic transitions sum to zero; thus illuminating an atom with this wavelength of light then produces no net energy shift. This provides a strict constraint on the transition dipole matrix elements without the complication and inaccuracy of other methods.

Using a novel atom-laser based technique we have made the first measurement of the the tune-out wavelength in metastable helium between the  $3^{3}P_{1,2,3}$ and $2^{3}P_{1,2,3}$ states at 413.07(2)nm which compares well with the predicted value\cite{Mitroy2013} of 413.02(9). We have additionally developed many of the methods necessary to improve this measurement to the 100fm level of accuracy where it will form the most accurate determination of transition rate information ever made in He* and provide a stringent test for atomic QED simulations. We believe this measurement to be one of the most sensitive ever made of an optical dipole potential, able to detect changes in potentials of $\sim200pK$ and is widely applicable to other species and areas of atom optics.
 \\[1cm]

% include statements effectively insert the contents of the named file.
% They are not necessary, but are useful for organising your work.
% They always start a  new page.

\include{thanks}  % include the contents of the file thanks.tex

\tableofcontents
%\listoffigures % makes a list of figures
%\listoftables  % uncomment this if you have tables

\pagenumbering{arabic} % switch to Arabic numerals for page numbers
\setcounter{page}{1}  % set page number to 1
\setcounter{chapter}{0}

\include{introduction}

\include{chapter1}

\include{Chapter2}

\include{chapter3}

\include{chapter4}

\include{chapter5}

%refrences
\printbibliography

\appendix 
\include{appendix}

\end{document}

%% file: thanks.tex
\chapter*{Acknowledgements}
Dedicated to the late professor Jim Mitroy who provided the inspiration and invaluable guidance for this project. 
This project would never have been possible without the fantastic team I have been lucky enough to work with, all of which deserve my deepest thanks.
To Dr Andrew Truscott who has been the driving force behind this experiment as the head of the He* BEC group. His deep physical insight and rigorous application of the scientific method has been invaluable in working with such a complicated experiment. 
To Dr Rob Dall whose experimental knowledge, time and patience for his students is seemingly unlimited. His enthusiasm in the lab made for a fun and engaging experience of even in the most dire of times.
To Professor Kenneth Baldwin who provided a clarity of purpose and action in the experiment. Involving me in the writing of a ARC discovery project proposal has been a invaluable experience into the mechanics of research.
To Roman Khakimov who has been a pleasure to work alongside. He has taught me patience and thoroughness in the face of an unforgiving experiment.
To Colin Dedman whose electrical knowledge is unmatched giving the term `The Dedman Limit' for a devices performance after visiting his workshop. His collaboration on the temperature stabilized enclosure has been outstanding and has provided us with an environment devoid of drifts and my first publication (see \cite{0957-0233-26-2-027002} ).
To the technical staff Ross Tranter and Stephen Battisson who are always happy to help and are able to turn a complex idea into reality with a quick chat.
To my previous supervisors without whom I would never have made it here:
Dr Mark Baker who showed me the wondrous world of atom optics and in so doing taught me more than I could think possible. 
Doctor Michael Bromley who taught me how to program properly and showed me the power of the NLSE.
To my parents who have had continuous love and support for me in all my endeavours.
To my year 11 and 12 physics teacher Dr Peter Bancroft who motivated me to pursue physics as a career through his wonderful teaching style and cool ties.
And finally Bill `the science guy' Nye who created a deep childhood interest in science and the natural world through his wonderful and often wacky TV show.

%% file: introduction.tex
\chapter{Introduction}
%[!!!Grading Goals]\\
%--to demonstrate that you can succinctly phrase your research and engage reader\\

%[!!! outline ]\\
%brief overview of the project\\
%---why its important\\
%---the structure of the thesis\\

\section{QED-`The Jewel of Physics'}
%!!!--- maybe change wording of this first sentence to not match abstract\\
%!!--motivate and explain the tune-out better\\
The workhorse of modern atomic physics-\textbf{q}uantum \textbf{e}lectro\textbf{d}ynamics (QED)-is one of the most widely tested theories in physics, producing predictions that are remarkably accurate. Based on early research by Paul Dirac in 1920 to understand the dynamics of absorption and emission in atomic systems, it was later expanded to unite quantum mechanics with special relativity to describe in detail the interaction of light and electrically charged particles. Although initial calculations were at odds with experiment \cite{Lamb1947}, the work of Hans Bethe in 1947 paved the way to producing accurate predictions of experiments through an ingenious process, known as renormalization. Following decades of work, J. Schwinger, R. Feynman, and S. Tomonaga were awarded the 1965 Nobel Prize in physics for producing a self-consistent theory that allowed observables to be rigorously calculated. This spurred a metrology race between theory and experiment to produce more accurate predictions and measurements of various QED phenomena that has continued to the present day.

 Excellent agreement with QED theory has been demonstrated for a number of phenomena incuding the anomalous magnetic dipole moment \cite{Hanneke2011}, Lamb shift \cite{Nebel2012}, positronium lifetime \cite{Odom2006}, and measurements in our own group of metastable helium\footnote{A highly exited state of helium with a long lifetime, see \autoref{metastablehe}.} atomic lifetimes \cite{Hodgman2009}. Advances in QED not only provide utility in atomic physics; using the template developed in QED, \textbf{q}uantum \textbf{c}hromo\textbf{d}ynamics (QCD) was developed, describing the interactions between quarks and gluons through the strong force. Able to provide accurate predictions in atomic regimes from chemistry and biology \cite{Engel2007} to astrophysics \cite{Pickering2011} and star evolution\cite{Haseeb2014}, it is an invaluable scientific tool. Such was the success of QED that Richard Feynman gave it the title `the jewel of physics'. 
 
In the quest to test QED at new extremes the `proton size puzzle' emerged.  Here the size of the proton measured from muonic hydrogen spectroscopy \cite{Antognini2013,Antognini2011} disagrees significantly (7 standard deviations)\footnote{ One in $2.5 \cdot 10^{12}$ chance of being a purely statistical fluctuation} with that measured in traditional electron scattering experiments. In muonic spectroscopy the electron in hydrogen is replaced with a muon (an elementary particle with the same charge as an electron but a mass ~200 times heavier), causing a shift in the energy levels of the atom which can be used to measure the size of the proton. This discrepancy with the prediction of QED has as yet defied any convincing explanation, and has inspired a new metrology race to test QED in more complex systems \cite{Karshenboim2005,Drake2008}. In particular helium, which provides an ideal 3-body testbed \cite{Mitroy2013} where advanced spectroscopy and future muonic tests \cite{Nebel2012} hope to illuminate the problem further. 

\section{Measurement}

While helium energy level measurements performed in the past to test QED predictions have come to agreement at fractional uncertainty of $10^{-9}$  \cite{CancioPastor2012}, both theory and experiment have not exceeded far beyond $10^{-4}$ \cite{Schmidt2007} for transition matrix elements. These describe how the multiple energy levels\footnote{Atomic energy levels will be used herein to refer to energy eigenstates.} of an atom interact, and are far more sensitive to subtle effects in QED (such as relativistic, compound nucleus and finite mass effects). In order to challenge QED further, we will attempt to constrain transition matrix elements of metastable helium to the highest accuracy thus far. By then comparing with high precision calculations, which can be done to far greater accuracy than in more complex atoms, we may put QED to a further stringent test. 

Traditionally these experimental values would have been derived from lifetime measurements of an excited atomic state. This approach is practically limited in accuracy to the $10^{-3}$ level \cite{Gomez2004}. Furthermore if multiple transitions exist then the relative strength of each (branching ratios) must also be determined, severely limiting the accuracy of the method. The energy level shift an atom experiences in an applied light field as a function of wavelength (AC Stark Shift) has also been used; however as the shift is in direct proportion to the light field intensity any uncertainty here directly translates to the measurement uncertainty and has so far been limited to the $10^{-3}$ level \cite{Sahoo2009}. 
 
To overcome this limitation we will measure the helium tune-out wavelength, which occurs when the energy level shifts (AC Stark Shifts) of the ground state from each atomic transitions sum to zero, thus illuminating an atom with this wavelength of light produces no net energy shift. Thus it provides substantial information on the dynamic polarizability of the atom, and in turn transition matrix elements with far less uncertainty than other methods. This is due to measuring a zero effect which only requires a stable light intensity, not accurate knowledge thereof. This accuracy in determining transition matrix elements makes it an excellent test of QED in atomic systems. Towards this end have measured the 413nm tune-out wavelength from the metastable $2^{3}S_{1}$ state to between the $ 3^{3}P$ and $ 2^{3}$P transitions, which is expected to be sensitive to finite mass and relativistic effects \cite{Mitroy2013}. A measurement here with an relative uncertainty of $2\cdot10^{-7}$ (100fm) would translate to the most precise measurement of a transition matrix element ever made for helium \cite{Mitroy2013}, and would challenge QED atomic structure theory at the $10^{-6}$ level.

\section{Experiment}
The process to measure a tune-out wavelength is relatively straight forward. First one must measure the small ground state energy level shifts experienced by the atoms in the presence of a light field around the tune-out wavelength. By then minimizing the measured shift by changing the wavelength of the applied light field we produce a measurement of the tune-out wavelength.  A key concern is the energy scale of these shifts are often more than ten orders of magnitude less than the thermal energy  of atoms at room temperature making a measurement all but impossible. To make a reasonable measurement then requires the coldest possible atoms. Our metastable helium (He*) Bose Einstein Condensate (BEC) facility is uniquely suited to this task. Here the atomic species of interest can be studied at temperatures of a few millionths of a degree kelvin allowing for a high sensitivity.

While measuring the He* tune-out wavelength to an accuracy where it will challenge QED atomic structure theory is a long term goal of the group, such a significant task is beyond the scope of an honours project. The goal of this project was to make the first measurement of the tune-out wavelength in metastable helium in order to develop techniques and expertise to eventually reach this long term goal. This first measurement will also serve to motivate further theoretical work to produce a more accurate prediction of the tune-out wavelength.
\section{Overview}
In \autoref{chBackground} we will examine the theory of QED, atomic polarizability and laser cooling in order to understand the motivation and methodology of such a measurement. Following this \autoref{Apparatus} gives a overview of the experiment. Then the additions to it needed for this work are detailed in \autoref{secmeasure}. In particular the implementation of a new laser system (\autoref{probe laser}), an alignment technique for small potentials (\autoref{sec:OpticsFocusing}) and a novel atom laser based potential measurement technique (\autoref{sec:theoryatomlaser}). This work culminates in \autoref{results} with the first measurement of the 413nm tune-out wavelength in metastable helium with a focus on possible errors (\autoref{toerror}). Finally we conclude by examining the implications of this measurement in \autoref{Conclusion} with particular attention on higher precision measurements to be undertaken in future work  (\autoref{FutureWork})

%% file: chapter1.tex
\chapter{Background Theory and Literature}
\label{chBackground}
%Grading Goals\\
%---show that you understand the motivation of your project\\
%---where your work fits into the larger field of research\\
%---critically analysed the background literature\\
%---that you understand the theory behind the experiment(qed,laser cooling)\\
%---give the reader enough knoledge to understand what and how i did things\\

\section{QED and Atomic Structure theory}
%[outline]\\
%-explain that aomic physics is well developed\\ 
%----motivated a lot of physics
%----used as a testing ground
%-give the reader enough understanding of atomic strucutre to continue 
%----quantum numbers spectroscopic format
%-explain that QED has improved atomic theory\\
%-unified with weak\\
%-signs the theory is incomplete

%-that it is well tested in simple systems\\
%-not as well tested in complex atomic systems\\
%----hard to do the theory
%----alkalis are intractible
%----muonic species\cite{Pohl2010}\cite{Antognini2013}
%-discrepancies have been demonstrated\\
%-how we will measure it
%	--precise measurment of the tune-out wavelength
%	--better than other measurement of levels ect
%	--3 body is far better for theory

\textbf{Q}uantum \textbf{e}lectro\textbf{d}ynamics (QED) is the description of charged particles and electromagnetism in a way that is consistent with quantum mechanics and relativity. While quantum mechanics gives the physics of the small and special relativity gives the physics of the very fast the two meet in the realm of atomic physics. The accurate description of which has motivated much of 21st century physics by providing a stringent testing ground for fundamental physical theories regarding light and matter. 

A great deal of progress has been made from the first predictive model of the atom, with each iteration adding another layer to our model to improve consistency with experimental results. The first of these was the Bohr model which dictated that electrons orbit the nucleus of an atom with angular momenta in integer multiples (n=1,2,3...) of the reduced Plank constant ($\hbar$)\cite{Bohr1913}. Next came quantum mechanics to explain the non-uniform spacing of these levels with non relativistic solutions to the Schr\"{o}dinger equation for a simple charged particle model \cite{debroglie1925,Schroedinger1926}
\begin{equation}
i \hbar \frac{\partial}{\partial t } \Psi=\hat{H}\Psi .
\end{equation} 
These solutions define the modern spectroscopic notation and the beginning of modern atomic physics where energy levels correspond to electronic wave-function solutions. In a manner analogous to the three body problem, we must now give up the the hope for exact solutions for all but the most basic (hydrogen) of systems.
 By using the previously discarded Dirac equation it was possible to produce a version of quantum mechanics that was Lorentz invariant and thus compatible with relativity.  When applied to atomic physics this relativistic formulation allowed the fine structure of an atom to be reproduced through inclusion of spin-orbit coupling, relativistic kinetic energy and the zitterbewegung of charges. 
 
However two problems remained. First if one used Fermi's golden rule to calculate the decay rate\footnote{Given for a initial state $i$ and final state $f$ by $T_{i\rightarrow f}=\frac{2\pi}{\hbar} |\bra{f}\hat{H'}\ket{i}|^2 \rho$ where $\hat{H'}$ is the interaction Hamiltonian and $\rho$ is the density of states \cite{dirac1927}.} of an exited state into the ground state the calculation would yield zero. This calculated infinite lifetime was directly contrary to observation.
	Secondly there was a small discrepancy between predictions and observations of the energy levels of a hydrogen atom. In particular the Dirac equation predicted that the $^{2}S_{1/2}$ and $^{2}P_{1/2}$ orbitals should have the same energy while experiments showed a small difference (about 1000MHz) \cite{Lamb1947}. These discrepancies were only resolved with a full description of the quantised nature of light and charged particles, in particular the interaction between them, known as a Quantum Field Theory which treats particles of light and matter as an excitation on an underlying physical field.
The first of this kind of theory was Dirac's work `Emission and Absorption of Radiation' \cite{dirac1927} where he used perturbation theory to solve for the interaction between a canonical quantised classical electromagnetic field and an atom. The true brilliance here was the ability to create and destroy a photon through emission or absorption. This was refined by Fermi to produce a theory that was able to describe a continuum of emission in a relativistic manner and hope was given to the idea that with a proper treatment one could unite quantum mechanics and relativity to explain the interaction of charged particles and light. However disaster struck when Bloch \cite{Bloch1937} showed that under higher orders in perturbation theory the self energy of an electron diverged to infinity.
Inspired by previous non-relativistic work, Tomonaga, Schwinder, Feynman, Dyson were able to realize that this obviously non-physical result was primarily a mistake in the interpretation of the theory\cite{Peskin1995}. The properties of particles measured in experiments are not the bare mass and charge but in fact the dressed parameters corresponding to the mass and charge after self and vacuum interactions to all orders. This process is known as on-shell renormalization  \cite{Peskin1995,youtubeqed}. Combined with a process of regularization to predict experimental parameters from calculations QED was now able to accurately predict reality.

By using the newly renormalized QED the problems of the Lamb shift and spontaneous emission were realized to be entirely due to interactions with the vacuum. This is best explained by the electric field of the vacuum having a vanishing expectation value $\expt{E}=0$ but a non zero variance $\expt{(\Delta E)^{2}}\neq0$  \cite{Yokoyama1995}. The development of QED has led directly to other areas of physics such as the creation of \textbf{q}uantum \textbf{c}hromo\textbf{d}ynamics (QCD) describing the interaction of quarks and gluons. Further developments also lead to the unification of electromagnetism and the weak interaction into a single electroweak force at large energies. The elegance of the theory and its unparalleled ability to produce meaningful experimental results has made QED the tool of atomic physicists to this day.

\subsection{Precision Tests}
 The success of QED in the 1970's led to a meteorology race in physics between the theoretical calculation and the experimental measurements of various parameters which has continued up to the current day \cite{Karshenboim2005}. As with any theory QED cannot predict parameters `ab-initio',  one must input measured physical constants in order to produce predictions or conversely use measurements to produce a measure of some physical constant. Thus one does not test the theory directly but the consistency of the predictions and results of experiments spanning a wide range of fields \cite{Drake2008}. As QED is not a theory that can be solved exactly any relation of the above must be calculated to some finite order in perturbation theory along with any approximations made giving a `theoretical' error. For simple systems modern mathematical techniques and computation can keep this well below the `physical' uncertainties for simple systems. However when complex systems are involved such as atoms of  Rb, Cs, K (often used in precision tests) the problem becomes far more difficult and more approximations must be made.
 
The most important physical constant in QED is the fine structure constant. While originally used in the Bohr model of the atom, in QED it gives the coupling constant between photons and electrons. It may be expressed as the combination of many physical constants however the most common is simply:
\begin{equation}
\alpha_{fine} =\frac{1}{4\pi\epsilon_0}\frac{e^{2}}{\hbar c}    .
\end{equation}
 To avoid confusion with the dynamic polarizability ($\alpha(\omega)$) later in this chapter the fine structure constant will always be addressed with the fine subscript. Its value is most often quoted as the reciprocal for historical reasons the CODTA \cite{Mohr2012} value of which is
\begin{equation}
 \alpha_{fine}^{-1} = 137.035\,999\,074(44).
 \footnote{Here brackets indicate the uncertainty, usual defined as the standard deviation, in the least significant digit.}
\end{equation}
with a fractional uncertainty of $0.32\cdot10^{-9}$. Its prevalence in QED means that at some order almost all processes depend on its value and often tests in simple systems are evaluated by comparing the derived fine structure constant. An accurate determination of this constant is a long-standing goal of experimental physics as it would allow for more precise theoretical predictions for complex systems. A excellent review of the current state of the art experiments for determining $ \alpha_{fine}$ can be found in  \cite{ZelevinskyPhd,Karshenboim2005}.

Currently the most precise value of the fine structure constant comes  from the Harvard group's  measurement of the anomalous magnetic dipole moment at a relative uncertainty in the derived fine structure constant $\alpha_{fine}^{-1} $ of  $0.37\cdot10^{-9}$ \cite{Odom2006,Hanneke2008,Hanneke2011}. Here a single electron is confined in a cryogenic penning trap, by measuring the ratio of spin and cyclotron frequencies an accurate determination of the magnetic moment and in turn the fine structure constant can be made. Further analysis of the Harvard result was performed by \cite{Aoyama2012} using high order QED and other contributions enabling the uncertainty to be reduced to $0.25\cdot10^{-9}$. While this measurement is sufficient for almost all applications we cannot rely on a single experiment as unknown systematic shifts could lead to a large error.

The field of atom optics is no stranger to these high precision tests, in fact the next best measurement comes from a determination of the recoil velocity of rubidium 87 at a relative uncertainty of $0.66\cdot10^{-9}$ \cite{Clade2006,Bouchendira2011}. This was achieved by accelerating atoms with many photons (500) of momentum ($\hbar k$) and measuring the resultant velocity. Then by expressing  $\alpha_{fine}$ in terms of the mass of an electron and a rubidium 87 atom an accurate measurement of the fine structure constant was performed.  It must be stressed that while involving a highly complicated atom this test does not depend in any way on the internal structure of the atom only its total mass.

Further tests of the fine structure constant have also been undertaken in a wide range of simple systems such as positronium lifetime \cite{Odom2006} to measurements in our own group of metastable helium lifetimes \cite{Hodgman2009}. Although these techniques fail to provide the precision of the above methods they give a good `sanity check'. The marked success here is the relative agreement of measurements from diverse methods which are entirely within their respective error bounds. This achievement is quickly overthrown in more complex atomic systems where there is a trend of disagreement in evidence.

Using the fine structure interval (energy difference of the $2^{3}P_{0,1,2}$ states) in metastable helium was proposed as a way to measure the fine structure constant with a low limiting uncertainty. When measured by \cite{ZelevinskyPhd,Zelevinsky2005} it was discovered that two theoretical models for translating the measured interval to a value for $\alpha_{fine}$ were both incorrect producing a significant disagreement with each other and established vales. Attempts to reconcile theory with experiment has as yet failed with one author going as far as to conclude that his theoretical predictions `...disagree significantly with the experimental values, indicating an outstanding problem in bound state QED.'\cite{Pachucki2006}. While such a dramatic conclusion should be met with a degree of scepticism it certainly warrants investigation.

%For more complex systems QED is often tested by comparing a theoretically predicted value using the fine structure constant found above to an experimentally measured parameter, as such it can be difficult to compare different measurements directly.

A further disagreement to emerge is known as the proton size puzzle. Here the radius of a proton as measured by spectroscopy of muonic hydrogen differs by 7 standard deviations\footnote{ One in $2.5 \cdot 10^{12}$ chance of being a purely statistical fluctuation.} \cite{Antognini2013,Pohl2010} from that measured in traditional proton-electron scattering experiments. A (negative) muon (denoted $\mu^{-}$) is a fundamental particle with the same charge as an electron but 200 times more massive, and similarly to the electron also has a positive antiparticle (denoted $\mu^{+}$). Both are unstable decaying after $2.2\mu s$ into an electron, electron antineutrino, and a muon neutrino. In muon spectroscopy the negative version of the muon is produced in a particle accelerator, slowed and then used to replace an electron in a atom. The energy levels of this muonic atom are then measured using high speed spectroscopy before the muon deays. Using this technique the Lamb shift in muonic hydrogen ($p \mu^{-}$) was measured and by using state of the art QED simulation the size of the proton was then extracted \cite{Antognini2011}. The large discrepancy between the muonic hydrogen spectroscopy and proton electron scattering measurements has as yet defied any convincing explanation. To shed light on the issue tests of muonic helium have been proposed \cite{Nebel2012} and will soon be carried out at the Paul-Scherrer-Institute in Switzerland \cite{Pohl2010}.

Experiments measuring energy levels in metastable helium have also shown inconsistencies. Two experiments measuring the size of the nucleus disagree significantly with each other and theory. Here the change in an energy level between isotopes ($^{3}He$ and $^{4}He$) is measured and used to then calculate the size difference of the nucleus. Two groups have used separate transition wavelength measurements to this same end; one \cite{Rooij2011} measuring the  $2^{3}S_{1} \rightarrow 2^{1}S_{0}$ transition at 1557nm  and the other  \cite{CancioPastor2012} measuring  the $2^{3}S_{1} \rightarrow 2^{3}P_n$ transition at  1083nm coming to a disagreeing at four standard deviations from one another.
 Attempts to shed light on this discrepancy by indirectly measuring the ionization energy through a forbidden transition ( $2^{3}S_{1} \rightarrow 2^{1}P_1$ 887nm) produced further disagreement with theory \cite{Notermans2014}.
 
These significant deviations of experiment from what is one of the best tested theories in physics is particularly worrisome. Two possibilities exist for this discrepancy \cite{Pachucki2006}.
\begin{enumerate}[label=(\alph*)]
\item QED is fundamentally incomplete and only approximates reality.
\item The techniques and approximations used in calculating parameters for many body systems are inaccurate.
\end{enumerate}
From the excellent agreement in simple systems (b) is far more likely to be the cause of these discrepancies. This is little solace as theoretical calculations attempt to predict the error of their approximations by $n^{th}$ order extrapolations \cite{Karshenboim2005} , thus unknown errors are likely present. To rectify this simulations must be done with the minimum of approximations and compared with some high accuracy measurement. This provide an invaluable guide to what is or is not valid and to develop the theory to produce accurate predictions in these systems. High precision tests of QED in atomic systems have predominantly concerned the energy levels of the atom \cite{Drake2008}, neglecting the transition dipole matrix elements that describe the motion of electrons between them. Such a test would form a critical comparison of predictions in many body QED atomic theory. 

%dont know if this hould be here
% However measuring some parameter in a complex atom system (such as $^{238}U}$) would provide little benefit as simulating it accurately with high order QED would take a unfeasable amount of computational resources. The natural choice is then the simplest atom that cannot be solved exactly, helium and as seen above this is the weapon of choice for most QED tests.

% to measurements in our own group of Metastable helium lifetimes ($10^{-2}$) \cite{Hodgman2009}.
%more tests that i dont think are needed
%mass spectrometery tests QED and CPT test \cite{Blaum2006}
%antihydrogen for parity violation \cite{Amole2014}
%dc polarizability of helium \cite{Schmidt2007}
%Lamb shift ($10^{-7}$) \cite{Nebel2012}
%parity nonconversion in barrium using light shift \cite{Sahoo2009}
%lifetime in rb \cite{Gomez2004} parity nonconversion 
%weak from rb using nonparity \cite{Gomez2004}
%measuremnt of the weak charge in cesium \cite{Vasilyev2002} then improved in \cite{Porsev2009}
\section{Atomic Polarizabilty}
\label{atompolz}

%show that polarizaibility is difficult theoreticaly all energy levels\\
%show that your measurment will help with this problem\\
%not testing fundamental qed but the implemnetation\\
%define criteria for a measurement technique\\

%optical dipole traps for neutral atoms \cite{Grimm2000}

%In order to test the predictions of QED to new extremes we will test the atomic struc

One of the most difficult parameters to theoretically and experimental determine in atomic physics are the transition dipole matrix elements. While the energy level structure of helium determined by experiment and theory is in agreement below the $10^{-9}$ level\cite{CancioPastor2012} the transition dipole matrix elements are only known at the $\sim 10^{-3}$ level \cite{Schmidt2007}. The latter is arguably more important as it measures what atomic transitions are possible and at what rate they occur. 

 To accurately predict these values one must calculate the electronic wave-function of each atomic state with all the complexity of QED and then take the overlap integral of two states with the dipole operator (or any other relevant transition operator). This is a difficult theoretical and computational task although having an accurate model here would have tremendous implication to resolving discrepancies of QED in many body systems.
  Helium is the ideal test-bed for such a simulation as the three particles present\footnote{For all but the most incredibly advanced simulations the structure of the nucleus is neglected giving ($2e +$neucleus) 3 body problem for helium \cite{Mitroy2010}.} allows for fewer approximations to be made, as would unavoidably need to be done for the far more complex alkali atoms\footnote{The time complexity of a many body problem scales as $O(N^2)$ making helium $2^{3}/2^{38} \sim 10^{10}$ times faster than Rubidium for a full QED  simulation.}. We aim to provide a stringent test to such a simulation by accurate measurement or constraints of the transition dipole matrix elements of helium. 
  
 Traditionally this would be done by populating some exited atomic state via an optical transition and then measuring the excited state population here as a function of time \cite{Gomez2004} giving the lifetime. While this gives the exited state lifetime, to calculate the transition dipole matrix elements the ratio of the decay into other states (or `branching ratio') must also be known. This is then determined by measurement of population in all possible decay channel levels as a function of time. The combination of many population measurements which are themselves limited in accuracy\footnote{In the best measurement of its kind  \cite{Gomez2004} the lifetimes of the $7S_{1/2}$ and 6p manifold of Rb were measured at relative uncertianty of $4.5\cdot10^{-3}$ and $10^{-2}$ respectively. } means this method has been practically limited to the $10^{-3}$ level of accuracy in transition dipole matrix elements \cite{Gomez2004}.
 
To go beyond this limit we can  measure the transition dipole matrix elements using the change in atomic structure in presence of an applied electric field. While an atom may initially have no electric dipole moment; under the influence of an electric field one may be induced, causing the atom to be repelled from or attracted to that field. This process is known as polarization, the strength of which is dependent on the transition dipole matrix elements of the atom.

In order to develop a high precision measurement using this polarizability we will examine a simple atomic model using first order perturbation theory as in \cite{Pethick2002}.
 We define the polarizability $\alpha$ as the degree of production of the electic dipole moment $\textbf{d}$ in response to an electric field $\bm{\varepsilon}$.
\begin{equation} 
\expt{\bm{d}}=\alpha \bm{\varepsilon}
\label{dipolemoment}
\end{equation}
%From a simple analysis of the static polarizability of hydrogen then a natural choice of units emerges  such that $\alpha_{cgs}=\frac{\alpha{si}}{4 \pi \epsilon_{0}}$ where $\alpha_{cgs}\sim a_{0}^{3}$ and is often quoted in multiples thereof.
The interaction Hamiltonian of some field with an atomic dipole is easily seen to be that of electron point charges with vector position $\bm{r}$ relative to the nucleus
\begin{equation} 
H'=-\bm{d}\cdot\bm{\varepsilon}=-e \sum\limits_{j} \bm{r}_{j}\cdot\bm{\varepsilon}.
\label{interaction haml}
\end{equation}
We can then approximate to a very good degree that the permanent dipole moment of an atom is negligible as only the parity violating weak force can cause atomic states to deviate from eigenstates of the parity operator. Given some periodic electric field with angular frequency $\omega$ applied in some direction with unit vector $\bm{v}$ and magnitude $\varepsilon_0$ the interaction Hamiltonian is simply
\begin{equation}
H'=-\bm{d}\cdot \bm{v} \varepsilon_0 sin(\omega t).
\end{equation}
Here we have neglected the quantization of light by assuming that the light field intensity is large compared to the photon energy. The original atomic state may be represented as combination of electronic eigenstates $u_n$ with expansion coefficients $a_{n}$ and energy $E_{n}$ represented by
\begin{equation}
\psi=\sum_n u_n a_n e^{-i E_n t/\hbar}.
\end{equation}
The state mixing under this interaction Hamiltonian is then governed by the  Schr\"{o}dinger equation%not sure if it should be nk or nm here
\begin{equation}
i\hbar \frac{\partial a_n}{\partial t}=\sum_k \bra{n}H'\ket{m} a_k(t) e^{i t (E_n-E_k)/\hbar} \quad.
\label{evolution}
\end{equation}
Given an initial pure state in state m such that $a_n=\delta_{nm}$ for $t=0$ we find the time dependence of $a_n $ where $n \neq m$ by integrating for some time with the Hamiltonian above
\begin{equation}
a_n=\frac{\bra{n}\bm{d}\cdot \bm{v}\ket{m}}{2\hbar}\left(    \frac{e^{i t((E_{n}-E_{k})/\hbar+\omega )}-1} {(E_{n}-E_{k})/\hbar +\omega}+ \frac{e^{i t((E_{n}-E_{k})/\hbar-\omega )}-1} {(E_{n}-E_{k})/\hbar -\omega}  \right).
\label{firstorder}
\end{equation} 
We then retrieve an expression for the initial state population $a_m $ by substituting \ref{firstorder} in to \ref{evolution}. To simplify the extraction of the shift in energy of the initial state $m$ we use $a_m=e^{-it\delta E/\hbar}$ and take the expectation value over one isolation period ($\delta E/\hbar$) in time
\begin{equation}
\expt{\Delta E} =-\frac{\varepsilon_0^{2}}{4\hbar}\sum_{n\neq m}\left(\frac{1}{(E_n-E_m)/\hbar+\omega}+\frac{1}{(E_n-E_m)/\hbar-\omega}\right)|\bra{n}\bm{d}\cdot \bm{v}\ket{m}|^{2}.
\label{big}
\end{equation} 
The term here of $\bra{n}\bm{d}\cdot \bm{v}\ket{m}$ describes possibility of a transition from stare m to n through an electric dipole transition and is known as a transition dipole matrix element. Comparing \autoref{big} with the form of the energy of an induced dipole in some electric field $\varepsilon$
\begin{equation}
\label{epolz}
\Delta E=-\frac{1}{2}\alpha \expt{ \varepsilon^{2}},
\end{equation}
we may then find the polarizability of the atom in state $m$ as a function of the frequency of the driving field
\begin{equation}
\alpha(\omega) =\sum_{n\neq m}\frac{2(E_n-E_m)}{(E_n-E_m)^{2}-(\hbar\omega)^{2}}|\bra{n}\bm{d}\cdot \bm{v}\ket{m}|^{2}.
\end{equation}
A shorthand is to define the oscillator strength as a more convent form to express the transition dipole matrix elements
\begin{equation}
f_{kl}^{\bm{i}}=\frac{2 m_{e} (E_{k}-E_{l})}{e^2 \hbar^2}|\bra{k}\bm{d}\cdot\bm{i}\ket{l}|^2.
\end{equation}
The values of these oscillator strengths are either predicted using a simulation of the electronic eigenstates of the atom or are measured in experiments. This gives a compact expression for the polarizability of an atom in state m with an applied oscillating electric field of frequency $\omega$. 
\begin{equation}
\label{dpolz}
\alpha(\omega) =\frac{e^{2}}{m_e}\sum_{n\neq m}\frac{f_{nm}^{\bm{v}}}{\omega^{2}_{nm}-\omega^{2}}.
\end{equation}

\begin{figure}
\begin{center}  % center environment centers the graphic on the page
  \includegraphics[height=7cm]{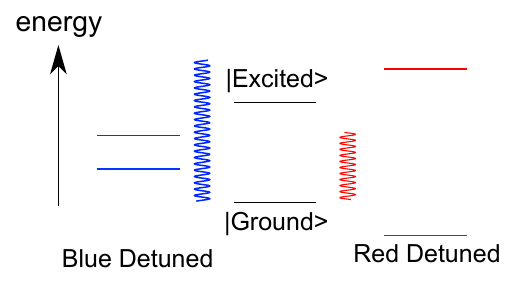}
  \end{center}
  \caption{The A.C Stark shift in a two level system under an applied light field with frequency less than (right) and greater than (left) the transition frequency.}
  \label{starkshift}
\end{figure}

This shift in the energy level of the atom is known as the Stark shift and when $\omega$ is comparable to the transition frequencies of an atom is known as the AC Stark shift \footnote{For static fields where $\omega=0$ this is known as the DC stark shift.}.  For a simple two level atomic system it can seen from \autoref{dpolz} and \autoref{epolz} that applying a electric field with frequency greater than the transition  between levels (blue detuned) produces an increase in the ground state energy while a lower frequency (red detuned) produces an decrease (see \autoref{starkshift}). In most atomic systems this oscillating electric field in the form of a coherent light field provided by a laser. The atom also experinces a shift in the excited state energy in the opposite direction (see \ref{starkshift}) due the now negative $(E_{n}-E_{m})$ term in the oscillator strength. The energy level shift of the atom then depends on which electronic state it occupies, for further treatment we assume it is in the ground state.

 By using a red-detuned laser beam an attractive potential can be made from the decrease in the ground state energy of the atom which may be used to confine atoms with a low temperature, known as a Optical Dipole Trap (ODT) or alternatively a Far Off-Resonance Trap (FORT) \cite{Grimm2000}. Neglected in the above theory is the excitation and subsequent emission of a photon that produces a scattering force which dominates the atoms behaviour near a transition, for this reason almost all dipole traps are operated with many THz of detuning from any transitions where the scattering effect is negligible. We will assume this condition for the rest of the section.

\begin{figure}
\begin{center}  % center environment centers the graphic on the page
  \includegraphics[height=9cm]{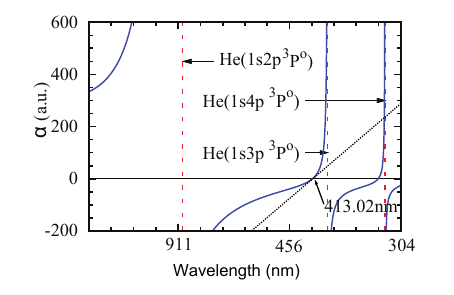}
  \end{center}
  \caption{The dynamic polarizability (blue) of He* with wavelength showing the 1083nm and 389nm transitions (vertical red lines) and the tune out wavelength between them. Also shown is the linear approximation of the potential about the tune out (black dashed). Modified from \cite{Mitroy2013}.}
  \label{dpolzhe}
\end{figure}

In reality atoms are never as simple as a two level system, instead the polarizability is comprised of the sum over all possible transitions (see \autoref{dpolz}). If the electric field frequency is in-between two transitions then the contributions from each transition are of opposite sign and the net polarizability is reduced.  An excellent example of this effect is seen in \autoref{dpolzhe}.
%When the field is at the frequency of a transition there is a infinite discontinuity in the polarizability.

If we aim to test predictions of the oscillator strengths (and in turn transition dipole matrix elements) then the most obvious approach would be to simply apply some probe beam and measure the energy level shift of the atom to give the polarizability $\alpha(\omega)$ as a function of $\omega_{probe}$ and fit for the oscillator strength. However, this suffers from serious practical limitations giving rise to significant errors, the greatest of which is the inability to independently measure intensity at the atomic sample, thus only the relative polarizability could be determined. Additionally producing a linear measurement of polarizability over many orders of magnitude is beyond the realm of most techniques. Finally lasers tend to have fairly limited frequency ranges compared to transition spacings, thus this measurement would require multiple laser systems and in turn be impracticably expensive. 

These limitations are overcome by measuring the point between two atomic transitions where the dynamic polarizability crosses zero, going from net repulsive to net attractive. If one measures the position of this crossing then the value is not affected by non-linearity, additionally the intensity only determines the slope of this crossing \footnote{To first order in perturbation theory as derived above this crossing is intensity independent, however to higher orders a small dependence arises known as Hyperpolarizability which will be negligible throughout this work.} so merely stability in the applied beam is needed. This is known as the tune-out wavelength\footnote{Also known as a Magic-zero wavelength, not to be confused with a `magic wavelength' used in atomic clocks.}, this is defined by the wavelength when contributions to the polarizability from all transitions sum to zero. Thus applying a laser with this frequency to the atoms would cause no potential shift and in turn no force `tuning out' the dipole interaction

\begin{figure}
\begin{center}  % center environment centers the graphic on the page
    \includegraphics[height=7cm]{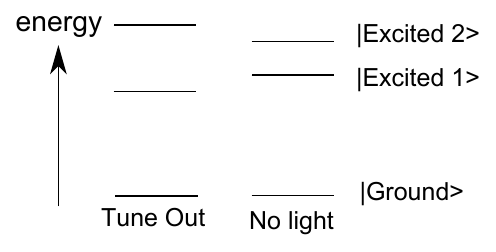}
  \end{center}
  \caption{The AC Stark shift in a three level with an applied field at the tune-out wavelength. The exited states are shifted although as there is no population in these levels the atom experiences no change in energy.}
  \label{tuneoutshift}
\end{figure}

\begin{equation}
0 =\frac{e^{2}}{m_e}\sum_{n\neq m}\frac{f_{nm}^i}{\omega^{2}_{nm}-\omega^{2}}.
\end{equation}

The behaviour about this zero can then approximated by a first order Taylor expansion as seen in \autoref{dpolzhe}. While zero does not provide a direct measurement of the transition dipole matrix elements it provides a constraint that can be very far more accurately determined. By combining multiple tune-out measurements a multi point constraint can then be put on any predictions of the transition matrix elements from QED simulations \cite{Mitroy2010}. Additionally in a system where two transitions dominate the polarizability it can be used to determine the relative strengths of the transitions far more accurately than other methods \cite{Cheng2013}. 
The name tune-out comes from the original proposal for a multi-species optical lattices where one can create a potential that would  `tune-out' for one species due to the vanishing polarizability \cite{LeBlanc2007} thus allowing indepent manipulataion. Tune-out wavelengths were quickly identified as a high precision test of atomic physics. The most accurate measurement to date is by \cite{Holmgren2012,HolmgrenPhd} using an atomic beam interferometer to determine a tune-out wavelength for potassium atoms at 768.9712(15)nm . To  perform a tune-out measurement a probe beam is applied to atoms of interest and the resulting energy shift measured, the wavelength of the probe beam is then varied a small distance around this tune-out wavelength and a linear regression used to find the zero point. In most measurements it is the sensitivity to this energy shift that is the liming factor in measurement and as this varies between tune outs distinct measurements cannot be easily compared.

Further all tune-out wavelength measurements are not equally valuable for testing QED, some are well determined by simple oscillator strength ratios while others are sensitive to hard-to-model relativistic corrections and vacuum interactions.  To this end J. Mitroy and L. Tang \cite{Mitroy2013} have produced a calculation of the tune-out wavelengths in helium which shows that the tune-out between the $3^{3}P_{1,2,3}$ and $2^{3}P_{1,2,3}$ from the $2^{3}S_{1}$ ground state at 413.02(9)nm (see \autoref{dpolzhe}) can provide an excellent test of these more interesting phenomena. Additionally they set out a criteria that a measurement here at the 100fm\footnote{Some common units for wavelength : 1 nanometer=$1nm=10^{-9}m$, 1 picometer= $1pm=10^{-12}m$, 1 femtometer = $1fm= 10^{-15}m$ .} level of accuracy `... would constitute the most precise measurement of transition
rate information ever made for helium'. The gradient of the polarizability with detuning about the tune-out wavelength is also given such that we can determine the potential sensitivity needed to correspond to a given wavelength uncertainty. Converted into SI units for potential this is given as:
\begin{equation}
\frac{U(\omega)}{\Delta \lambda}=5.9*10^{-39} \frac{J m^{2}}{nm_{TO} W}=2.8*10^{-16} \frac{K m^{2}}{nm_{TO} W}
\label{potential}
\end{equation}
the units are energy (joules on the left and kelvin on the right)\footnote{Here we have found the temperature of a non interacting gas that would be needed to escape this potential  $E=\frac{3}{2} k_{b} T$. This forms a convenient comparison of energy scales with ultra cold atoms.}  per (intensity times detuning in nanometers). It should be noted that here a beam red-detuned from the tune out is repulsive while a blue-detuned one is attractive opposite to the case about a transition. To provide a frame of reference for this potential an irradiance of $\sim 10^8 W m^{-2}$ is a reasonable for a focused beam, at 1pm detuning (a tenth of the accuracy in the criteria above) this would give a potential of a mere $28pK$. By any standard this is a remarkably small potential to measure with an experimental apparatus. To provide any hope of detection we require some of the coldest atoms in the universe, a million times colder than interstellar space.

%predicted for Be,Mg,Ca,Sr \cite{Cheng2013}
%for K \cite{Jiang2013}
%polarizabiity in cesium \cite{LeKien2013}
%Review by mitroy \cite{Mitroy2010}

\section{Ultra-Cold Atoms} 
In our metastable helium (He*) Bose Einstein condensate (BEC) apparatus we routinely produce ultra cold atomic clouds below $1\mu K$ by combining laser and evaporative cooling. This allows the minuscule potentials described above to be detected through more comparable energy scales. The near elimination of (the otherwise significant $\sim pm$) Doppler broadening  and the high density ($\sim 10^{21} m^{-3} $)/purity$(\approx1$) of metastable helium in our apparatus make it well suited to performing this measurement. These atoms are cooled to such extremes in order to reach a unique state of matter known as a Bose Einstein condensate the study which is an active field of research in and of itself. Four our purposes it forms the ideal measurement tool though its thermal statistics and coherence. In order to understand the apparatus used we will briefly examine the theory of laser cooling and BEC's. Futher detail can be found in \cite{Pethick2002,Metcalf,Phillips1998}. 

%talk about moticated for bec

\section{Laser Cooling}
\label{lasercooling}

%[outline]\\
%-how we get ultra-cold atoms (motivated by quest for bec)\\
%---Doppler cooling\\
%---explain how there is subdoppler cooling but we cant use it...\\
%---zeeman slower\\
%---MOT with beam description\\
%---magnetic trap\\
%-----in trap cooling is it just molasses?\\
%---RF evaporation\\
%nobel lecture on laser cooling 

Reaching ultra-cold ($\sim \mu K$) temperatures would not be possible without laser cooling which can decrease the temperature of a gas of atoms orders of magnitude with a relatively simple laser system. While in the previous section we have focused on the dipole force, the scattering force provides the basis for laser cooling that we rely on to produce ultracold atomic clouds necessary to create a BEC. This can be understood by the momentum transfer that occurs when an atom absorbs a photon, moving from its ground state to some exited state, from a resonant light field which is equal to $\bf{\Delta p}=\hbar \bf{k}$. The atom will then decay to the ground state after some time radiating a photon in a random direction once again changing is momentum although now in a random direction. Thus the average momentum transfer over many absorption-emission cycles is equal to that of the incident photon. This process is known as photon scattering and becomes a dominant force on the atoms when in a light field near a transition.

\subsection{Optical Molasses Cooling}
The simplest form of laser cooling is that of optical molasses cooling where two counter propagating laser beams with a frequency less than an atomic transition are applied to a dilute atomic gas. For a stationary atom both beams are off resonance and have an equally low chance of scattering and thus no net momentum in transferred to the atom over many photon absorption and emissions\cite{Phillips1998}. However if the atom is moving with some small velocity along the axis of the beams then the co-propagating beam is red-shifted further off resonance by the Doppler effect $-v/\lambda$ while the counter-propagating one is blue-shifted onto it  $v/\lambda$, giving an unequal scattering probability from the two beams \cite{Pethick2002}. This then produces a net momentum transfer countering the direction of motion in proportion to there velocity along the beams, cooling the atoms through a `frictional' force. As the atoms must be shifted onto resonance with the Doppler shift this only produces cooling for a small subset of the velocity distribution. However by starting with a large detuning from resonance to `catch' high temperature atoms and gradually reducing  the detuning with time the entire temperature distribution may be 'compressed' to a low temperature. However this method does not allow cooling indefinitely as the random emission of a photon each cycle produces a finite heating rate.  This heating process is in direct competition with the velocity dependent damping force and places a lower limit on the temperature that molasses cooling can achieve given by 
\begin{equation}
T_{Doppler}=\frac{h\gamma}{2 k_B}.
\end{equation}
This is known as the Doppler cooling limit\footnote{For metastable helium cooled on the 1083nm transition as described later in this work this limit is  $\sim 39 \mu K$ \cite{Baldwin2005}.} and is reached when the detuning of the two beams is half the natural linewidth of the transition (given by $\gamma$). By applying a pair of optical Molasses beams in three orthogonal axes cooling can then  be provided for each motional degree of freedom and total temperature of a gas can be quickly reduced. This is known as a 3D optical molasses and was one of the first laser cooling systems.
\subsection{Zeeman Slower}
The small velocity subset that can be cooled at any one time in optical molasses is prohibitive for cooling hot, continuous atomic beams with large axial velocity distributions as produced in our source of He*. To cool the entire subset simultaneously it is possible to compensate for the Doppler shift with a magnetic field using the Zeeman effect. When an atom is subject to a magnetic field the magnetic hyperfine sub-levels are shifted by $\Delta \omega_{Zeeman} =( \bm{\mu} \cdot \bm{B} )/\hbar$. By using circularly polarized light only the $\Delta m_{f}=\pm1$ transition is allowed through conservation of angular momentum.  Using this shift it is possible to compensate for the Doppler shift ($v \omega_{las}/c $) of cooled of atoms with a spatial dependent magnetic field keeping the atoms shifted into resonance ($\omega_{res}$) and continuing to cool.
\begin{equation}
\frac{ \bm{\mu} \cdot \bm{B} }{\hbar}- \frac{v \omega_{las}}{c} =\omega_{las}-\omega_{res}
\end{equation}

A Zeeman slower starts with a field and detuning such that some maximum velocity (capture velocity) atoms are on resonance at the beginning of the slower. The magnetic field , produced by a specially designed electromagnet, then changes as a function of length in order to keep these atoms on resonance and cooling even after slowing. Atoms that are slower than this capture velocity at the beginning proceed along the length until reaching a position where they are shifted onto resonance, after which cooling like above. By using a field profile such that the maximum capture velocity atoms experience a constant deceleration the length of the slower is minimized. A Zeeman slower can reduce a high velocity($\sim1000 s^{-1}$) atomic beam down to a mere $\sim80m s^{-1}$.
\subsection{Magneto Optical Trap}
While the resultant gas may be cooled with a 3D molasses cooling (described above) the random walk of the atoms over many cooling cycles eventually moves them out of the range of the laser beams providing the cooling and are subsequently lost from the process. To confine the atoms within the beams to reduce this loss we require a velocity dependent damping force (as in Molasses cooling) and and a position dependent restoring force, this is the principle of a Magneto Optical Trap (MOT). Using the Zeeman shift in a similar way to above we start with two opposing electromagnets (anti-hemholtz) which create a linear dependence on magnetic field strength with position in each axis which crosses zero in the center of the trap.

\begin{figure}[H]
\begin{center}  % center environment centers the graphic on the page
    \includegraphics[height=7cm]{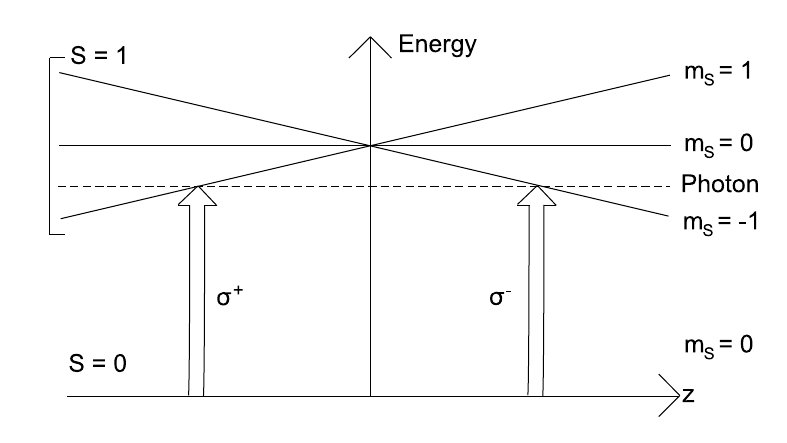}
  \end{center}
  \caption{Energy level diagram for a Magneto Optical Trap \cite{ManningHnrs}}
   \label{motenergy}
\end{figure}

 To simplify matters we will first consider a single axis, ignoring for a moment the motion of atoms. Here two red detuned counter propagating laser beams are applied with opposite circular polarizations. In opposite positions about the magnetic zero the Zeeman sub-levels of atoms are shifted onto resonance with the beam. Using circularly polarized  light (denoted $\sigma^{+}$ and $\sigma^{-}$) means that only one of these resonances is allowed for each beam through conservation of angular momentum. By choosing the polarization such at that this allowed resonances occurs for the beam pointing inwards a restoring force towards the ceneter is achieved (see \autoref{motenergy}). The motion of the atoms produces a Doppler shift as in the Molasses although it is now the combination with Zeeman shift that satisfies resonance 
\cite{Metcalf}. To this end atoms away from the center require a smaller Doppler shift (and corresponding velocity) to reach resonance and are pushed towards the center. Applying this technique in each axis produce a 3D MOT which is able to confine and cool atoms reaching high atom numbers. Often only single beam in each axis is input with the counter-propagating beam produced by passing the beam through a quarter wave plate and reflecting it back over the input with a mirror, reversing both direction and polarization. By modifying either the detuning of the beams or the magnetic field strength the density of the MOT can be controlled, often starting with lower density warm clouds and changing parameters to end with a high density cold cloud.
 
\section{Metastable Helium}
\label{metastablehe}
The large energy transition between the ground state and any excited states in helium means that the aforementioned laser cooling techniques are impractical with ground state helium as an XUV laser would be needed. To get around this we use a highly excited state (approximately 20eV) of helium which when undisturbed has a lifetime of 7870s \cite{Hodgman2009} owing to a doubly forbidden transition to the ground state. An atom in this state is known as metastable helium (He*). This long lived excited state forms an effective ground state which allows the use of widely available fibre lasers at 1083nm for laser cooling. Its low mass (4 u), low saturation power($0.17 mw\cdot cm^{-2}$) and high magnetic moment makes it is particularly suited for laser cooling and magnetic trapping.

\begin{figure}[H]
\begin{center}  % center environment centers the graphic on the page
    \includegraphics[height=9cm]{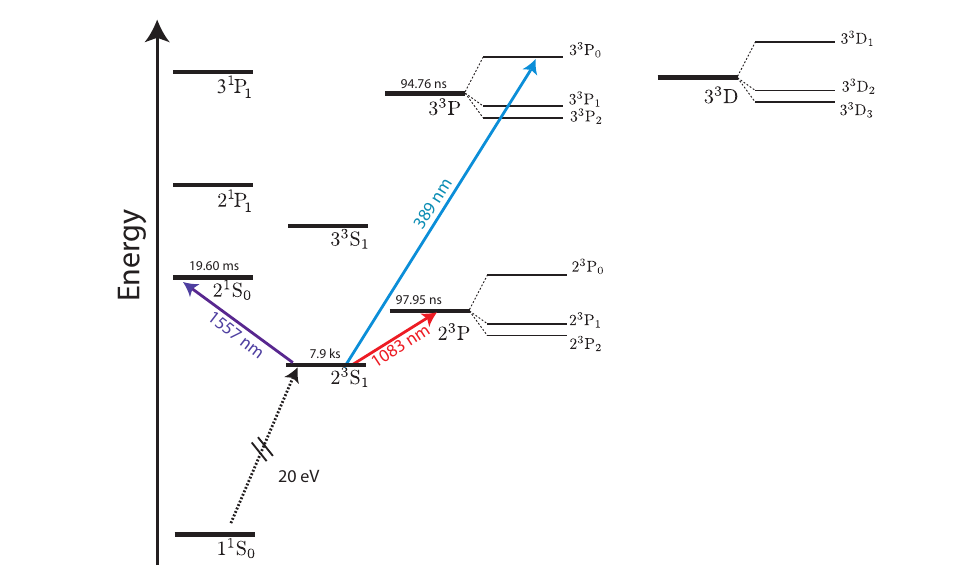}
  \end{center}
  \caption{The Energy level structure of He*, of note are the $3^{3}P$ and $2^{3}P$ levels which are the dominant contributions to the tune-out wavelength. From \cite{SimonetPhd}, energy levels not to scale.}
\end{figure}

  While long lived in isolation a process known as penning ionization can allow He* to exchange its electron with a nearby object for one with the opposite spin removing the spin flip requirement for a transition to the true ground state. The transition then proceeds rapidly and the energy is released as kinetic energy of the original electron (minus the ionization energy). This process means that any collisions of He* with He*, a neutral background gas or liberated electron is a dominant loss mechanism and must be avoided at all costs. 
\begin{equation}
He^{*}_{\downarrow}+He^{*}_{\uparrow} \longrightarrow He+He^{+}_{\uparrow}+e^{-}_{\downarrow}+\sim20eV-E_i
\end{equation}
The two body loss rate due to He* -He* collisions would then scale as the density of the gas squared and prevent attainment of a BEC. We reduce this loss in our experiment (where density is highest) by spin polarizing the atoms which prevents the above reaction by spin conservation rules
\begin{equation}
He^{*}_{\uparrow}+He^{*}_{\uparrow} \longrightarrow He^{*}_{\uparrow}+He^{*}_{\uparrow}.
\end{equation}
For such a collision to result in loss one the incident atoms would be required to transition to the opposite spin which for sufficient magnetic fields is improbable enough to suppress the reaction many orders of magnitude \cite{Dall2007}.
While at times a inconvenience penning ionization provides a relatively simple way of detecting single atoms. When incident on a metallic surface the atom undergoes a similar process known as Auger de-excitation which again releases an electron and $\sim20ev$ although now reduced by the work function of the material.
\begin{equation}
He^{*}_{\downarrow}+S_{\uparrow} \longrightarrow He_{\uparrow}+S^{+}+e^{-}_{\downarrow}+\sim20eV-W
\end{equation}
Here W denotes the work function of the material. This process allows for relative ease of detection of the atoms with high spatial and temporal resolution by using the ejected electron. This is the primary motivation for pursuing BEC with He*.

\section{Bose Einstein Condensation}
\label{bec}

A Bose Einstein Condensate is a unique state of matter characterized by macroscopic particle occupation of the same quantum state. Based on his work with Bose Einstein showed \cite{Pethick2002} that if a collection of bosons were cooled below a critical temperature($\sim100nK$) the particles would undergo a phase transformation and `condense' into the ground state of the system. This is a consequence of the indistinguishability of particles and the symmetric wave function of bosons. As this wave function consists of many particles in the same quantum state the fascinating world of quantum mechanics can then be probed directly in experiments.

 Initially it was thought that this state would be impossible to achieve as any atoms would liquefy or solidify well before the critical temperature was reached, halting further cooling relegating it to a theoretical curiosity \cite{Ketterle}. However it was proposed that by using a dilute gas of atoms in a vacuum cooling could continue well below this point as the three body collision necessary to form a `liquid' would be sufficiently suppressed. This regime is known as a weakly interacting dilute gas. This was the approach taken by Cornell and Wieman at the University of Colorado who used laser and evaporative cooling of rubidium to achieve the first BEC. This was quickly followed by Ketterle, et al. at MIT with all 3 going on to win the 2001 Nobel Prize in Physics. BEC's quickly became a prime area of physics research as they allow for otherwise impossible experiments into quantum mechanics.  Additionally BEC's offer a number of advantages in a diverse range of high precision meteorology experiments, from gravity measurements using atomic interferometers \cite{Hardman2014a} to measurements of the Casimir-Polder force using centre of mass motion of a condensate \cite{Obrecht2007}. For the measurement presented in this work we continue in this trend by using the out-coupling rate of an atom laser to measure an optical dipole potential. 
 
The full treatment of a interacting many body quantum system such as a BEC is intractably complex as the interaction between each pair of particles must be considered. By approximating the sum of these interaction as effective mean-field we can produce a reasonable description of the macroscopic behavior of the many body wave-function of a gas of weaky interacting bosons. This approach gives the time independent Gross-Pitaevskii equation for a dilute gas of particles with mass m, scattering length $a_s$, chemical potential $\mu$ in an applied potential V as
\begin{equation}
\left(\frac{\hbar^{2}}{2 m} \frac{\partial}{\partial \mathbf{r}^2} +V(\mathbf{r}) +\frac{4\pi \hbar^2 a_s}{m}|\Psi(\mathbf{r})|^2\right)\Psi(\mathbf{r})= \mu  \Psi(\mathbf{r}) \quad .
\end{equation}
This chemical potential is determined by normalization of the many body wavefunction to the number of particles present.
\begin{equation}
N=\int d \mathbf{r} |\Psi(\mathbf{r})|^2
\end{equation}
Its value can be interpreted as the potential energy needed to add a particle to the system.
This approximation can be further simplified by neglecting the kinetic energy of the system giving the Thomas-Fermi approximation for the density of a BEC
\begin{equation}
\rho(\mathbf{r})=|\Psi(\mathbf{r})|^2=\frac{m}{4\pi \hbar^2 a_s} (\mu -V(\mathbf{r})) \quad .
\end{equation}
This approximation is only valid when $\mu>V$ and must be zero elsewhere. As the kinetic energy has been neglected in this approach the approximation is only reasonable for weakly confined atoms at potential less than the kinetic energy. It however gives a useful approximation of the density of a cloud in a potential which can be seen to `fill it up' as if a liquid in a container. We use this approximation later in \autoref{sec:theoryatomlaser} to calculate the effect of a potential on a BEC.

\begin{figure}[H]
\begin{center}  % center environment centers the graphic on the page
    \includegraphics[height=4cm]{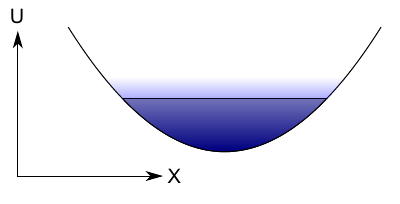}
  \end{center}
  \caption{The Thomas Fermi approximation of the density of a weakly interacting gas in a potential predicts that the potential is `filled' to the chemical potential.}
\end{figure}

The next chapter will describe the experimental system we used to generate such a BEC of metastable helium atoms.

%\begin{equation}
%H=\sum
%\end{equation}
 %However to aproximate the behavior 
%review of use \cite{Baldwin2005} \cite{Vassen2012}
%\section{Magnetic Traps}
%\label{magnetic trap}
%To create a BEC we cannot rely on laser cooling as minimum achievable temperature is orders of magnitude larger than the critical temperature. To this end evaporative cooling is used where the hottest atoms in a conservative trap are selectively removed, the remaining atoms then rethermalize through collisions to a lower average kinetic energy. In this experiment we utilize a 
%a conservative potential is used and the hotest 
%[outline]\\
%conservative\\
%stable\\
%rf outcoupling\\

%% file: chapter2.tex
\chapter{Apparatus}
\label{Apparatus}
%Grading Goals\\
%---show that i understand how the experiment works \\
%---explain enough to the reader so they understand what your doing\\
%---demonstrate understanding of the limitations of the experiment\\
%---show that this is a complex experiment that is a big ask to accomplish by individual
%-applying the above tecniques to produce a BEC of metastable helium is no small feat\\
%first done by []\\
%-realtively large and complicated to RB\\
%-current experment designed for studying correlation and quantum stuff\\
%-culmination of a decades work by truscot dall baldwin, hodgman,manning, hackimov, and many more.\\
%-able to produce 100katoms at 1uk in bec with amazing purity\\
%-takes 25~ seonds to make a bec known as a 'shot' or run\\
%-inside heppa hood to keep out dust and regulate temprature to the 100mk level\\

Applying the techniques discussed in the preceding chapter to create a BEC of He* is not a simple task. A number of difficulties unique to He* are involved, primarily the inefficiency of (He*) production resulting in a high gas load combined with Penning ionization creating a high loss rate. The work of Aspect\cite{Robert2001} and Cohen-Tannoudji \cite{PereiraDosSantos2001}  gave the the first BEC of He* in 2002. A decade of work by the ANU He* BEC team has produced an apparatus that can reliably make a large clouds ($\sim10^6$ atoms) of ultra cold ($\sim10^{-6} K$) metastable helium in a extremely high vacuum ($\sim10^{-10}$ pascals) environment every 30 seconds.

\begin{figure}[H]
  \begin{center}  % center environment centers the graphic on the page
    \includegraphics[height=10cm]{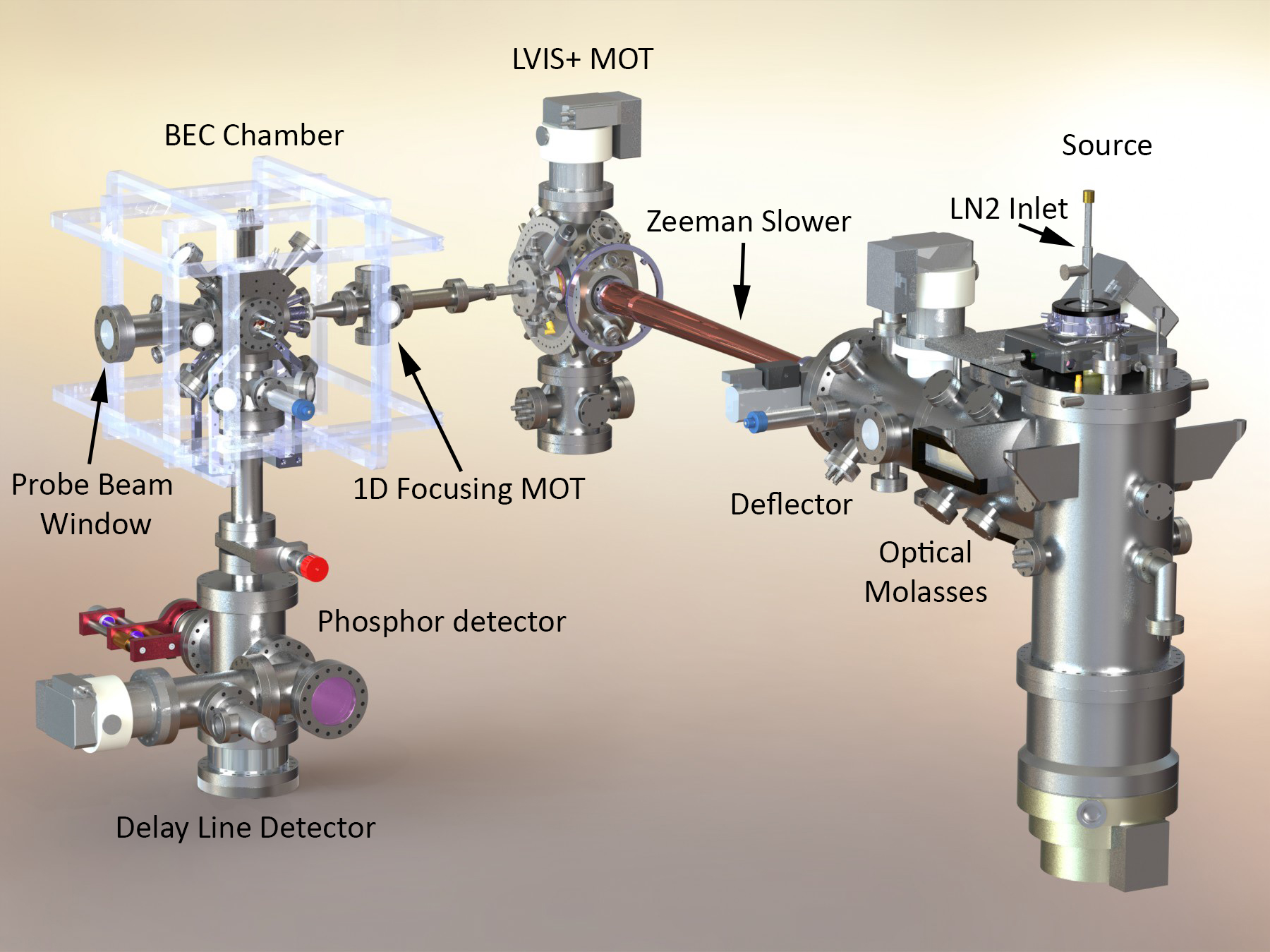}
  \end{center}
\caption{Schematic of the vacuum chamber that forms the heart of the Experimental Apparatus}
\label{ExpRig}
\end{figure}
This is a complicated apparatus which consists of many stages to produce the final result. While a detailed understanding mechanics of each stage is not necessary for this work, it is necessary to understand how the experiment operates to understand the subsequent advantages and limitations on later techniques. For further detail see \cite{ManningPhd,Baldwin2005,Dall2007,Swansson2004,Swansson2006,Hodgman2009,Dedman2007,RuGway2013}.

\section{Metastable helium DC discharge Source}	

To create a He* BEC we require a high density source of cold metastable helium with the minimum of unexcited atoms in a low pressure environment. We start with a helium gas reservoir at $10^3$ pascals which is cooled with liquid nitrogen to reduce the kinetic energy of the atoms. We then apply an 2kV electric discharge to this gas from this reservoir through an insulating aperture to a tungsten anode in a low pressure region at $10^{-5}$ pascals. The free electrons in this plasma are accelerated by the electric field and are able to excite helium to the $2^{3}S_{1}$ in a low pressure environment (past the aperture) where loss due to background collisions are  reduced \cite{Swansson2004}.
The cross section of other excitation  processes in the discharge such as non triplet excitation and ionization are far larger resulting in approximately $10^{-5}$ of the atoms in the desired state  $2^{3}S_{1}$.
The resultant mixture of He* and other He states then passes through a skimmer with provides a large pressure differential ($10^3$) between the source and the collimator while simultaneously selecting only the atoms with a small transverse velocity. Here the beam is collimated by a 2D optical molasses stage which further reduces the transverse velocity. The absence of interaction with the applied light fields reduces the transmittance of unwanted excited He states through an aperture at the end of the collimation stage. To further enhance differential pumping the collimated beam of  $2^{3}S_{1}$ He is optically deflected by a second molasses stage orentated at a few degrees to the original beam and then transmitted through an aperture to the Zeeman slower. The use of three differential pumping stages each with their own turbo-molecular vacuum pump results in the production of $10^{11} s^{-1}$ He* atoms at a background pressure of $10^{-7}$ pascals in the Zeeman slower.

\section{Zeeman Slower}

While the transverse velocity of the atomic beam from the source is relatively low the longitudinal velocity is approximately 700m/s \cite{Swansson2006} and must be reduced to the capture velocity of the first MOT which is an order of magnitude less. To provide such dramatic cooling we use a Zeeman slower as described in section \ref{lasercooling} which cools the atoms to approximately 80m/s which are then captured in the first MOT. 

\section{MOT}
\label{apparatusmot}

While the atoms at this point have passed through 3 differential pumping stages the pressure at the end of the Zeeman slower is   is still too high ($10^{-7}$ pascals) to produce a magnetic trap with a long lifetime. Additionally, to produce high stability trap our magnetic coils used to form a BEC are only 9mm apart giving a trap capture volume that is prohibitively small to efficiently load from the Zeeman slower. To this end we utilize two MOT stages, the first is a modified configuration known as a (LVIS+) which provides high differential pumping and efficient loading into the second MOT where our BEC is formed. 

The LVIS is formed by moving one of the the 3 quarter wave-plate and mirror combinations that reflect the 3 input beams in a typical MOT into the vacuum system with a small hole in the middle. The unbalanced scattering for atoms in the center of the MOT caused by the shadow of the hole in the reflection produces a reasonably collimated atomic beam that passes through this hole into the high vacuum BEC chamber on the other side. A further modification to the LVIS configuration is to add a separate push beam coaxial with the mirror hole which has independent detuning and power relative to the rest of the MOT to allow for control over the beam velocity and out coupling rate\cite{Swansson2007}.
 As the atoms that pass through the LVIS+ mirror must then travel almost a meter to the second MOT any transverse velocity would result in meagre capture rates due to the limited capture volume of the second MOT. To this end we use a simple two dimensional MOT placed halfway to refocus the beam and give a 5 fold increase in the transfer efficiency between MOT's. 
 
 The small (9mm) separation between the windows in the main BEC chamber combined with the dramatic loss of the metastable state at high densities due to penning ionization means that a normal MOT is not practical for this second stage. Instead we form a MOT in a pancake shape by using separate detuning for the beams that pass through the magnetic coils (horizontal). The magnetic field needed for this MOT is produced by operating a single anti-Helmholtz pair of coils of the BiQUIC\footnote{See \autoref{magtrap}.} trap. The loading into this MOT is additionally enhanced by the use of two additional beams at $15^{\circ}$ improving the capture velocity and the resultant atom number by a factor of two \cite{Dall2007} .

Once loading is complete we then decrease the detuning in the MOT beams producing a 5 fold density increase in the center. This however causes heating which is removed by switching off the magnetic field and applying a 3D molasses cooling stage.  

\section{Magnetic trap}
\label{magtrap}

To cool the cloud further we must now transfer to a conservative potential in the form of a magnetic trap. A number of options exist for producing a magnetic magnetic field gradient as needed however few are suited to our experimental constraints of large optical access, stability and strong gradients to produce effective `in trap cooling'.

Our \textbf{bi}-planar \textbf{q}uadrupole \textbf{I}offe \textbf{c}onfiguration (BiQUIC) trap consists of two pairs of anti-Helmholtz coils, the first pair provides a strong field gradient and the other produces the bias. By occupying only a single plane this configuration is far easier to incorporate into the titanium BEC chamber. To minimize the number of amp-turns with their resultant heating and large thermal time constants we place these coils as close as possible(9mm) inside re-entrant windows. This gives large trapping frequencies (50Hz transverse, 500Hz Axial) with only 36 amperes of current and 10 turns of wire. The relatively low resultant power dissipation then allows for cooling with water stabilized to within 10mK giving excellent shot to shot stability necessary for atom laser experiments.\cite{Dall2007}

While the current in the coils gives a magnetic stability of order $\mu G$ the mains (50Hz) electricity in the lab produces a magnetic field on the order of $mG$ which is prohibitive for creating a RF atom laser. To counter this we employ a magnetic `nuller' which consists of two electromagnet coils (1m in diameter) and magnetic sensors on each side of the BEC chamber. The current in each coil is controlled using proportional-integral-derivative feedback from the respective sensor in order to produce the desired set-point. A pair of these coils is applied in each axis to produce a DC magnetic stability of $\sim30\mu G$ and reduction in AC noise of $\sim 10^2$ \cite{Dedman2007}.

To produce a BEC in the magnetic trap we load the high density cold cloud from the final stage of the previous section by bringing on the coil currents to produce a weak and approximately isotropic trap (70Hz) with a large magnetic field bias. The attractive magnetic potential inevitably heats the atoms which is removed using a process known as in trap cooling, here two weak ($10^{-2} I_{sat}$) circularly polarized beams resonant with the Zeeman shifted transition at the center of the trap are applied. This produces a cooling similar to a 1D optical molasses where the low intensity and high magnetic bias decreases the chance of pumping to untrapped levels, unlike a MOT the process relies on rethermalization of the un-cooled axes. Once cooled to  $\sim200 \mu K$ the trap is further compressed resulting in a cigar like trap and a second in-trap cooling stage is applied reducing the temperature to $\sim 100\mu K$. A final compression stage is then employed resultant in the trap used for the rest of the experiment, with trap frequencies of 500Hz transverse and 50Hz axial.

\begin{figure}[H]
  \begin{center}  % center environment centers the graphic on the page
    \includegraphics[height=5cm]{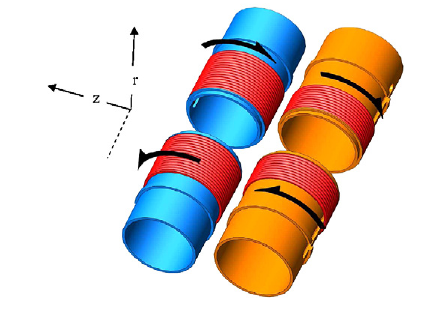}
  \end{center}
\caption{Schematic of the BiQUIC trap showing trapping (right) and bias coil pairs (left) \cite{Dall2007}.}
\label{biquic}
\end{figure}

As we cannot cool the atoms further using laser cooling we now rely on evaporative cooling. Here we selectively out-couple high temperature atoms from the trap with a radio frequency (RF) knife. This consists of a RF field which is applied so only those atoms with large kinetic energy can `climb' the magnetic potential producing a Zeeman spiting resonant with the RF field and transition into the un-trapped state, being lost from the trap. The atomic ensemble then rethermalizes through collisions resulting in a decrease in the average trap temprature. This `knife' is ramped from a large temperature to a cooler one over a period of  20 seconds following an approximately exponential trajectory. The BEC transition is reached at roughly $\sim 1\mu K$ although we continue evaporation further to produce a high purity condensate ($\sim98\%$) with $2\cdot10^{5}$ atoms.

\section{Atom Laser}
\label{atomlaser}

As the previously formed BEC is a coherent atomic cloud it is possible to create a coherent atomic beam by out-coupling condensate atoms from the trap. This can be done in a magnetic trap using an RF `knife' or in a optical trap by lowering the trap potential until atoms fall out of the trap. As such an atom laser has a number of uses in particular high precision measurement \cite{HainePhd,Cronin2009,RussellPhd,Bernard2011,Robins2013}. Creating a stable beam can be a significant challenge as any fluctuation in the initial trap translates to a change in the out-coupling rate and in turn intensity,energy and spatial profile fluctuations of the atomic beam. The aforementioned nuller helps to a great extent although the beam still has a peak in frequency space at 50 Hz.
\begin{figure}[H]
  \begin{center}  % center environment centers the graphic on the page
    \includegraphics[height=15cm]{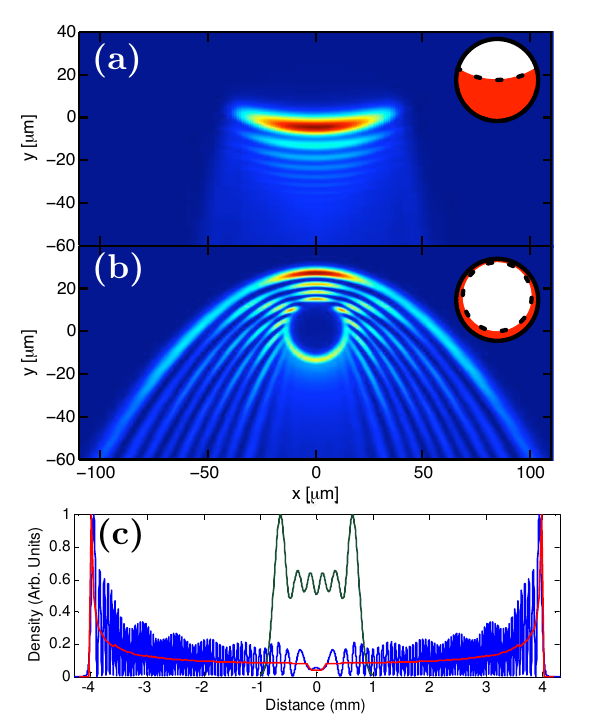}
  \end{center}
\caption{A simulation of the RF out-coupling of metastable helium and rubidium \cite{RussellPhd}, the large mean field potential causes a the fountain effect seen here.}
\label{RF outcoupling}
\end{figure}
The operation of an atom laser in He* is distinct from that of Rb primarily in the dominance of mean field interaction over the comparatively weak gravitational energy. This means that RF out-coupling surfaces nearly uniform in density, a fact that will be exploited later in the theoretical treatment of a atom laser based measurement (see \autoref{sec:theoryatomlaser}).

\section{Detection}

\begin{figure}[H]
  \begin{center}  % center environment centers the graphic on the page
    \includegraphics[height=10cm]{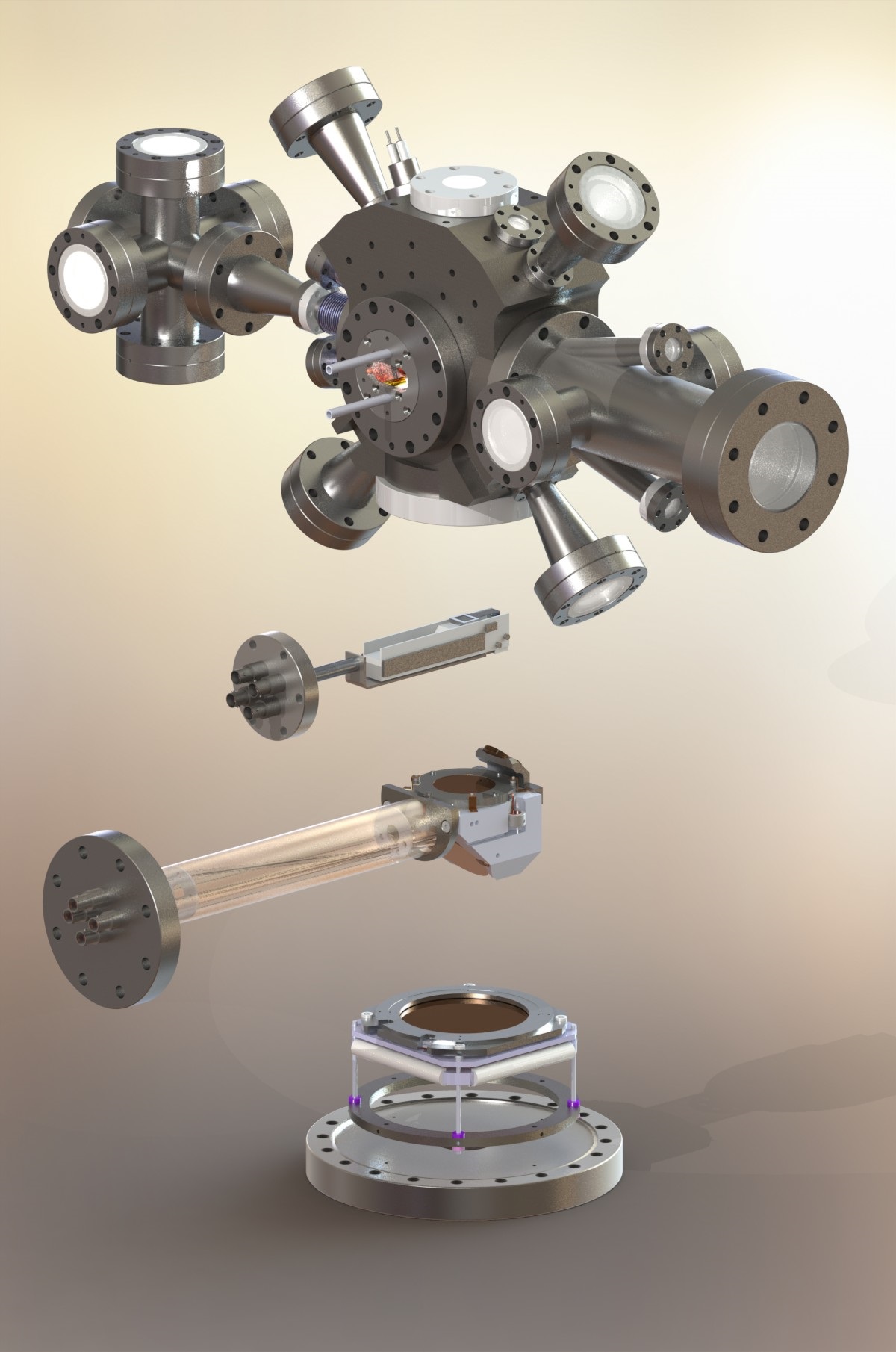}
  \end{center}
\caption{A schematic of the detectors in our experiment. Top centre: BEC chamber, Top Right: Probe beam window, In order downwards : ETP, MCP Phosphor, MCP DLD }
\label{DetectorRig}
\end{figure}

One of the primary advantages of He* over other species for condensation is the ability to  easily detect single atoms with reasonable quantum efficiency (25\%) using electron multiplying detectors. This is done by taking advantage of the nearly instantaneous decay of the metastable state through Auger de-excitation when in contact with a conductive surface. This process liberates a (20eV-Work function) electron from the surface of the detector that can amplified by acceleration with an applied electric field before colliding with subsequent low work function surfaces producing secondary electrons which are further amplified in turn ($\sim 10^{6}$ electrons). Our experiment uses three detectors based on this technique placed below the magnetic trap. 

\begin{figure}
\begin{center}
\includegraphics[height=7cm]{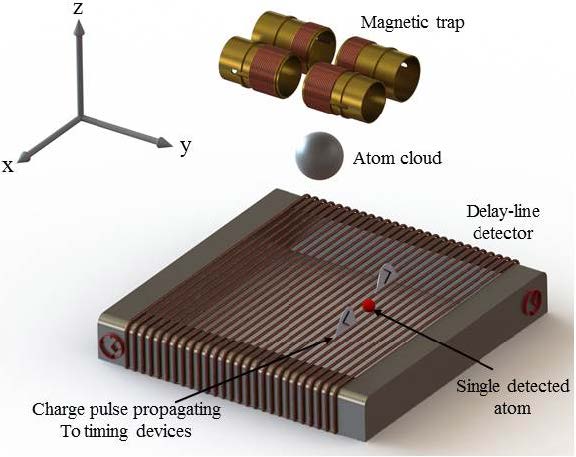}
\end{center}
\caption{delay line detector showing the induced charge pulse, with multi channel plate's omitted. The BiQUIC coils are seen center.}
\label{DLD}
\end{figure}

Unlike other experiments we use no absorption or fluorescence imaging and rely on dropping the atomic cloud onto a detector below.  The first is a discreet dynode electron multiplier which provides only arrival time information. The second, is a micro-channel plate detector (MCP) which is a type of two dimensional electron multiplier formed by a ceramic plate with a large density of small ($\sim10\mu m$) holes, by applying a voltage across the plate a continuous dynode electron multiplier is formed for each of the ($\sim 10^{6}$) holes. We use two of these plates, one feeding the other to produce a large signal ($10^{6}$ electrons) from a single incident atom, the output of which is accelerated onto a phosphor screen which converts the electrons to light and is imaged through a window onto a C.C.D. camera. The final detector is another double-MCP type that feeds onto a \textbf{d}elay \textbf{l}ine \textbf{d}etector (DLD). Here the amplified electron pulse strikes a pair of flat coiled wires inducing a voltage signal at each end (\autoref{DLD}) which is then fed into a digitizer to produce a measurement of the position and arrival time of the incident atom to within $\sim 100\mu m$ and $1 ns$ respectively. Two such wire loops are used one for the  x and y coordinate respectively.

%% file: chapter3.tex
\chapter{Precision Measurement}
\label{secmeasure}
%[Grading Goals]\\
%---show a good scientific process for selecting the method based on theory\\
%---demonstrate understanding of the limitations of the experiment\\
%---show that you obtained concepts and procedures independently from the literature and discussed them for study\\
%---demonstrate understanding why chose between alternative methods\\

A long term goal of our group is to use this apparatus to test fundamental atomic theory by measuring the 413~nm tune-out wavelength in He* at an accuracy better than 100fm \cite{Mitroy2013}. Measuring the tune-out wavelength relies on two simultaneous measurements; an accurate determination of the probe beam wavelength (resolution) and the small resultant shift in the ground state energy caused by this probe beam (sensitivity). Due to the scale of sensitivity and resolution needed for a measurement at this accuracy it is beyond the scope of an honours project. However we have endeavoured to make the first measurement this tune-out wavelength in metastable helium in order to develop techniques and expertise on the path towards a test of fundamental atomic theory.

 This work has made significant inroads in this regard, developing an ultra stable laser system and an atom laser based potential measurement technique with unrivalled sensitivity. This culminates in the first ever measurement of the He* tune-out wavelength at an accuracy only two of magnitude less than the 100fm criteria \cite{Mitroy2013}. Producing a laser system capable of reaching this ultimate goal required numerous technical developments primarily aimed at producing a highly accurate, tunable and stable laser wavelength. We have developed the techniques necessary for such a system including cavity and wavemeter locks, temperature stabilization, and calibration spectroscopy. Using these methods we believe an accuracy of order $10fm$ is achievable. Further we have developed a multi stage laser alignment method which is able to align very weak potentials with the atomic cloud to be probed. Using a combination of an atom laser and frequency space techniques we are able then able to measure a potential with sensitivity better than $200pK$. We believe that this method is the first of its kind and is widely applicable to other areas of atom optics.

\section{Probe Laser}
\label{probe laser}

The primary addition to the experiment during this work was a 413~nm laser system which forms the probe beam. The core of the system is a commercial hermetically sealed external cavity diode (Moglabs ECD004) laser (EDCL) in Littrow configuration with an anti-reflection coated laser diode. This is tunable in wavelength from 410~nm-415~nm critically including the predicted tune out wavelength, with approximately 20mw of output power. For more detail on the laser techniques used here see \cite{Hardman2014}. 

\begin{figure}[H]

	\begin{center}
	\includegraphics[height=9cm]{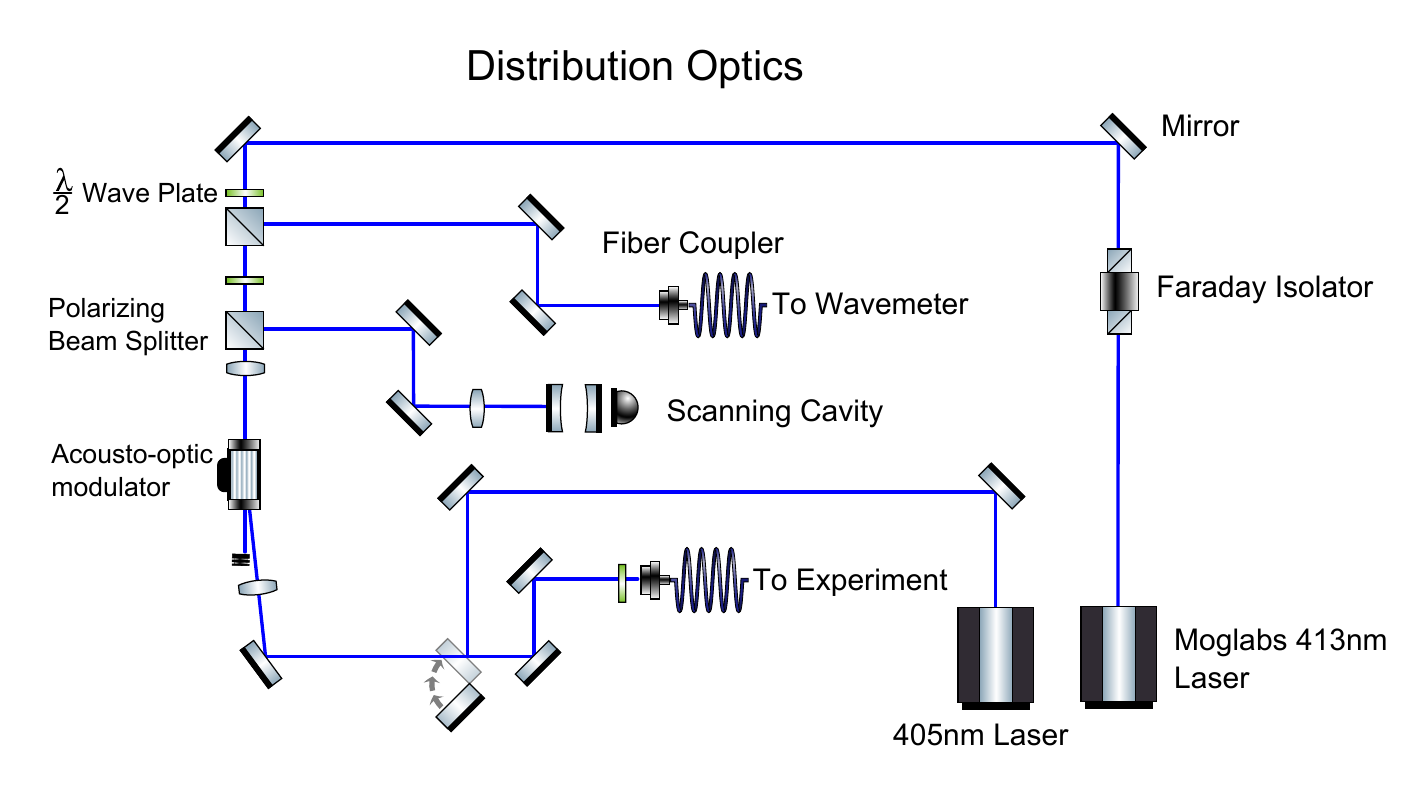}%.jpg for real pic
	\end{center}
	\caption{Distribution Optics. We are able to select either 405~nm or 413~nm laser light to send to the experiment for alignment purposes.}
	\label{distopt}
\end{figure}

A small amount of the light produced by the laser split from the main beam path and sent to two instruments, the first a wavemeter which precisely measures the wavelength of the light and the second a scanning Fabry Perot interferometer. This interferometer can be used to determine if the laser is `multi-mode' wherein the laser outputs multiple wavelengths of light simultaneously or alternatively as a stable wavelength reference (detailed in \autoref{wavemeterlock}).
 We pass the main laser beam through an acoustic optical modulator (AOM) to allow for intensity modulation and then couple the light into a single mode optical fibre. This eliminates the pointing error of the laser with wavelength tuning and produces a high purity Gaussian beam. The light then passes through to our focusing optics  where it is expanded before being focused into the chamber (\autoref{distopt}).

\subsection{Focusing Optics}
\label{sec:OpticsFocusing}
%[outline]\\
%most difficult part of project micron over half a meter is bloody hard (months)\\
%-needed to use a diffrent laser to even see effect (power and polz)\\
%-complicated by the chromatic shifts by changing\\
%-changed optic system many times\\
%----difficult to achive small spot[final fig]\\
%----initaly same paths\\
%----used alignemt camera\\
%-eventual found a technique that can reliably produce a good alignment (532 beasty ,405)\

 To produce a measurable potential (Equation \ref{potential}) and maximize the sensitivity of our measurement we must focus this probe beam to the smallest possible size to give the greatest intensity and in turn energy shift of the atoms. For a Gaussian beam like the ones used here the beam radius is defined by the waist denoted $w$. The potential for a given power P scales strongly with this waist
 \begin{equation}
 U_{max}\propto I_{max}\propto\frac{2P}{\pi w^{2}}
 \end{equation}
 and is thus is of great importance. This maximization is hindered by limited optical access in our experiment as we must introduce our probe beam through a window 500~mm from the BEC (see \autoref{ExpRig}). The diffraction limit\footnote{ As defined by the minimum possible waist from an initial beam diameter D focused with a lens of focal length F with wavelength $\lambda$, given by $\omega_{diff}= 1.22 \frac{F\lambda}{D}\quad.$} then dictates that for the largest practical optics size of 50~mm we can only reach $\omega \sim 8 \mu m$. Additionally alignment of this small spot with an atomic cloud of comparable size over such a large distance requires an exceptionally stable optics system.
Developing this system took many design iterations as it was found that aberrations which enlarge the focal spot are far more dominant for these smaller wavelengths .

\begin{figure}[H]
\centering
\subfloat[]{{\includegraphics[height=4.9cm]{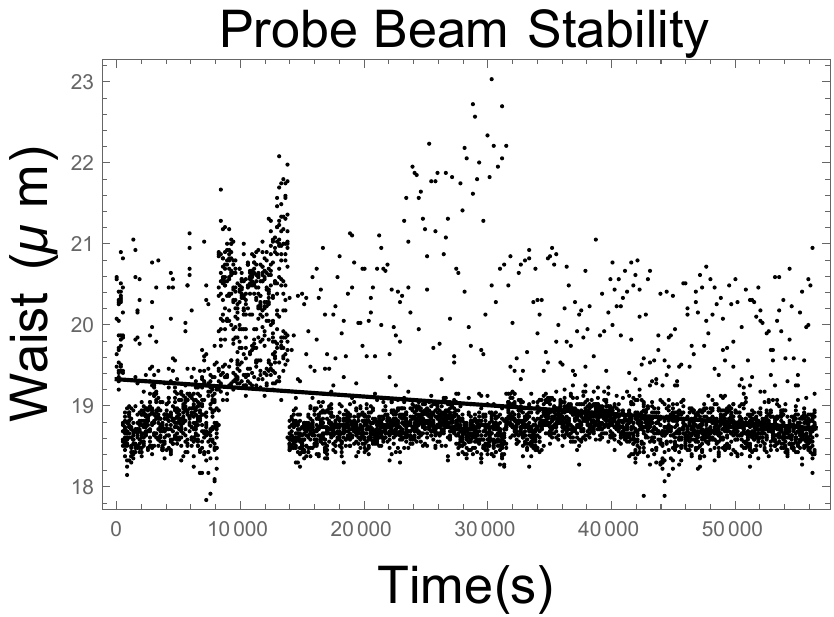} }}%
\qquad
\subfloat[]{{\includegraphics[height=4.9cm]{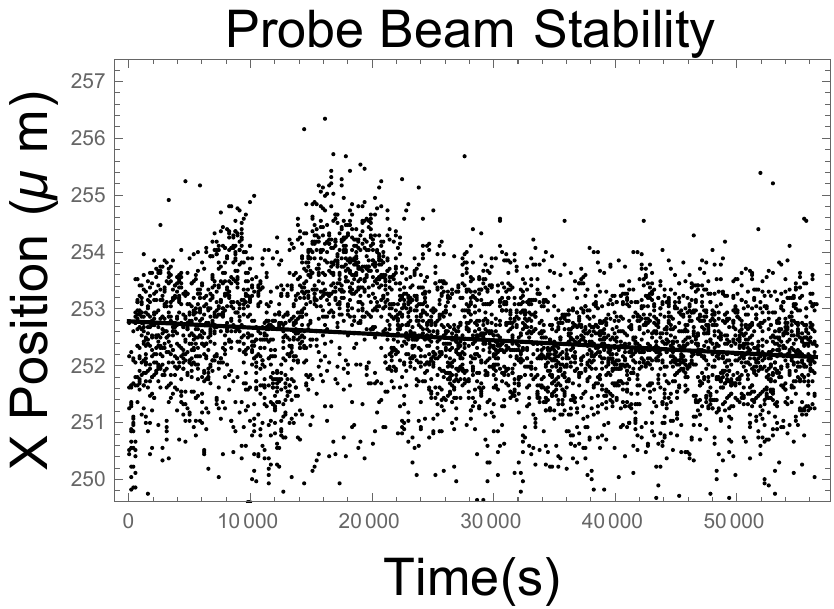} }}% 
\qquad
\subfloat[]{{\includegraphics[height=4.9cm]{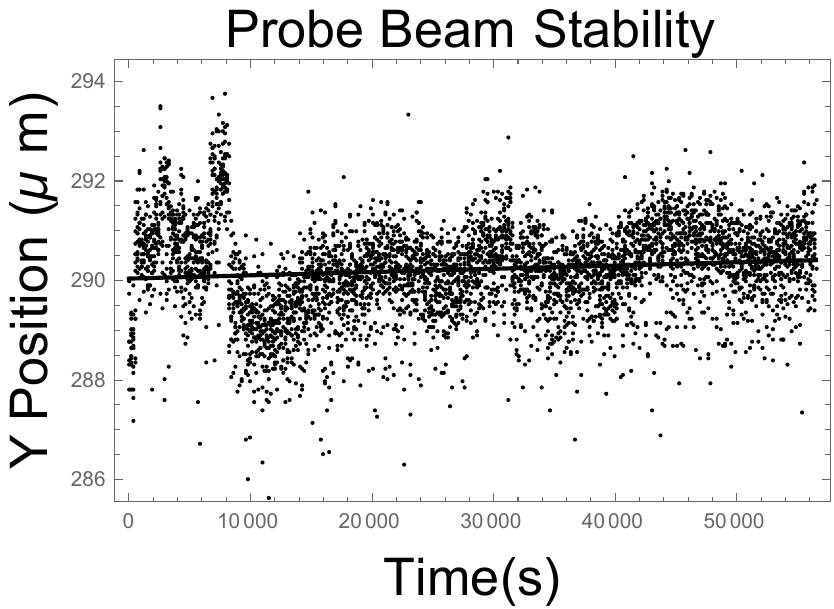} }}% 
\caption{Probe beam stability measurements. (a) Waist: standard deviation $0.68\,\mu m$, drift $0.93\,\mu m/day$. (b) Horizontal position: standard deviation $1.03\,\mu m$, drift $0.96\,\mu m /day$. (c) Vertical position: standard deviation $0.95\,\mu m$, drift $0.57\,\mu m/day$.}
\label{probestability}
\end{figure}

Our current focusing arrangement starts with a large diameter (6~mm thorlabs F810FC-543) fibre collimator which is then expanded with a 125~mm achromatic lens to approximately 30~mm before passing through a 750~mm and 400~mm achromatic lens forming a focus $\sim512$mm behind the input window. 
The position of the focused spot is adjusted by the last lens (400~mm) which is mounted on a three axis ultra fine flexure stage (Thorlabs MBT616D/M), this provides a means by which the beam can be aligned to the cloud with excellent stability (see figure \ref{probestability}). Large focal shifts can be made using the position of the 750~mm lens which is mounted on a linear translation stage. We purify the polarization from the fibre coupler by passing it through a polarizing beam splitter (PBS). The waist of the beam is determined by using a mirror on a removable kinematic mount that diverts the beam onto a CCD camera (point grey SCOR-20SO) with a $4.4\mu m$ pixel size placed in the equivalent location as the BEC outside the vacuum. To produce a more accurate measurement here we place a identical vacuum window in this beam path which serves to simulate any focal shifts due to the chamber window.

\begin{figure}[H]
\centering
\subfloat[]{{\includegraphics[height=4.9cm]{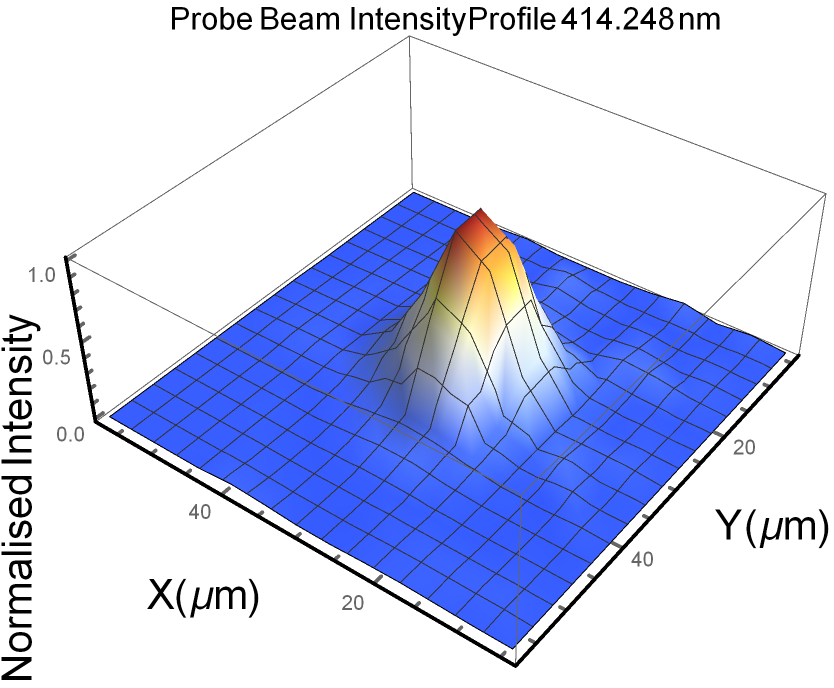} }}%
\qquad
\subfloat[]{{\includegraphics[height=4.9cm]{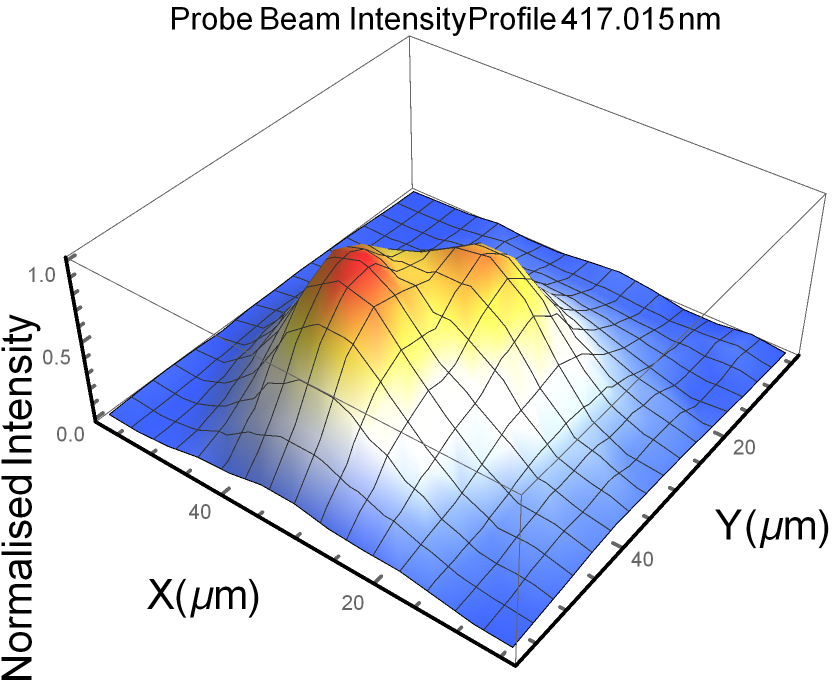} }}%
\quad
\subfloat[]{{\includegraphics[height=4.9cm]{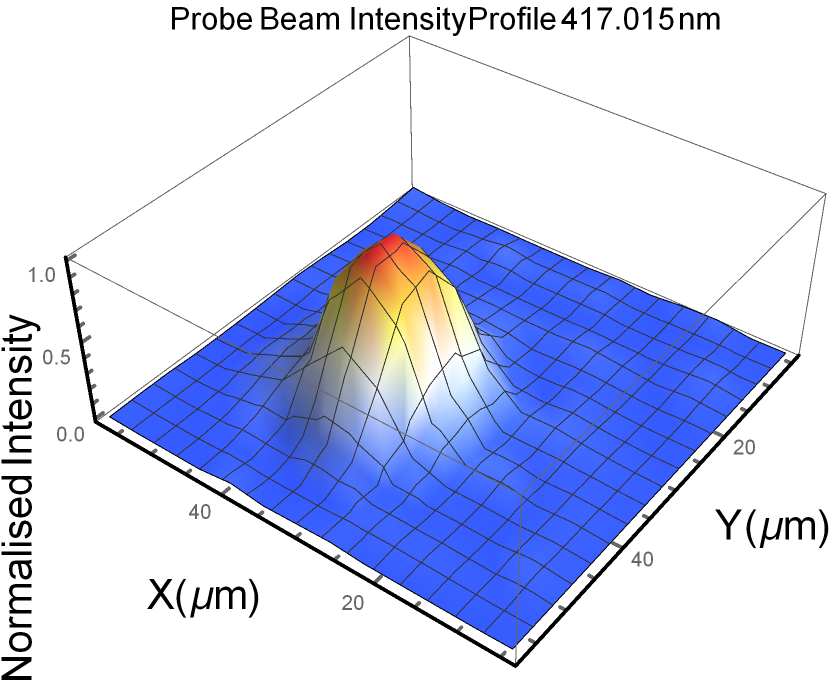} }}%
\caption{Probe beam intensity profiles as measured using the ccd for (a) 414.248~nm $\omega=9.8\pm0.1$ (b)  417.015~nm $\omega=22.7\pm0.3$. The peak intensity of (b) is then 19\% of the initial intensity(a). This shift can be corrected for by adjusting the focus as in  (c) giving $\omega=11.2\pm0.1$.}
\label{waist}
\end{figure}

 Using this camera we then adjusted the optics to reduce aberrations and minimize the spot producing a Gaussian-fitted beam waist that is nearly at the diffraction limit ($\omega_{measured}=9.8\mu m$ vs $\omega_{diff}=8 \mu m$) (See figure \ref{waist}). 
\begin{figure}[H]
	\begin{center}
	\includegraphics[height=10cm]{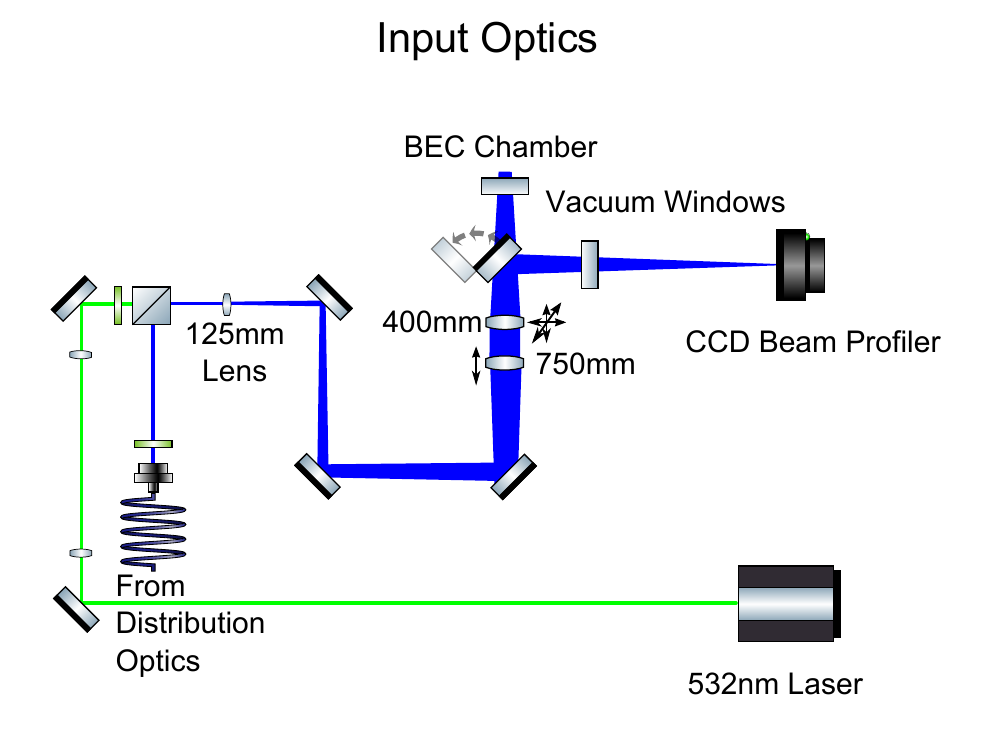}%.jpg for real
	\end{center}
	\caption{Input optics: The beam starts from the blue fibre at the bottom left and enters the chamber at the top right. The last mirror seen before the window is the removable mirror for CCD profiling. The 532~nm laser is used for alignment purposes (see\autoref{alingment}). }
	\label{inputoptics}
\end{figure}

To account for any drifts in the alignment with time we used this camera to sample the beam profile at regular intervals over a day. We then fit a Gaussian profile to the images and extract the position and waist as shown in \autoref{probestability}. The standard deviation and trend line were then determined for each measurement. While both drift and standard deviation were reasonable for all measurements the excursions present will likely become a source of error.

For the tune-out measurement we must change the wavelength used over a relatively large range, this can then produce a chromatic change in the focal lengths of lenses. While nominally achromatic doublet lenses were used for the focusing optics a large and undesirable dependence of the probe beam's focal length on wavelength was found. This was determined by using the camera to measure the intensity profile at two wavelengths of the probe laser which was then fit to a Gaussian profile to determine the beam waists (see \autoref{waist}). By using the expression for a Gaussian beam waist as function of axial distance from the focus (see \autoref{trapfreqodt}) the chromatic focal shift per change in wavelength was measured at $\frac{\delta z}{\delta \lambda}=0.55 \frac{mm}{nm}$.  The corresponding change in relative peak intensity for a given change in wavelength $\Delta \lambda$ and Rayleigh length\footnote{Here $\sim$0.73 mm.} $z_{R}$ can be found (see \autoref{trapfreqodt}) to be 
\begin{equation}
\frac{I_0}{I(\Delta \lambda)}=\frac{1}{1+\left(\frac{\frac{\delta z}{\delta \lambda}\Delta \lambda}{z_{R}}\right)^{2}}\quad .
\end{equation}
This shows that as the wavelength range is narrowed there is a significant decrease in the intensity change and corresponding error in measurement of the energy level shift.
We can then derive a criterion that to maintain the intensity within 5\% of initial intensity the change in wavelength must be less than 0.3~nm.

For large wavelength ranges the change in intensity can be reduced by refocusing the beam at each point using the position of the 750~mm lens as in \autoref{waist} (b) which improves peak intensity change from 19\% to 77\% of the initial value. Thus we propose a simple two step protocol for our measurement: first we perform a coarse measurement of polarizability over a large wavelength range, optimizing the focus at each wavelength. Then the polarizability is measured over a small range about the tune-out estimated in the first step, while keeping the focus fixed. As range of wavelengths are far smaller the chromatic shifts are negligible so focus adjustment is not needed avoiding the uncertainly introduced by this adjustment thus giving the minimum uncertainty in the intensity and in turn the tune-out wavelength.

\subsection{Alignment}
\label{alingment}

Aligning the probe beam with the atomic cloud to be interrogated was the most difficult aspect of this project taking many months to perfect. While the alignment techniques usually employed are developed for dipole trap beams with a potential 100's of $\mu K$ deep the probe beam can only produce at best a meagre $~30nK$.

\begin{figure}[H]
\centering
\subfloat[]{{\includegraphics[height=2cm]{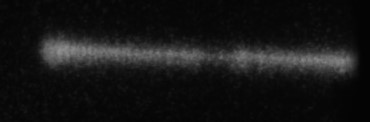} }}%
\qquad
\subfloat[]{{\includegraphics[height=2cm]{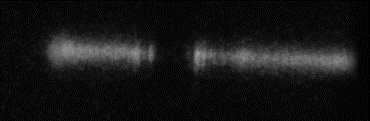} }}%
\caption{Atomic deflection(b) of atoms in an atom laser from a 100~mw 532~nm beam focused to 15~{\textmu}m. (a) is shown as a reference without the effect of a beam. }
\label{deflectionatomlas}
\end{figure}

Initially it was thought that this beam could be aligned in a similar manner to the 1550~nm trap beams that had previously used. Here one would align the beam through a target on the probe beam window and the hole in the LVIS mirror (see \autoref{apparatusmot}). A BEC would be then created, dropped and any disturbance in the far field profile such as a dark line (see \autoref{deflectionatomlas}) would then indicate the presence of the beam underneath the trap. If this was not observed the beam would be moved after each iteration of the BEC sequence, first stepping through a small horizontal range (1~mm), and once exhausted moving down a small distance (0.2~mm) and the process repeated. Once found, the time the beam was held on for after the trap switch off would be reduced and the signal optimized thus moving the beam towards the trap center. Repeating this process until the beam remained on for only a few ms after magnetic trap switch off then gives an alignment within a beam waist or so of the condensate.

 To optimize even further (with attractive potentials) we hold the beam on for many 100's of ms after trap switch off and measure the number of atoms trapped in the beam due to the dipole potential. As gravity is a small contribution to the net potential for helium optimizing transfer between two traps gives the optimal potential overlap\cite{SchulzPhd}.  This procedure usually takes a day from beginning to end for a 1550~nm dipole beam used previously in our laboratory.

\begin{figure}
  \begin{center}  % center environment centers the graphic on the page
    \includegraphics[height=12cm]{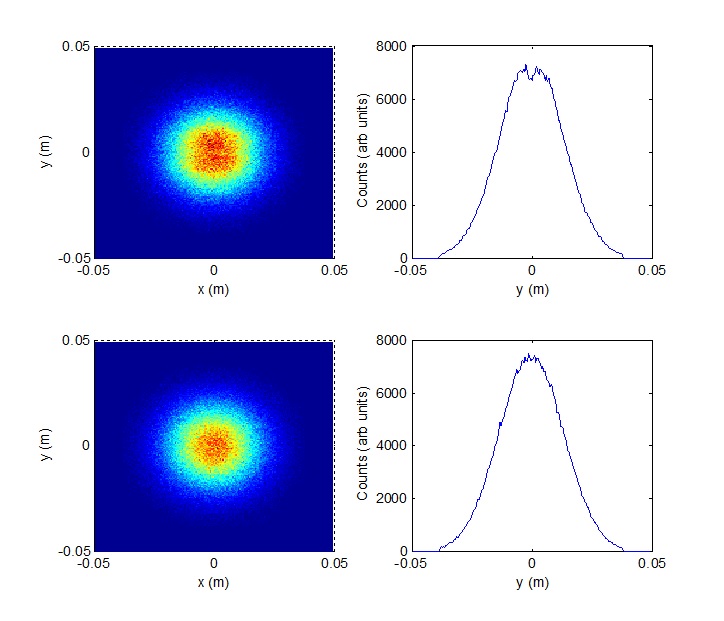}
  \end{center}
\caption{Monte Carlo simulation of atomic deflection dude to the probe beam ($\lambda=410$~nm, $P=50$~mw, $\omega_{0}=10$~{\textmu}m) $25$~{\textmu}m (upper quadrant) and  $250$~{\textmu}m (lower quadrant) below the initial BEC. The left is a density map of the atoms as they would hit the detector while the right is an integral through the x axis to see the change in density due to the probe beam. }
\label{AtomLaser}
\end{figure}

This process was attempted many times for the probe beam, operating at its largest detuning and thus potential (410.6~nm), however no disturbance was observed due to the probe beam. Using a Monte Carlo simulation (see \autoref{AtomLaser}) for the motion of non-interacting particles in the presence of the probe beam potential is was soon discovered that this deflection technique would at best require the beam to be displaced by less than $25 \mu m$ below, $25 \mu m$ horizontally, and $0.5mm$ in the focus relative to the BEC. 

As the initial alignment through the target and LVIS mirror could be off by approximately 1.5~mm in each axis (x, y and the focus (z)) and a sequence to create a BEC takes approximately 30 seconds this would take
\begin{equation}
\frac{1.5mm}{0.5mm}\frac{1.5mm}{0.025mm} \frac{1.5mm}{0.025mm}\cdot30s=90\ hours
\end{equation}
to align.

This time combined with the weak signal means that such a simplistic approach is not practical in this experiment. However, by using a more powerful, strongly interacting laser to initially align the position and then switching over to the probe laser one could practically align the beam in a far shorter time due to the much greater signal.  The chromatic shift measured above complicates alignment further as one could not simply swap between a well-aligned far detuned laser  and the probe laser without loosing the alignment altogether. This meant that while providing a deep potential 1550~nm, 1083~nm and 780~nm lasers were unable to provide much help as the focal length shifts were unable to be compensated for. After many iterations of development including 532~nm, 450~nm and 405~nm lasers a alignment solution was finally developed.
 
Here we introduce a separate 532~nm laser\footnote{Laser Quantum Finesse 10~w Ti-Sapphire pump laser.}  (see \autoref{inputoptics})  through the PBS seen to the top left of \autoref{inputoptics} while simultaneously a 405~nm laser is input through the Fiber using the flip mirror in \autoref{probestability}. By adjusting the input size and divergence of the 532~nm beam before the PBS while using the CCD camera it is possible to focus both wavelengths to the same spatial location. The initial alignment of the 532~nm beam is (as described for a 1550~nm beam as above) a straight forward process with up to 10~w of power.

As it is not possible to trap in the repulsive 532~nm beam the alignment (once within $~10\mu m$) is optimized by maximizing the distortion of the atomic cloud on the detector. Next the 532~nm is blocked and the number of trapped atoms in the 405~nm beam is optimized. This still gives a 5~nm jump (405~nm-410~nm) to the closest wavelength of the probe laser. By heating the (nominally) 405~nm laser diode considerably ($65\degree C$) it is possible to reach  a wavelength of 410~nm, by then optimizing the trapping it is then simple to switch to the final Moglabs probe laser.

\begin{figure}[H]
  \begin{center}  % center environment centers the graphic on the page
    \includegraphics[height=7cm]{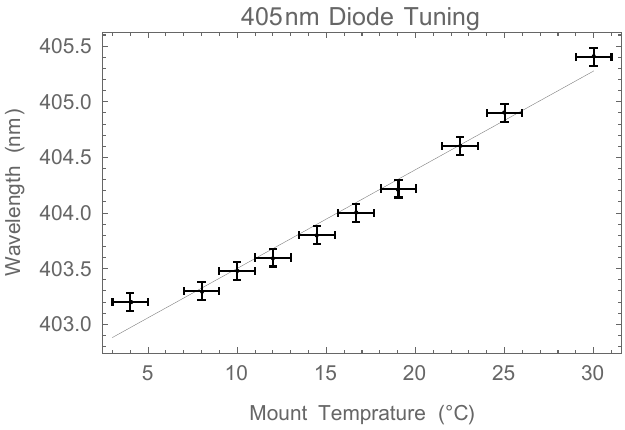}
  \end{center}
\caption{Temperature based wavelength tuning. Line of best fit (402.62$\pm$0.08)~nm + T$\degree C$ (12.3$\pm$0.4)$\cdot 10^{-2} \frac{nm}{\degree C}$ .} 
\label{CavTemp}
\end{figure}
We then repeat the optimization of the trapped atom number using the narrow linewidth Moglabs probe laser. After which the system is aligned and ready for the main measurement. This ladder of laser wavelengths means that it is reasonable to to go from no observed effect to an aligned probe beam within a day of work.

\subsection{Wavemeter and Locking System}
\label{wavemeterlock}

The accurate determination of the probe beam wavelength is essential for the tune-out wavelength measurement. The wavemeter used in the work presented here is a commercial device (Moglabs MWM001) based on \cite{White2012}. It is a relatively simple design that relies on a diffraction grating to produce a change in angle of a beam with wavelength, which is measured by position a CCD and gives a resolution of 100fm. It is calibrated using a ten point calibration to a (He-Ne calibrated) High Finesse WS-7 wavelength meter (absolute accuracy $35fm$). By comparing the the two wavemeters we can give an estimate of the uncertainty at ($\sim200fm$). We then measured the drift of the wavemeter by repeating the comparison the next day, this revealed a relatively large drift of the wave meter at $\sim300fm$. In the results presented in (section  \ref{results}) we relied only on this measurement of wavelength as other errors dominated.

However in our ultimate goal to produce the best measurement of the transition rate ever made a resolution of better than a 100fm is needed, here drifts of the laser over the time-scale of the experiment become significant. We have thus developed a ultra stable tunable laser system for the probe beam, based on a cavity lock with wavemeter feedback. 

\begin{figure}
  \begin{center}  % center environment centers the graphic on the page
    \includegraphics[height=7cm]{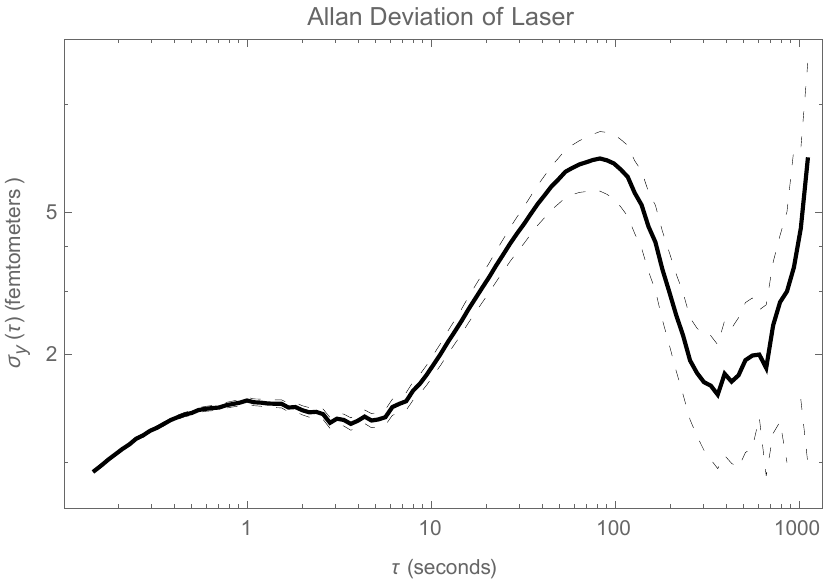}
  \end{center}
\caption{Moglab Laser stability as measured with High Finesse WS-7, dashed lines indicate uncertainty.}
\label{unlockedStab}
\end{figure}

\begin{figure}[H]
  \begin{center}  % center environment centers the graphic on the page
    \includegraphics[height=7cm]{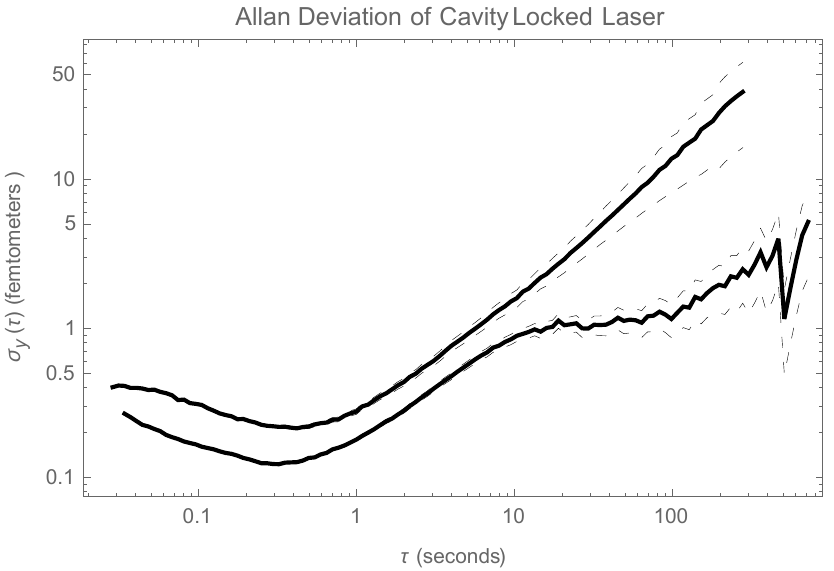}%the version in Ghz allanbothghz.pdf
  \end{center}
\caption{Laser stability with cavity stabilization as measured with High Finesse WS-7. Upper trace is with a small change in temperature $\backsim 0.2\degree C$ of reference cavity causing a large increase in the long term drift. Lower trace is at ambient conditions. Dashed line indicate the uncertainty.}
\label{CavStab}
\end{figure}

Here we measure the power transmitted through the a cavity formed by two high reflectivity mirrors separated by a low thermal coefficient metal (Invar). When the resonance condition of the cavity is met by the input laser wavelength the circulating cavity power builds and light is transmitted through the cavity which is detected with a photodiode. By positioning the laser frequency to be on the edge of this resonance (`side of fringe') we are able to produce a large change in photo diode voltage with a small change in wavelength. A proportional-integral-derivative (PID) controller (integrated in the Moglabs driver) allows for stable feedback to this signal to stabilize the laser wavelength to the cavity resonance. While relatively crude in comparison to `top of fringe methods' this was able to improve the stability by an order of magnitude for short($<100s$) time scales as seen in fig \ref{unlockedStab} and \ref{CavStab}.

A particularly promising approach to correct for any long term drifts of this cavity is to use feedback from a wave-meter to adjust the cavity length by varying the piezo voltage. This scheme would combine the short term stability of the cavity lock with the long term stability of the wavemeter. This scanning Fabry-Perot interferometer (Thorlabs SA200-3B) is re-purposed from its original role as a laser spectrum analyser which we use when adjusting the wavelength of the Moglabs laser to ensure single mode operation. As such it contains a piezoelectric modulator inside the cavity that may be used to adjust its length and corresponding resonance many GHz with a simple voltage input. To investigate the feasibility of this approach we locked the laser as above to the `side of fringe' varying the piezoelectric transducer voltage while measuring the resulting wavelength with a WS 7 wavemeter.
\begin{figure}[H]
\centering
\subfloat[]{{\includegraphics[height=5cm]{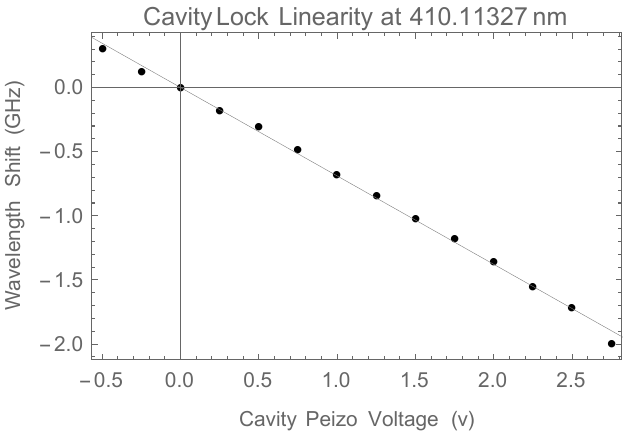} }}%
\qquad
\subfloat[]{{\includegraphics[height=5cm]{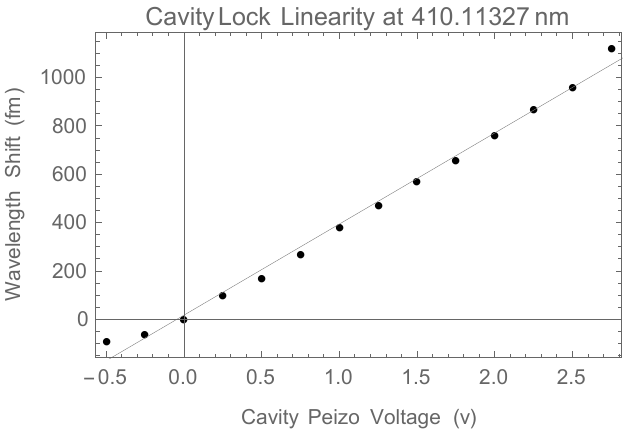} }}%
\label{fig:example}% 
\caption{Cavity lock linearity 386.725~fm/v or 0.68931~GHz/v. A wide ($\sim$2~GHz) tuning range is possible due to the large mode hop free range of the Moglabs laser when used with current feedback. Error bars in measurement are smaller than plot markers.}
\label{CavLinearity}
\end{figure}
The maximum stable slew rate was found to be approximately $100 mV s^{-1}$ which is well above the drift rate in \autoref{CavStab}. The dual (cavity-wavementer) lock was not implemented as  a second independent wavemeter is needed to properly evaluate the performance of the lock loop. This system will ultimately be required in future high accuracy experiments.  Using this technique the laser wavelength stability would be limited only to the long term stability of the wavemeter.

The dominant source of such drifts in such a high precision laser system are due to temperature instability of the environment causing minute expansion and contraction of materials. To suppress this we investigated a temperature stabilization system as submitted for publication in Journal of Instrumentation.  This was able to significantly decrease the temperature change in the laboratory to within 10mK and reduce the drift of a EDCL by a factor of $\sim5$. We intend to apply this technique to our wavemeter to minimize drifts and improve the ultimate accuracy of the system.

\subsection{Wavelength Calibration}

While temperature stabilisation can go a long way to reducing the drift an absolute wavelength reference is needed to give a reasonable accuracy and reduce any systematic errors.  To this end we wish to use atomic transitions to provide a high stability wavelength reference. For this reference we require strong ground state transitions close to the tune-out wavelength as provided by thulium (409.419~nm, 410.584~nm)\cite{Sukachev2010,Chebakov2009} and indium (410.1745~nm)\cite{Morton2000}. Thulium is particularity useful in that it has been laser cooled and thus has been used for laser stabilisation before\cite{Sukachev2010,Chebakov2009,Sukachev2010a,Kolachevsky2007} although there is some disagreement on the reported wavelength \cite{Wickliffe1997,Jiang1993}. This scheme would provide a 3 point calibration of the wavemeter within close proximity ($\sim$3~nm) of the tune-out wavelength. 

\begin{figure}
  \begin{center}  % center environment centers the graphic on the page
    \includegraphics[height=7cm]{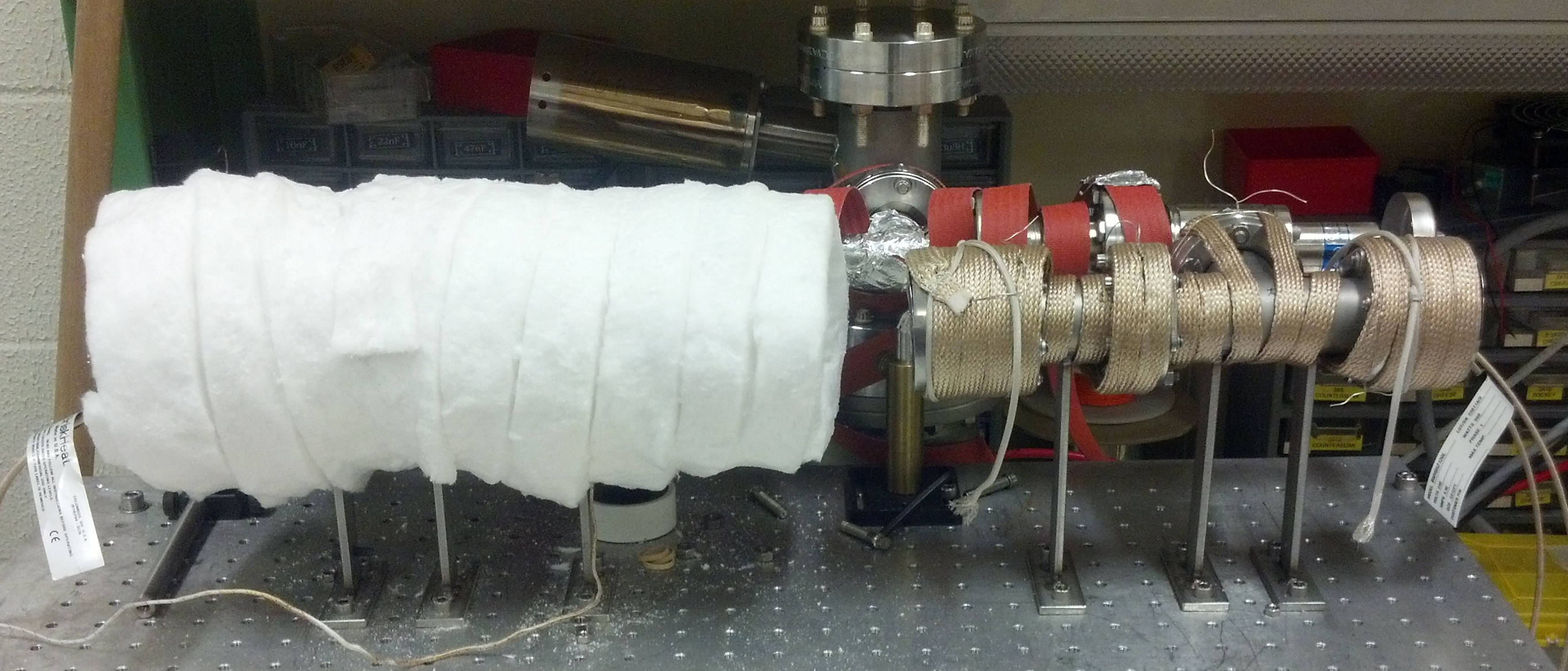}
  \end{center}
\caption{Calibration cells showing the insulation (left) which is wrapped over the high temperature heater tape (left).}
\label{cellsbare}
\end{figure}

While convenient in wavelength these metals have very low vapour pressure at room temperature, requiring heating to upwards of $500\degree C$ to produce a vapour density useful for spectroscopy. To this end we built two high temperature spectroscopy cells (see figure \ref{cellsbare}) out of stainless steel vacuum connectors, one containing thulium metal and the other containing indium foil. The cells were heated by high temperature heater tape ($\sim 1000\degree C$) wrapped around the outside (see right of figure \ref{cellsbare}), insulated with glass wool (see left of figure \ref{cellsbare}) and pumped to an ultimate vacuum of $2*10^{-5}$pascals by a turbo-molecular pump backed by a rotary vane pump.

\begin{figure}[H]
  \begin{center}  % center environment centers the graphic on the page
    \includegraphics[height=7cm]{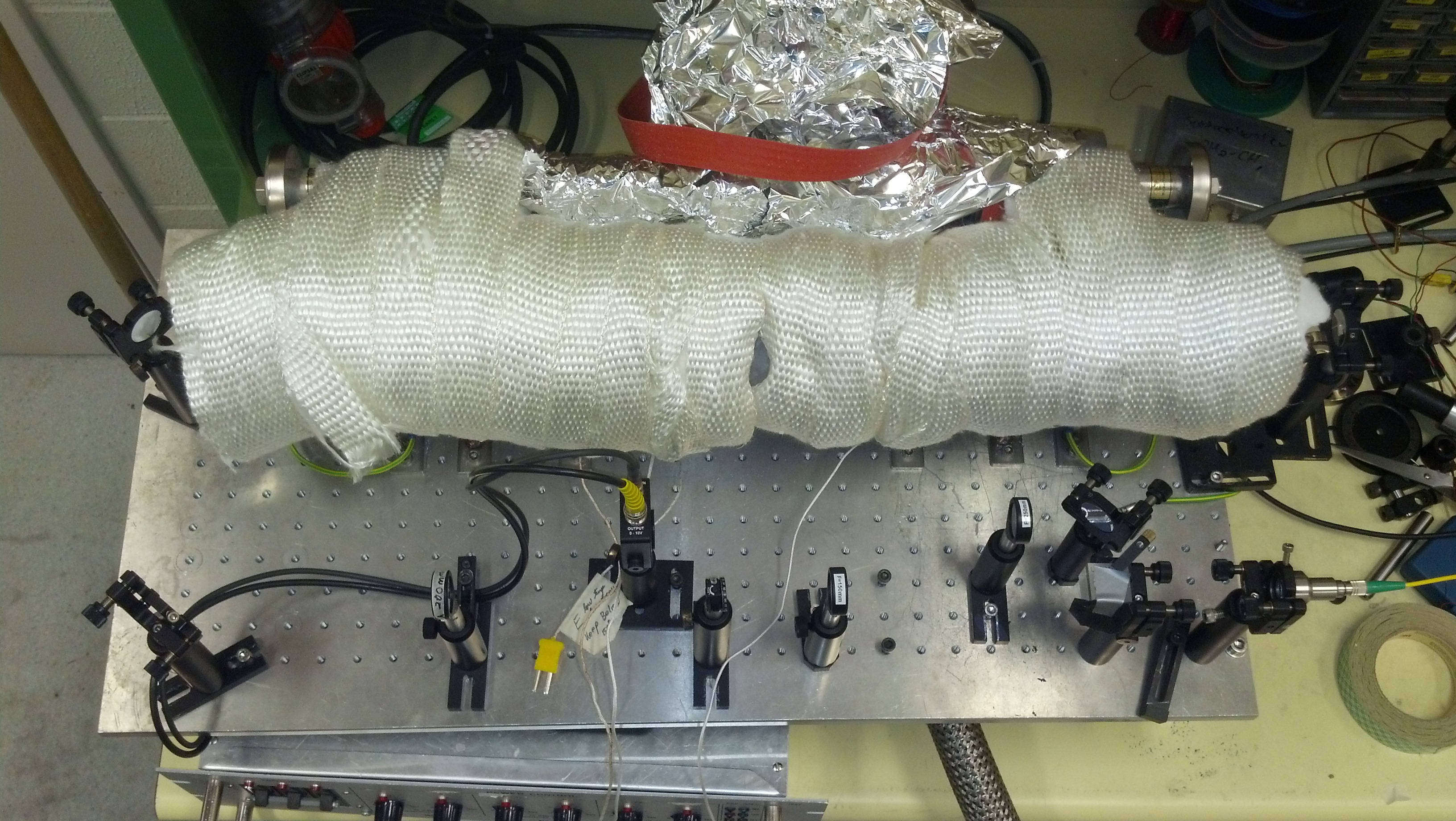}
  \end{center}
\caption{The final Calibration Cells with saturated absorption spectroscopy optics, high temperature insulation (front) and low temperature insulation (back).}
\label{cellsoptics}
\end{figure}

To reduce the (3~GHz) broadening due to the thermal velocity of the atoms that would otherwise prevent a high accuracy calibration we use a technique known as saturated absorption spectroscopy \cite{Metcalf}. Here we apply two counter propagating laser beams with the same frequency. When they are sufficiently close to the atomic transition some subset of atoms will be Doppler shifted onto resonance with each of the beams due to their velocity along the laser beam axis. When the beams are off resonance (for an atom at rest) these subsets are different and each beam is absorbed by some amount. However when the frequency is on resonance (for an atom at rest) then each beam addresses the same atomic subset that has zero velocity along the beam axis. The absorption of one beam by the atoms causes a decrease in the population of atoms in the ground state and that are able to absorb the other beam. This saturation of the absorption is used to recover a signal that approaches the zero temperature absorption spectra. We will use a simple implementation that passes a beam through the cells, reflects it over the same path as the original beam and then measures the beam power with a photo-diode (\autoref{cellsoptics}). By modulating the input beam frequency and using a lock in amplifier with the photo diode signal a derivative signal can be produced which crosses zero at the peak of the saturation dip. This signal can be used with a PID feedback loop to `lock' a laser with high stability to the calibration transition wavelength. This method is used in the main experiment to lock the cooling laser to the relevant atomic transition in He*.

The first step of setting up a spectroscopy system is to simply observe absorption due to the presence of the species of interest, however this was not achieved in these cells. First the ambiguity over the thulium transition wavelength means that one cannot simply dial up a wavelength but must instead hunt through a small range, second the uncertainty in the temperature (and in turn pressure) necessary to see a reasonable absorption signal is large. The temperature issue was further complicated  by operating the vacuum windows outside their rated temperature range. Precautions were taken in order to reduce the thermal stress on the windows by providing a uniform heating (see the right of \autoref{cellsbare}) and changing the temperature gradually. Unfortunately these did not prevent the windows from imploding at $580\degree$C below the $600\degree$C used for the thulium spectroscopy oven in \cite{Sukachev2010,Sukachev2010a}. Time constraints prevented the cells from being redesigned for higher temperature operation. However this will need to be resolved to achieve the ultimate accuracy of the experiment.

\section{Potential Measurement}

%outline\\
%-once we have the experiment above how do we measure the tune-out
%-many methods investigated although few piratical (refractive index,parmetric heating , trap frequency)

With a stable, well focused, tunable probe beam the critical next step in an accurate measurement of the tune-out wavelength is the ability to measure the small dynamic polarizability (and resultant energy shift) of the atoms around the tune-out wavelength. This has many possible solutions, from measuring the force applied due to a spatial dependent potential to a direct measurement using the phase rotation rate of a condensate in an interferometer. The combination of the small probe beam power delivered into the chamber($\sim3mW$) and the small gradient of the polarizability with detuning (see \autoref{potential}) means that to reach a reasonable level of uncertainty in this measurement requires a potential sensitivity on the order of pK and forms the limiting uncertainty in our measurement.

Using ultra cold atoms for high precision meteorology has become relatively commonplace due to the diverse and flexible array of techniques that have been developed.
Although limited work has been done towards directly measuring small optical dipole potentials. The closest that we are aware of is the smallest force ever measured \cite{Schreppler2014} at $\sim10^{-24} N$ and the measurement of the Casimir-Polder force gradient \cite{ObrechtPhd,Obrecht2007,Harber2005} at $\sim10^{-22} N/m$. For comparison reasonable experimental parameters (at 10pm detuning) would give a optical dipole force from the probe beam of\footnote{Using the maximum gradient of an pptical dipole potential with 3mw beam focused to a  $10\mu m $ waist.} $\sim10^{-27} N$ and gradient in force of $\sim10^{-23} N/m$. While neither is a direct potential measurement the fixed relation between them and the potential from a Gaussian beam (see \autoref{trapfreqodt}) means that they have promise if improved in sensitivity.  To this end we experimented with parametric heating \cite{Savard1997} and direct trap frequency measurement approaches although both were found to be insufficiently sensitive.

\subsection{Atom Laser Outcoupling}
\label{sec:theoryatomlaser}

Based on experience from these techniques we have developed and implemented an novel atom laser based potential measurement technique in order to measure these minuscule potentials around the tune-out wavelength. Here we continuously out-couple atoms from a magnetic trap using a RF knife (as described in \autoref{atomlaser}) measuring the rate on a detector below. When the focused probe beam near the tune-out wavelength is overlapped with the magnetic trap the net potential changes shape, which in turn modifies the density at the RF out-coupling surface and the subsequent out-coupling rate. We find that for a suitable choice of parameters this effect is highly sensitive and linear forming an excellent potential measurement technique.

\begin{figure}[H]
  \begin{center}  % center environment centers the graphic on the page
    \includegraphics[height=4cm]{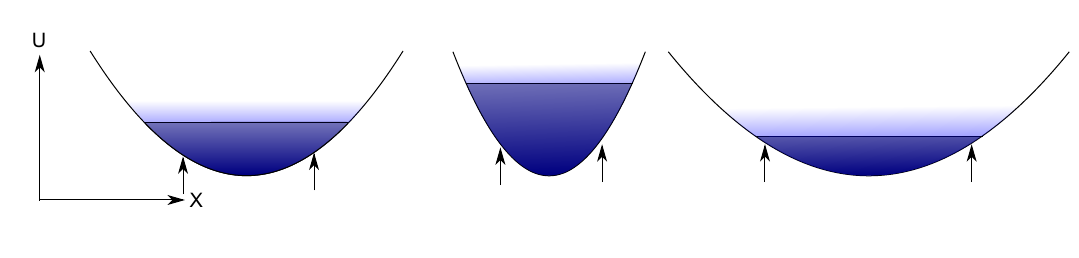}
  \end{center}
\caption{Phase space (potential-position) representation of the technique. The change in density (here represented as depth) from an unperturbed potential (left) is seen for a attractive (center) and repulsive (right) potentials at the RF out-coupling positions represented by arrows.}
\label{AtomLaserPhaseSpace}
\end{figure}

To investigate this method we can construct a simple model using the Thomas Fermi approximation and a two level ($m_f=0,1$) atom based on approaches in \cite{HainePhd}. We may approximate the magnetic trap as a three dimensional harmonic oscillator with a radial ($r$) and transverse ($z$) trapping frequency. We neglect gravity due to the small contribution to the potential relative to trapping energy.
\begin{equation}
U=\frac{1}{2}\omega^{2}_{rnet} m r^2+\frac{1}{2}\omega^{2}_{z} m z^2.
\label{hopotential}
\end{equation}
The chemical potential for a BEC as can be  found for a three dimensional trap with cylindrical symmetry \cite{Pethick2002} to be
\begin{equation}
\mu =\frac{1}{2} \left( 15^{2} N^{2} a^{2} m \hbar^{4} \omega^{2}_{z} \omega^{4}_{r} \right)^{1/5}
\end{equation}
To find the density of the BEC we use the Thomas-Fermi approximation 
\begin{equation}
\rho=\frac{1}{g}(\mu -U)
\end{equation}
where $g=4\pi \hbar^{2} a/m$  \cite{Pethick2002}.
The outcoupling rate is proportional to the surface integral of this density across the isomagnetic surface given by:
\begin{equation}
U=\frac{1}{2}\omega^{2}_{r0} m r^2+\frac{1}{2}\omega^{2}_{z} m z^2.
\end{equation}
Here we have taken the low power limit of the RF field and neglected coupling of atoms back into the potential once out-coupled. This  can be rearranged to give the out-coupling radius as a function of z position
\begin{equation}
R_{out}(z)=\sqrt{\frac{U_{mag}-\frac{1}{2}\omega^{2}_{z} m z^2 }{\frac{1}{2}\omega^{2}_{r0} m}},
\end{equation}
however this only holds when less than the Thomas-Fermi radius in the Z direction
\begin{equation}
z\leq \sqrt{\frac{2 U_{mag}}{m \omega_z^{2}}}.
\end{equation} 
The radial symmetry here allows for a simple surface integral to be conducted involving only the z coordinate to give the infinitesimal of the out-coupling density $d\eta$.
\begin{equation}
\label{bigint}
d\eta=2\int^{ \sqrt{\frac{2 U_{mag}}{m \omega_z^{2}}}}_{0} \rho(R_{out}(z),z) \pi R_{out}(z) dz
\end{equation}
In the approximation that the probe beam is larger than the condensate, the change in potential can be represented as a simple change in the radial trapping frequency.
\begin{equation}
\omega^{2}_{rnet}=\omega^{2}_{r0}+\omega^{2}_{probe}\quad
\end{equation}
The trap frequency for the probe beam is given from the derivation in \autoref{trapfreqodt} where $\alpha$ is given by the linear approximation about the tune out (\autoref{alphasi}).
\begin{equation}
\omega_{probe}=\sqrt{\frac{1}{2m \epsilon_{0} c} Re(\alpha) I_{0} \frac{4}{w_{0}^{2}}}
\label{trap freq}
\end{equation}

The proportional change in the out-coupling rate can then be found by computing the integral in \autoref{bigint} and then taking the difference between the case when $\omega^{2}_{probe}=0$ and that given by \autoref{trap freq} divided by the $\omega^{2}_{probe}=0$ case. This is a complicated procedure that we carried out using symbolic computation (code included in appendix \ref{outcouplingmath}). It must be stressed here that the value $d\eta$ is not the direct out-coupling rate, to produce such a prediction we would need to integrate over the out-coupling `width' determined by the magnetic sub-level transition linewidth.

Using the symbolic result with reasonable experimental parameters\footnote{$\omega_{r}=500\cdot 2\pi$, $\omega_{z}=50\cdot 2\pi$, $N=9*10^5$, $P=3mW$, $w_{probe}=10\mu m$} we can examine the influence of the probe beam on the out-coupling rate. As expected an out-coupling closest to the Thomas Fermi radius provides the greatest signal. The main motivation for this theoretical examination is to determine the linearity of the technique with detuning. 
\begin{figure}[H]
  \begin{center}  % center environment centers the graphic on the page
    \includegraphics[height=7cm]{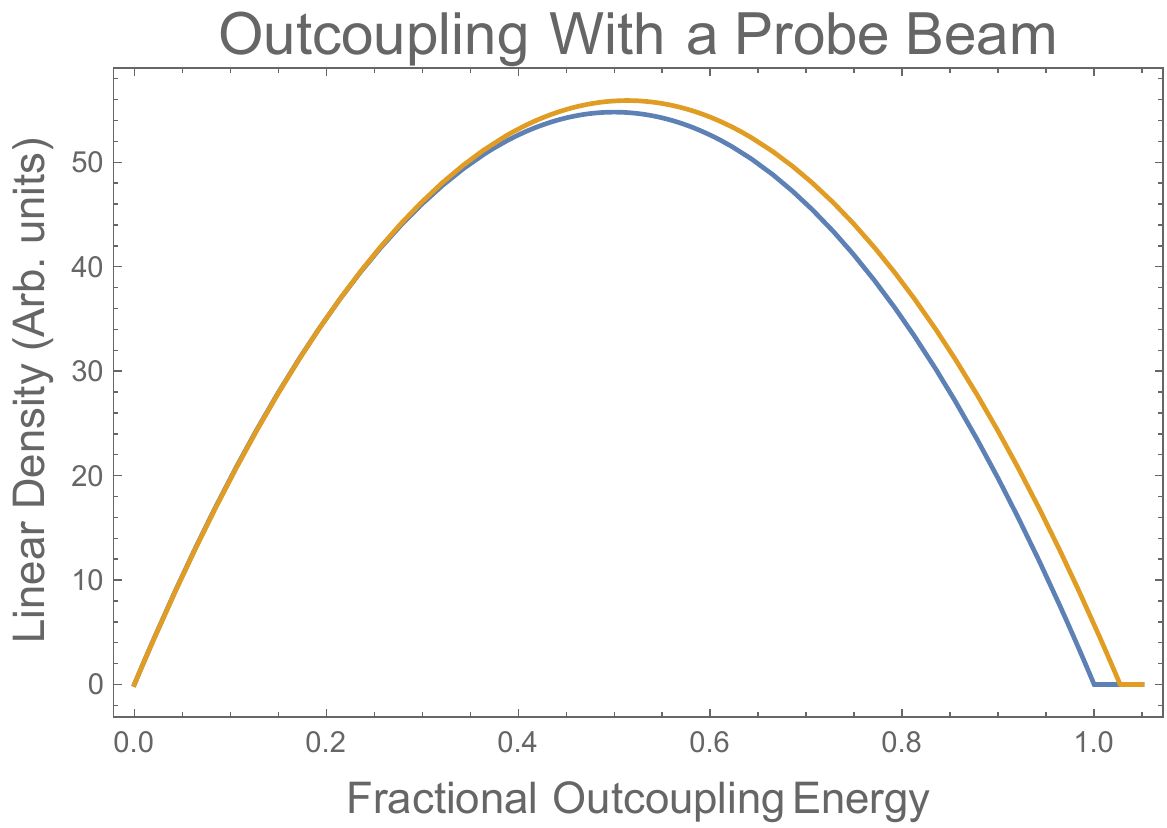}
  \end{center}
\caption{The predicted out coupling number as a function of the magnetic isosurface energy with (blue) and without (orange) the applied probe beam at 1nm detuning from the tune out.}
\end{figure}

\begin{figure}[H]
  \begin{center}  % center environment centers the graphic on the page
    \includegraphics[height=7cm]{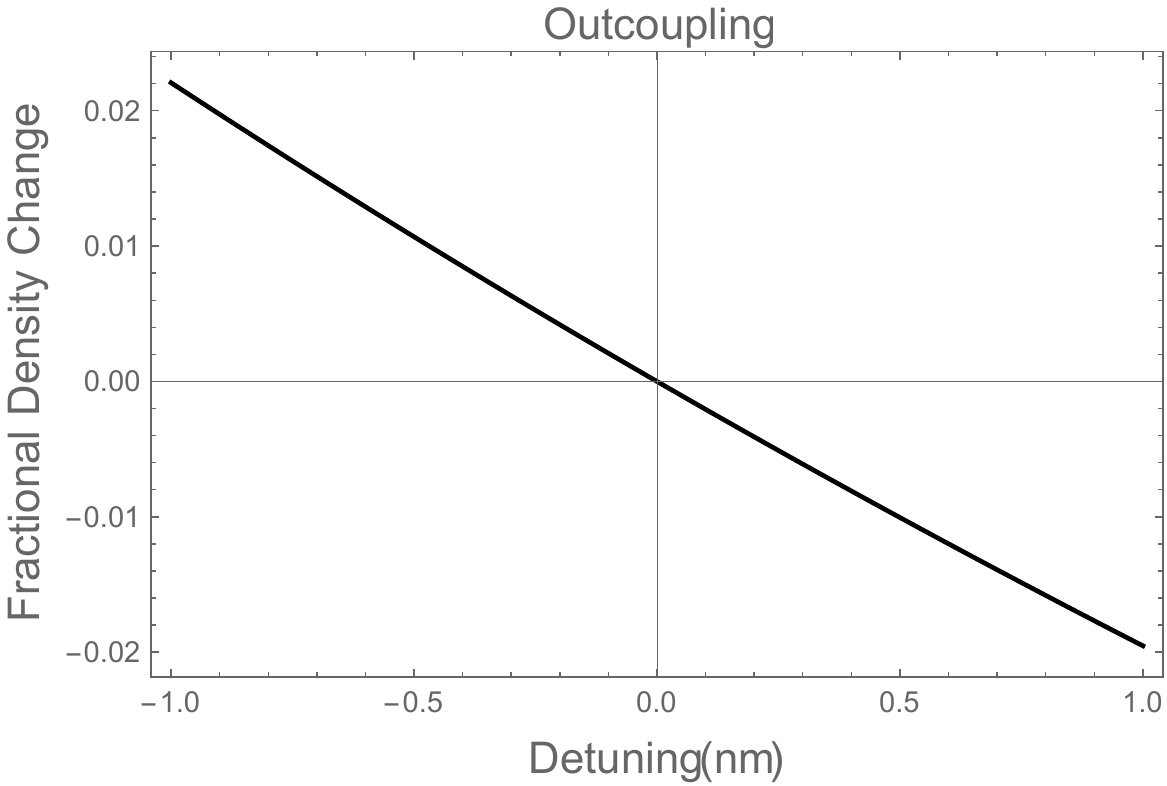}
  \end{center}
\caption{The relative change in out-coupling number as the probe beam detuning is changed. Here the out-coupling energy is chosen to maximize the rate at 0.5. }
\label{atomlaserlinearity}
\end{figure}

The relative change in out-coupling seen in \autoref{atomlaserlinearity} is approximately linear with the probe beam detuning. This combined with its relatively high sensitivity makes it the ideal candidate for our measurement of the tune-out wavelength. The fundamentally limiting factor in such a measurement is shot noise in the the measurement of the relative change in out-coupling rate given by $\sqrt{N}$ where N is the number of particles considered.

In the following chapter we describe the use of this out-coupling technique to sensitively measure the He* tune-out wavelength.

%% file: chapter4.tex
\chapter{Results and Discussion}
\label{results}
%[Grading Goals]\\
%---demonstrate technical mastery \\
%---demonstrate understanding of the limitations of the experiment\\
%---demonstrate ability to work independently???\\
%---persistence under difficulty\\

To make a measurement of the  $3^{3}P$ to $2^{3}P$  tune-out we combined a perturbing laser system with the atom laser based potential measurement technique.  We found that the sensitivity of the system was initially quite poor, so we developed a resonant-lock-in technique to improve the sensitivity by many orders of magnitude. Using this improvement we were then able to make a measurement of the tune-out wavelength at an uncertainty of 20pm which was predominantly limited by statistical uncertainty.

\section{Atom Laser Sensitivity}
\begin{figure}
  \begin{center}  % center environment centers the graphic on the page
    \includegraphics[height=7cm]{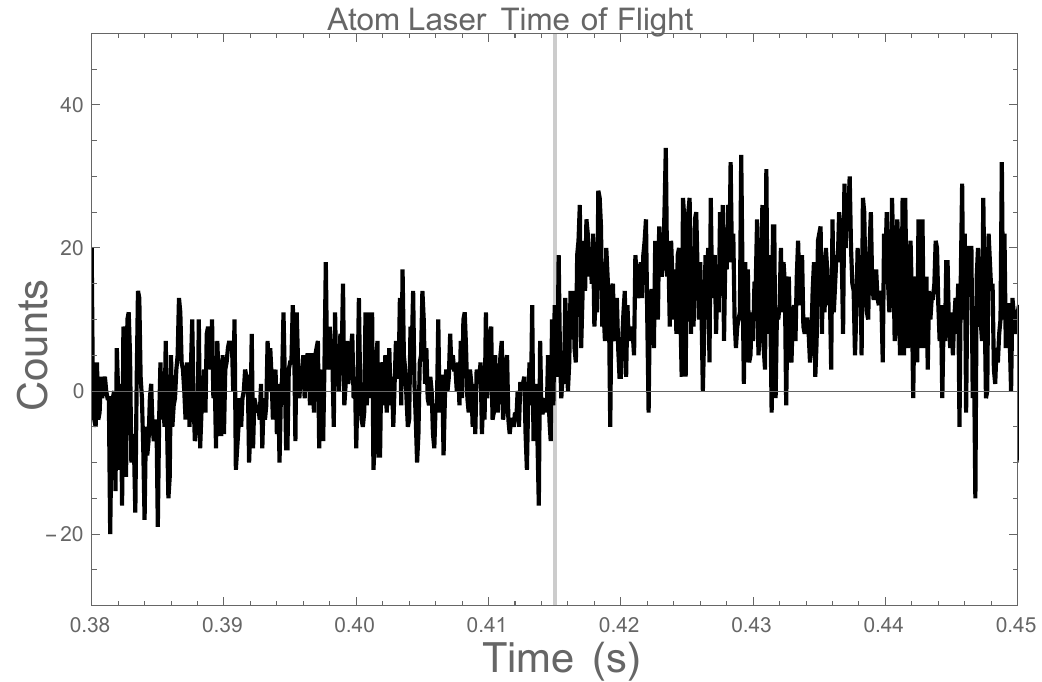}
  \end{center}
\caption{Effect of a probe beam (64mW at 405nm) on the out-coupling rate of the atom laser. Average out-coupling rate 100 counts per bin. The vertical line denotes when the probe beam is switched off.}
\label{DCatomlaser}
\end{figure}

To  determine the sensitivity of the atom laser technique detailed (see \autoref{sec:theoryatomlaser}) we first prototyped this system using a 405nm probe beam laser combined with the multi stage alignment technique (described in \autoref{alingment}). In this measurement we apply the probe beam to the magnetic trap and continuously sample the density at the out-coupling surface with an atom laser. By switching the applied 405nm beam off halfway through the out-coupling we then determine the change in the density. To produce a reasonable signal the time of flight data from 10 such runs was taken and the background subtracted by taking another ten with the 405nm beam off the entire time. The time of flight profile of this average is shown in figure \ref{DCatomlaser}.

 Here we can see significant change in the atom laser signal of $\sim$20 counts per bin caused by the probe beam after background subtraction. For comparison between methods we will define the sensitivity as the power (in mW) times detuning from the tune-out wavelength (in nm) divided by the signal to noise ratio for a single shot
\begin{equation}
\mathrm{Sensitivity}=\frac{P\Delta \lambda}{\mathrm{SNR}}.
\end{equation} 
  This value can be interpreted as the minimum detuning that could be resolved in a single shot with a single mW of beam power. For the above method this gives a sensitivity of $\sim 100$ mW$\cdot$nm, far to large to be practically useful for a measurement. Additionally this method is particularly prone to drifts in the out-coupling number as a function of time, this would cause the net difference between the two sides.
  
 Inspired in part by the lock in technique used in \cite{Viering2012} to align a MOT, we then used a modulation of the probe beam power to create a periodic dependence in the out-coupling number. The effect could be analysed by taking the real argument of the discrete Fourier transform of the out-coupling signal. By modulation at a frequency away from any background frequency components (50Hz and harmonics thereof) we were then able to routinely measure a potential with many orders of magnitude improvement in sensitivity.  This type of modulation and frequency space filtering is known as a lock-in technique. By taking the Fourier transform we unfortunately remove any sign(repulsive/attractive) information of the polarizability, although this will not affect the minimization method used to find the tune-out wavelength. As the dynamics of the condensate have a characteristic time-scale on the order of the inverse trap frequency when the modulation frequency of the probe beam far slower it reasonable to expect that the theoretical description in \autoref{sec:theoryatomlaser} still holds.
 
\begin{figure}
\centering
\subfloat[]{{\includegraphics[height=4.5cm]{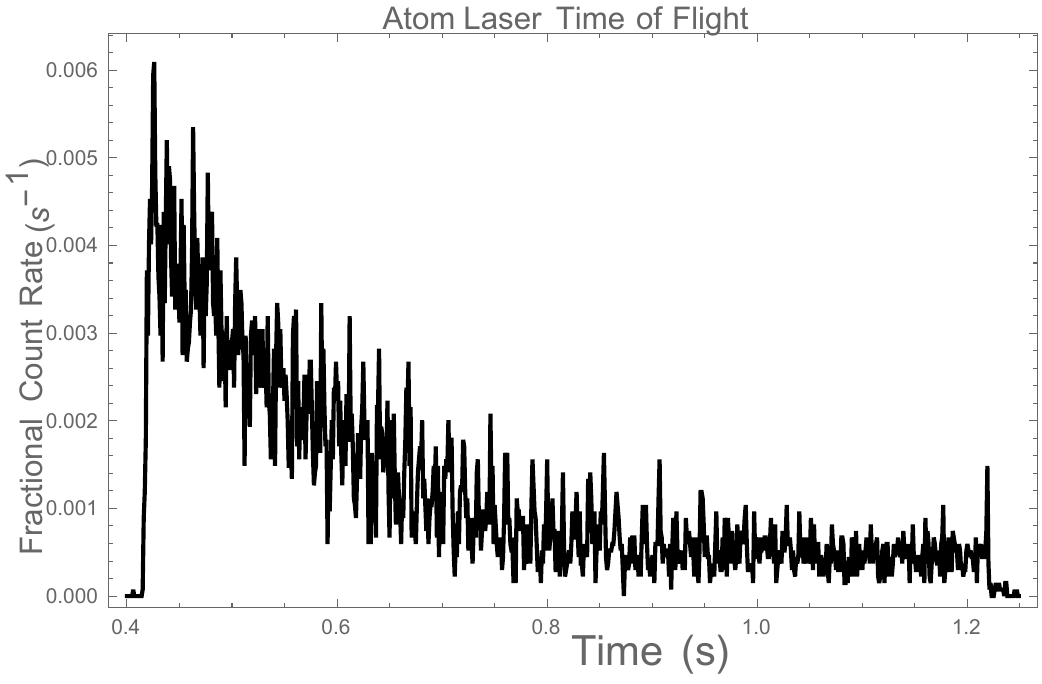} }}%
\qquad
\subfloat[]{{\includegraphics[height=4.5cm]{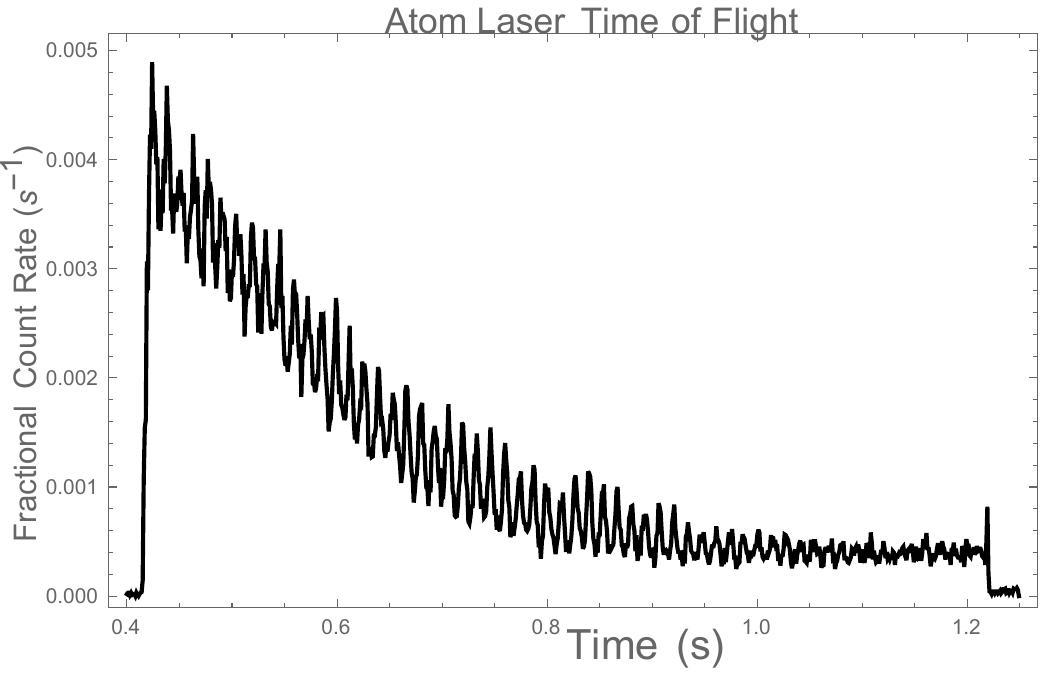} }}% 
\caption{Time of flight of an atom laser for a single shot(a) and 10 shots of TOF averaging (b).}
\label{AtomLasertof}
\end{figure}

To test this method we again used a 405nm diode laser although now with a 75Hz modulation on the drive current. This produced a very obvious signal in the time of flight plot. To test the sensitivity of the technique the power was reduced until no effect could be seen from a single shot in the time of flight plot (\autoref{AtomLasertof}) which was at 1.2mw of modulation.

\begin{figure}
\centering
\subfloat[]{{\includegraphics[height=4.5cm]{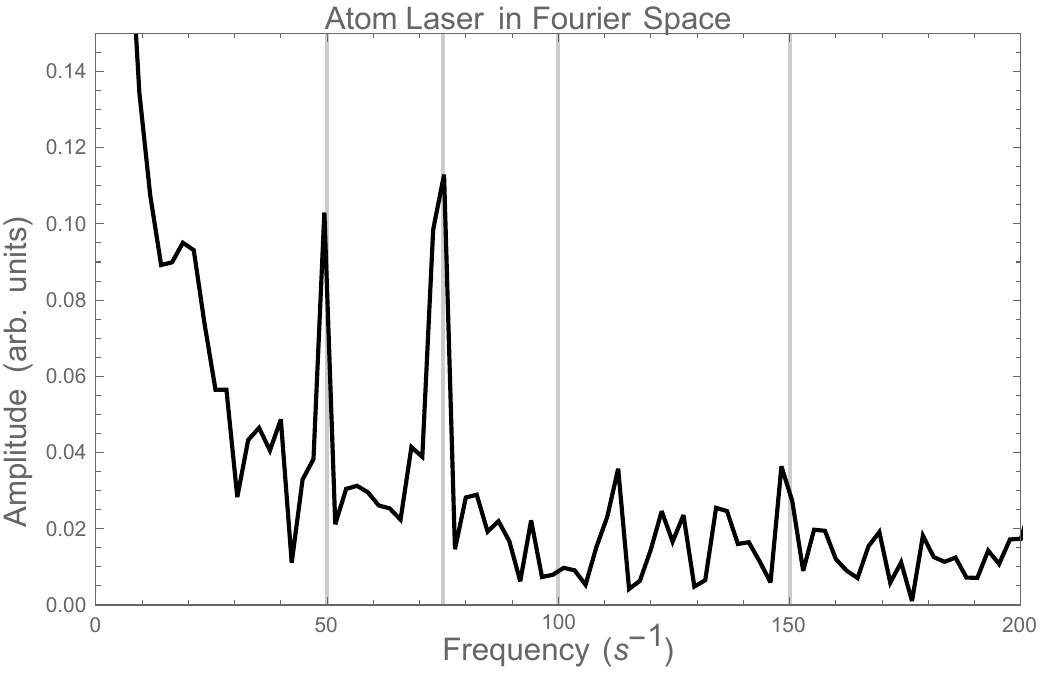} }}%
\qquad
\subfloat[]{{\includegraphics[height=4.5cm]{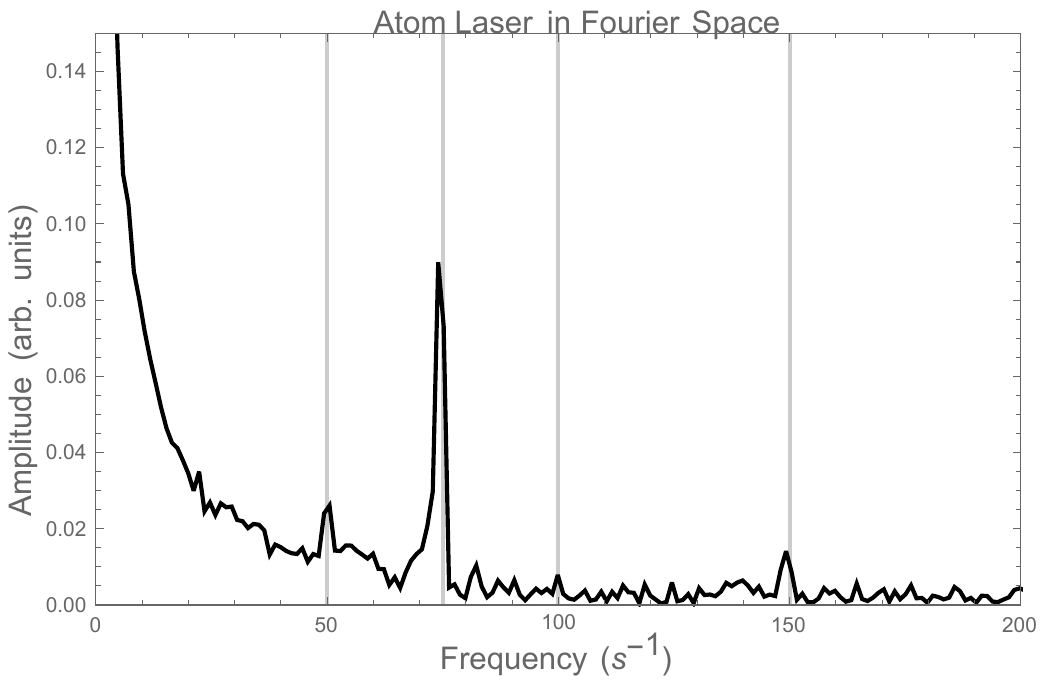} }}%
\caption{The Fourier space of an atom laser for a) a single shot and  b) 10 shots of TOF averaging. Vertical lines show the 50Hz (mains noise) and 75hz (modulation) frequencies of interest and their harmonics.}
\label{AtomLaserSpectrum}
\end{figure}

By taking the Fourier transform of this time of flight of the atom laser the effect of the probe beam is readily apparent (\autoref{AtomLaserSpectrum}). As the 50Hz background is not phase locked with the experimental sequence as the modulation is, averaging over may shots nearly eliminates this component in Fourier space and produces a far clearer signal in the time of flight. The Fourier spectrum was converted into a single value measurement by taking an average over small region around the modulation frequency. To measure the uncertainty in this value each shot was analysed separately and the standard deviation taken. The sensitivity of this techniques is then $\sim 1$ mw$\cdot$nm, two orders of magnitude better than the DC case. As this method is relatively insensitive to the out-coupling rate it is then possible to ramp the out-coupling frequency with time in order to improve the signal level.

To improve the signal by a further order of magnitude we modulate the probe beam near the transverse trap frequency where a resonant enhancement can be produced. This regime is not covered by the theory described in the previous section, the rigorous treatment of which would likely require the use of the Gross–Pitaevskii equation to describe these resonant dynamics. We believe the excitation of collective oscillations is responsible. To this end we use a modulation equal to the trap frequency at  $\omega=2 \pi 491 s^{-1}$ (as determined by the greatest signal) on the probe beam.

\begin{figure}[h]
  \begin{center}  % center environment centers the graphic on the page
    \includegraphics[height=7cm]{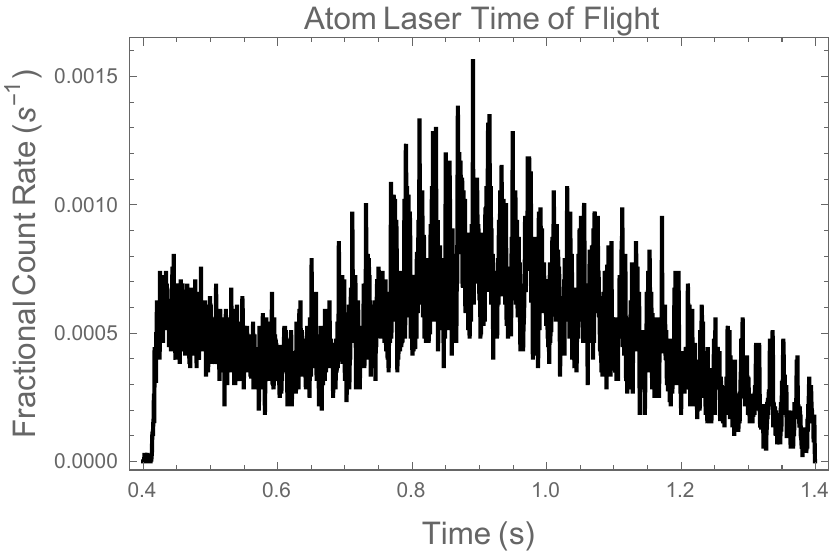}
  \end{center}
\caption{Time of Flight of a single shot with a 414.8nm, 3.3mw probe beam and resonant modulation at 491Hz.}
\end{figure}

\begin{figure}[h]
\centering
\subfloat[]{{\includegraphics[height=5cm]{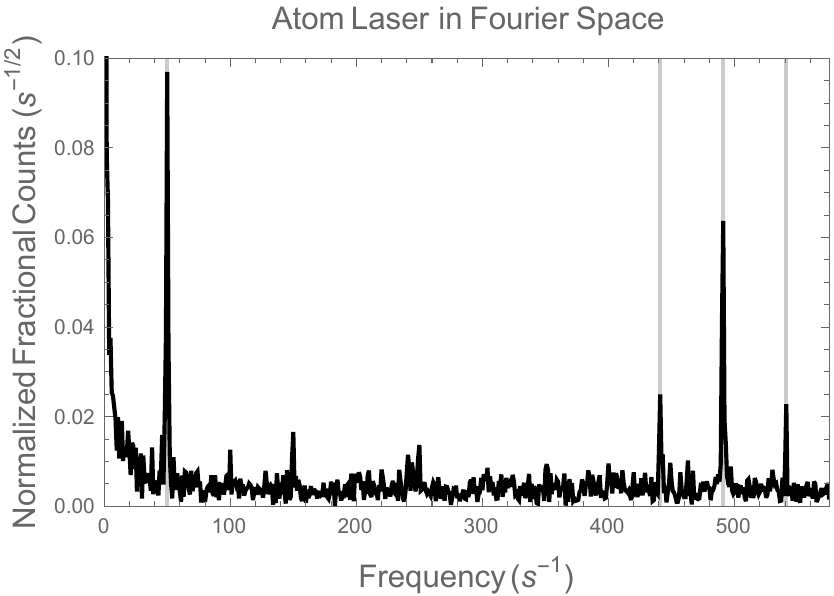} }}%
\qquad
\subfloat[]{{\includegraphics[height=5cm]{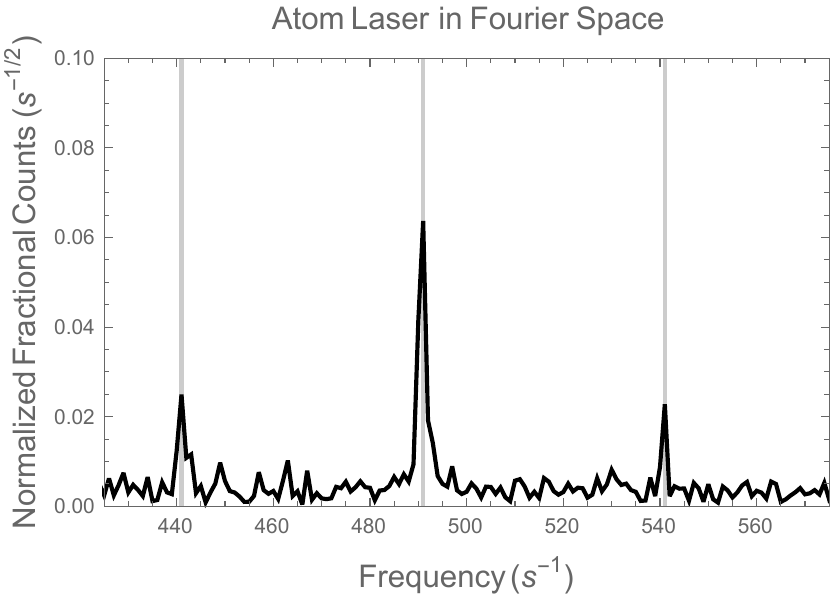} }}%
\caption{The Fourier space of an atom laser for a single shot showing a) the entire range and b) a smaller region around the modulation frequency. Vertical lines show the 50Hz(mains noise), 491Hz modulation with 50Hz side-bands. These side bands are evidence of a non-linear process mixing the modulation frequency with the AC noise in the out coupling rate.}
\label{resmodatomlas}% 
\end{figure}
\begin{figure}[H]
  \begin{center}  % center environment centers the graphic on the page
    \includegraphics[height=7cm]{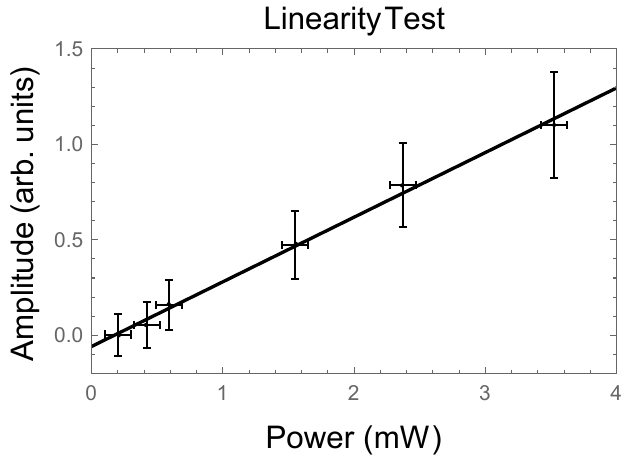}
  \end{center}
\caption{Linearity of the resonant modulation technique in power, at a probe beam wavelength of 415.2nm. Errors in power and amplitude are determined by statistical uncertainty. }
\label{linearity}
\end{figure}

%talk about how this is a super sensitive measurement
This resonant modulation combined with a more aggressive out-coupling then produces a sensitivity $\sim 0.1$ mw$\cdot$nm which is sufficient to produce a measurement of the tune-out wavelength  with a reasonable accuracy. We measured the linearity of this measurement technique by changing the power at a fixed detuning. The resultant dependence (see \autoref{linearity}) is linear within errors showing that the Fourier space amplitude is directly related to the energy shift of the ground state.

\section{Tune-Out Wavelength}
We used this resonant lock-in technique to produce the first ever measurement of the the $3^{3}P$ to $2^{3}P$ tune-out wavelength in He*. We first took measurements over the lasers tuning range, changing the wavelength in half-nanometre steps in order to reduce the region of interest for a more precise tests. The chromatic shifts of the focusing optics with wavelength (\autoref{probe laser}) required the focus to be re-adjusted at each new wavelength. This produces a large uncertainty in the magnitude of the measured values, although as indicated previously this does not change the zero crossing location only the slope on either side.

\begin{figure}[H]
  \begin{center}  % center environment centers the graphic on the page
    \includegraphics[height=9cm]{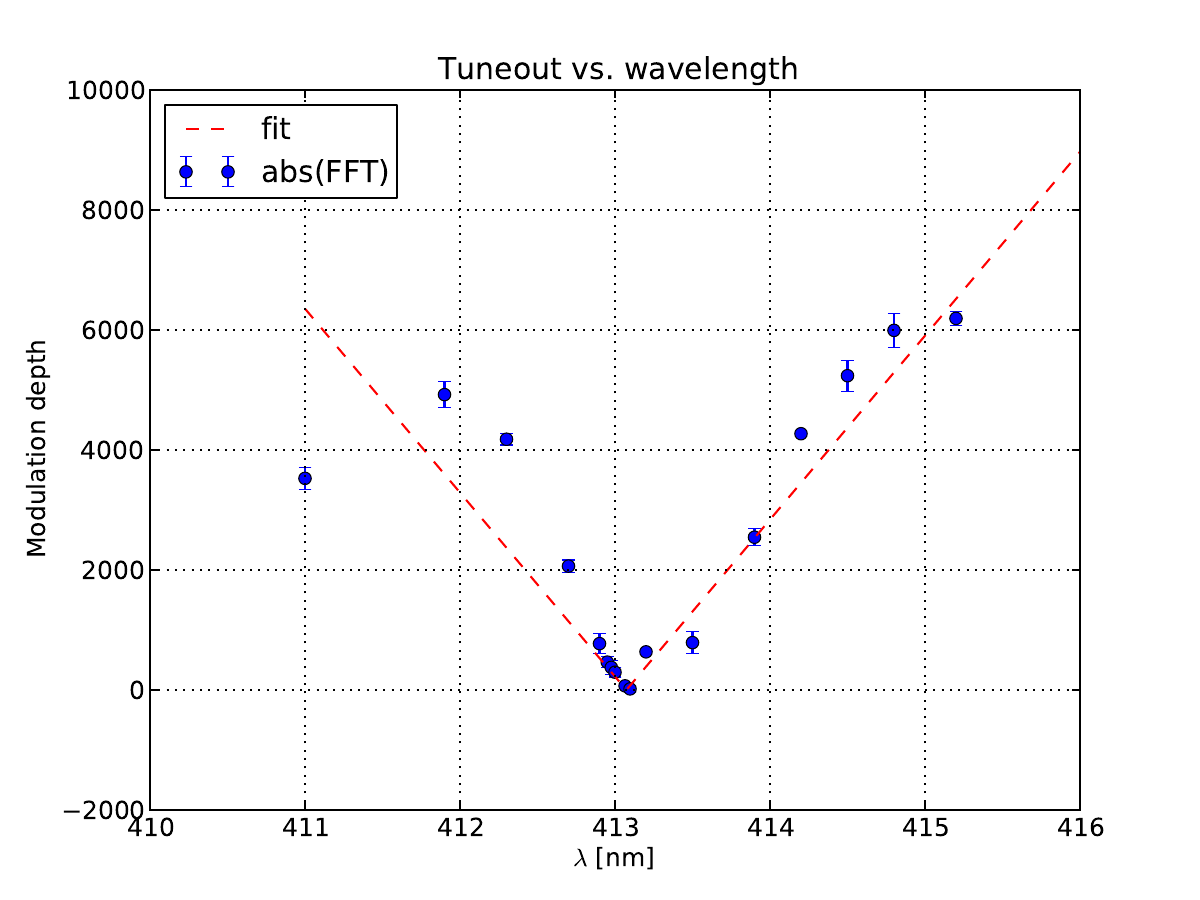}
  \end{center}
\caption{Initial measurement of the tune-out with error bars showing statistical errors. This error is dwarfed by the large intensity change due to chromatic shifts of the focal length which we estimate from \autoref{waist} at $\sim 30\%$ of the measurement value.}
\end{figure}

By then reducing the scan region to $\sim0.3nm$ around the tune-out wavelength as determined from the above measurements, the error in the intensity due to chromatic shifts of the focal length is reduced significantly.
\begin{figure}[H]
  \begin{center}  % center environment centers the graphic on the page
    \includegraphics[height=9cm]{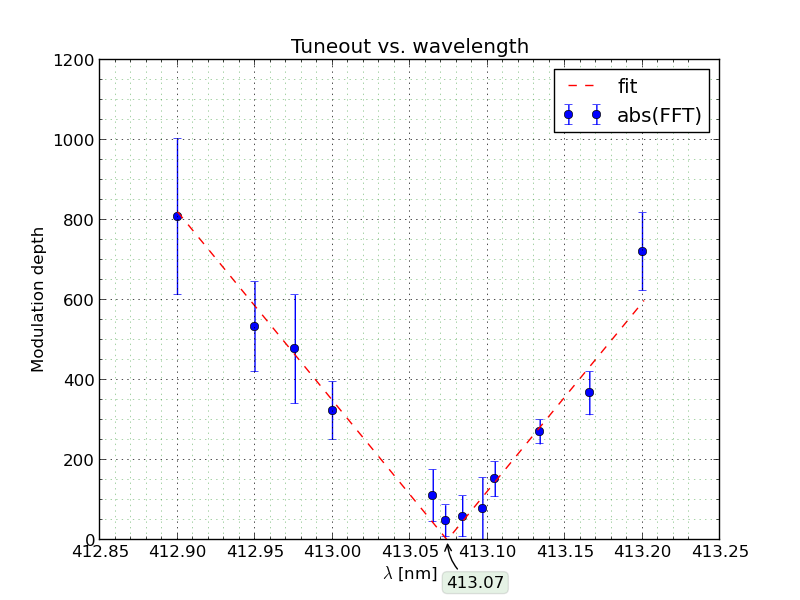}
  \end{center}
\caption{Narrowing the region of interest then allows for a far more accurate measurement of the tune out. Error bars are determined by statistical error (standard deviation) in the measurement amplitude over many samples and then scaled by the square root of the number of samples. The zero crossing is determined by a  linear fit to be 413.07(2)nm.}
\end{figure}

Here we produce our best measurement of the tune-out wavelength at 413.07(2) nm. This measurement is the first of its kind in helium and lays the ground work for future measurements to test atomic structure theory. With a value that is consistent with the theoretical prediction by \cite{Mitroy2013} at 413.02(9)nm, we have good validation of both our experimental methods and the theoretical predictions. We also believe this is one of the most sensitive measurements of an optical dipole potential. Taking 0.01nm as the minimum distinguishable signal, using the gradient of the dynamic polarizability around the tune-out as given by \cite{Mitroy2013} combined with our measurement of the beam power and waist we determine the potential sensitivity to be $\sim200$~pk. To our knowledge this is the most sensitive measurement of an optical dipole potential to date\footnote{A distinction must be made here that we only claim a high sensitivity and not a high accuracy as we have no calibration against a well known potential or force.}. Further using the beam waist and potential depth we can derive the corresponding maximum force imparted by the potential to the atoms. At our lowest detuning this corresponds to a sensitivity in force at $\sim4\cdot10^{-28} N$ which compares well\footnote{See 1} against the record \cite{Schreppler2014} for a direct force measurement at $\sim10^{-24} N$.

The sensitivity of this measurement technique means it may have a wide variety of uses in other atomic systems. One such example would be a Rb atom laser where the out-coupling signal would be derived from absorption imaging of the out-coupled beam. The same lock in technique could then be applied for detecting small potentials (that can be varied in time). This could be used for a improved \cite{Herold2012} tune out measurement in Rb or for measuring the Casimir-Polder force \cite{Obrecht2007}, among others.

\section{Error analysis}
\label{toerror}

The error in this measurement is dominated by the statistical error of the Fourier space amplitude at the driving frequency. We believe that this short term error is predominantly due to the shot noise in frequency space and could be improved by a factor of $\sqrt[]{N}$ where N is the number of samples. To this end the measurements closest to the tune-out were done with many hundreds of `shots' and such significant improvements would require an unrealistic amount of time. The medium to long term error in this measurement is expected to be due to the positional stability of the probe beam which has been shown to vary $\sim 10\mu m$ and drift (see figure \ref{probestability}) on the order of $1\mu m/ day$. However as the beam has a waist on the order of $10\mu m$ we expect the corresponding change in intensity to be less than a few percent over time-scales smaller than a few days. A small contribution to this uncertainty was the power delivered into the chamber, this was manually set at the beginning of each set of data and monitored throughout giving an uncertainty on the order of \%5. In future this would be improved with a feedback system. 
 
 The uncertainty in the wavelength measured using the Moglabs wavemeter was estimated at 500~fm by comparison with a much higher accuracy wave-meter, as such its contribution to the uncertainty for this measurement may neglected. A particularly important consideration to the error analysis is the wavelength purity of the laser. If some fraction of the power was outside the main laser mode then the tune-out could be easily shifted in proportion detuning and power of this spurious mode. To this end we have used the Moglabs wavemeter as a high resolution spectrometer, we were unable to measure any power outside the main peak in frequency space when the laser was adjusted properly. To further this investigation we intend to enhance the dynamic range and corresponding sensitivity by exposing the spectrometer CCD at multiple exposures. 

Due to experimental methods used the prediction of the tune out by Mitroy and Tang \cite{Mitroy2013} does not correspond directly to our experiment. The most obvious of these is the Zeeman shift from confining the atoms in a magnetic potential. This splits the magnetic sub levels by $\sim$1~MHz, while in the calculation no magnetic field was present. It may be possible to eliminate this effect experientially by measuring the dependence of the measured tune-out wavelength with magnetic Field and extrapolating to zero.

 Additionally the spin polarization in this trap changes the relative pumping of transitions from the un-polarized case as was calculated. These combined effects could account for the 0.05~nm discrepancy between prediction and measurement. Additionally these predictions took the limit of a low probe beam intensity where higher order contributions known as hyperpolarizability are negligible. For our current probe beam intensity this limit is reasonable although if a higher laser power was used these may need to be examined. Further theoretical work is therefore needed for the experimental conditions in our experiment.

The electrostatic interaction of the confined atoms may cause a perturbation in the energy levels leading to another source of error. The order of magnitude for this effect can be easily given by the mean field potential in the cloud which is on the order of $\mu K$ for this experiment. Converting this to frequency gives some approximation of the shift in energy levels at $\sim$20~ kHz, and is insignificant in this and envisioned work.

%% file: chapter5.tex
\chapter{Conclusion}
\label{Conclusion}
By developing a novel atom-laser-based potential measurement technique we have been able to make the first measurement of the tune-out wavelength in He* between the $3^{3}P$ and $2^{3}P$ states at 413.07(2)~nm. In doing so we have measured some of the smallest optical dipole potentials ever measured and produced a validation of theory with a prediction of  and experiment. We have additionally developed the tools necessary for the next generation measurement which will be able to make a stringent test of atomic structure predictions. The potential measurement technique we have developed is also widely applicable to a number of other problems in atoms optics.

\section{Future Work}
\label{FutureWork}
%[goals]\\
%---outlined future research directions that are feasible and show originality

While we have not tested QED in this measurement we have developed a number of the techniques necessary to do so while gaining invaluable insights into the further development that are required. With this measurement we hope to encourage further high accuracy calculations to be done which we will be able to test with further measurements. A number of improvements have been planed to this experiment.

\subsection{Laser System}
The simplest and most beneficial modification would be an increase in probe beam power. The low laser power, diffraction efficiency of the AOM and coupling efficiency conspired to leave only 3.3~mw to enter the vacuum chamber and interact with the atoms. If this power was improved by a hundred fold the sensitivity of our measurement would improve by the same factor allowing us to reach the criteria \cite{Mitroy2013} of 100~fm accuracy to test QED.
While a narrow linewidth titanium sapphire laser laser (600-900nm) could be frequency doubled in a nonlinear optical crystal to produce upwards of a watt of widely tunable light such a laser system would cost upwards of 150,000\$ (AU) \cite{m2laser}. Alternatively a fibre laser at 1650nm could be frequency quadrupled (through two doubling stages), although the lack of high power systems at the original wavelength combined with a limited doubling efficiency also makes this method impractical.

A pragmatic option would be to injection lock a heated 405~nm diode. Here a seed laser in the form of our probe laser is sent into the diode by overlapping it with the output, the injected light is then amplified suppressing the normal lasing modes and a high power version of the seed laser is emitted. This process requires the cavity mode of the free running laser's wavelength to be as close to the injection wavelength as possible and to this end we could heat a 405~nm diode to its limit. If this fails then we could attempt to add an anti reflective coating to the diode in order to suppress the cavity mode and allow injection at lower temperatures.  The high output power $>1$~w obtainable with lasers at this wavelength mean that orders of magnitude improvement in power may be possible.

To match this improvement the focusing system for the probe beam would need to be improved. In particular it must be redesigned with the goal of all but eliminating the chromatic shifts that have produced large errors in the focal position. To this end we will simulate the lens system in an optical modelling program.

By using a dichroic mirror which reflects one wavelength and lets another pass unimpeded we can then introduce our probe beam into the chamber on the vertical axis.  This would then reduce the distance to the BEC allowing for a smaller spot size while also improving the atom laser based potential measurement. Here the small axial trapping frequency of the magnetic trap would produce better sensitivity to the radial trap frequency of the probe beam. \\ 

In order to take advantage of this improvement in sensitivity we will then require a far more accurate and stable laser system than used in this measurement as developed in section \ref{wavemeterlock}.  To this end we will redesign the calibration spectroscopy cells to tolerate far higher temperatures and provide us with a stable wavelength reference. To reduce the temperature the windows experience we will use a single long tube, the spectroscopy metals (thulium and indium) will be placed in the center which will be heated while the ends will be water cooled. This design was originality discarded due to concerns that the metal would condense on the windows reducing the transmission.

Using these techniques we hope to produce a measurement of the tune-out wavelength with an accuracy better than a 100fm.

\subsection{1083~nm Tune-Out}
While the tune-out wavelength between the $3^{3}P$ and $2^{3}P$ states is most sensitive to more interesting effects (finite mass, relativistic, etc.), it provides only a single constraint to the transition dipole matrix elements. By measuring the tune-out in the multiplets we would then provide a multi point constraint on any predictions. In many ways this measurement would be far easier to perform. First the polarizability (and corresponding potential) for a given detuning is nearly a million times greater allowing for easy alignment and spectacular sensitivity. Secondly the $2^{3}P$ 1083~nm transition is easily accessed using commercial fibre lasers with high power (many watt) outputs. Additionally the relatively close transitions to the tune-outs (GHz) mean that an offset locking system is possible which allows for laser frequency accuracy better than the MHz level. Here a `local oscillator' laser is locked using saturation spectroscopy to to the known transition.  The relative frequency difference between this and a probe laser is measured by overlapping a beam from each and measuring the resulting beat frequency on a photo-diode. This signal is then sent to a frequency voltage converter and the output used for feedback to the probe laser. This would allow for these tune to be measured at an accuracy at or below the fm level. 

\subsection{Interferometer}
\begin{figure}[H]
  \begin{center}  % center environment centers the graphic on the page
    \includegraphics[height=7cm]{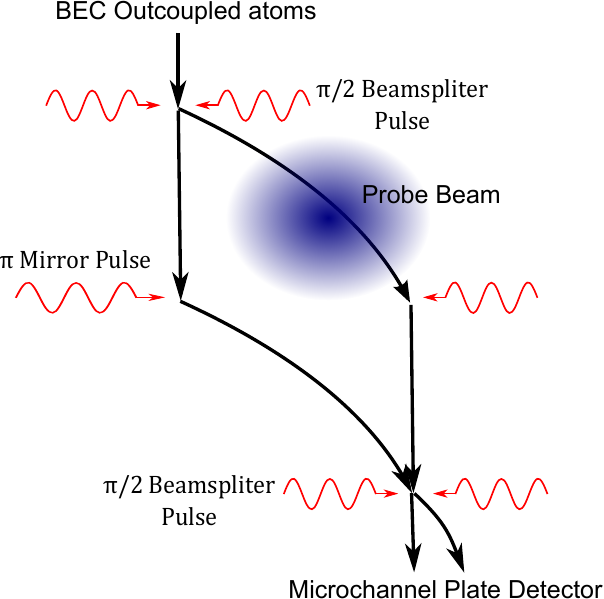}
  \end{center}
\caption{An atomic Mach-Zender Interferometer. Bragg beams which induce a phase shift in the atoms are shown in red. The probe beam in blue causes a phase shift in the right arm in proportion to wavelength and intensity.}
\label{interfer2}
\end{figure}

Further experiments may seek to improve on the 413~nm tune-out measurement by use of an atomic Mach-Zender interferometer which is analogous to the light based experiment \cite{AltinPhd}. By using a bragg pulse \cite{Kozuma1999} it is possible to manipulate atoms in an analogous manner to mirrors and beam splitters. Here instead of using a refractive index to change the phase evolution rate in one of the arms the probe beam is used to create a potential that performs the same effect (\autoref{interfer2}). The relative phase difference of the two arms are then converted to a population difference in the output ports which can then be measured and the difference minimized to find the tune-out wavelength. This technique has been applied to a number of areas atom optics from the tune-out measurement in potassium \cite{Holmgren2012,HolmgrenPhd} to precision gravimetry \cite{Altin2013}. It is particularly appealing for our case as the atoms are free from any perturbing effects of a trapping potential on the atomic structure and an uncertainty limited only by wavelength measurement may be achievable. \cite{Scully1993,Cronin2009}

The main difficulty with this type of experiment is to produce a large (integrated) potential-time difference between the arms, this can be challenging due to the shape of any applied beam and the atoms path. To this end we will also investigate an in-trap or guided \cite{McDonald2013} interferometer which may allow for far longer interaction times.

\section{Concluding Statement}
Using the future work detailed here we aim to measure the tune out wavelength at the femtometer level of accuracy and rigorously test QED atomic theory in He*.

%% file: appendix.tex
\chapter{Appendix}

\section{Trap Frequency of an ODT}
\label{trapfreqodt}
We often wish to approximate the potential of an optical dipole potential by a harmonic oscillator. To do so we find the second derivative in potential as a function of R and Z. The optical dipole potential due to a Gaussian beam is given by \cite{Grimm2000}
\begin{equation} 
U_{dip}=-\frac{1}{2\epsilon_{0} c} Re(\alpha) I_{0}\left(\frac{w_0}{w(z)}\right)^{2} e^{-\frac{2 r^{2}}{w(z)^{2}}}.
\label{trap freq appendix}
\end{equation}
where the waist as a function of axial position is defined as
\begin{equation}
w(z)=w_{0}  \sqrt{1+(\frac{z \lambda}{\pi w_0^{2}})^{2}}
\end{equation}
we then take the second derivative in both x and z about the origin.
\begin{equation}
\frac{\partial U_{dip}}{\partial r}=\frac{1}{2\epsilon_{0} c} Re(\alpha) I_{0}\left(\frac{w_0}{w(z)}\right)^{2} \frac{4 r}{w(z)^{2}} e^{-\frac{2 r^{2}}{w(z)^{2}}}
\end{equation}

\begin{equation}
\frac{\partial^{2} U_{dip}}{\partial r^{2}}=\frac{1}{2\epsilon_{0} c} Re(\alpha) I_{0}\left(\frac{w_0}{w(z)}\right)^{2} \frac{4}{w(z)^{2}} e^{-\frac{2 r^{2}}{w(z)^{2}}}+\frac{1}{2\epsilon_{0} c} Re(\alpha) I_{0}\left(\frac{w_0}{w(z)}\right)^{2} \frac{8 r^{2}}{w(z)^{4}} e^{-\frac{2 r^{2}}{w(z)^{2}}}
\end{equation}

\begin{equation}
\frac{\partial^{2} U}{\partial r^{2}} \biggr\rvert_{x=0,z=0}=\frac{1}{2\epsilon_{0} c} Re(\alpha) I_{0} \frac{4}{w_{0}^{2}}
\end{equation}

\begin{equation}
\frac{\partial^{2} U}{\partial z^{2}} \biggr\rvert_{x=0,z=0}=\frac{1}{2\epsilon_{0} c} Re(\alpha) I_{0} \frac{2}{z_{R}^{2}}
\end{equation}

The trap frequency can then be found from this second derivative as 

\begin{equation}
\omega=\sqrt{\frac{\frac{\partial^{2} U}{\partial x^{2}} \biggr\rvert_{\frac{\partial U}{\partial x}=0} }{m}}.
\end{equation}

For r and z this then evaluates to
\begin{equation}
\omega_r=\sqrt{\frac{1}{2m \epsilon_{0} c} Re(\alpha) I_{0} \frac{4}{w_{0}^{2}}}
\end{equation}

\begin{equation}
\omega_z=\sqrt{\frac{1}{2m \epsilon_{0} c} Re(\alpha) I_{0} \frac{2}{z_{R}^{2}}}
\end{equation}

The sum of two trapping frequencies can be seen from \autoref{trap freq} to be
\begin{equation} 
\omega_{Net}=\sqrt{\omega_{Trap}^{2}+\omega_{Probe}^{2}}.
\end{equation}

\section{Polarizability Conversion}
To convert the expression for the dynamic polarizability near the tune out wavelength as given in \cite{Mitroy2013} as $ \frac{\partial \alpha}{\partial \lambda}=-1.913 a_{0}^{3} (CGS)$ to SI. We first convert into units of volume polarizability using the value for $a_{0}$ \cite{NIST}.
To convert this to si units we then use \cite{Mitroy2010}
\[ 4 \pi \varepsilon \alpha'=\alpha .\]
With
\[\varepsilon=8.85\dot 10^{-12}\]
\[a_{0}=0.529\cdot 10^{-10} \]
To give the SI value of
\begin{equation}
\label{alphasi}
\frac{\partial \alpha}{\partial \lambda}=3.15 \cdot 10^{-50} \: C \cdot m \cdot v^{-1} .
\end{equation}

The resultant potential can then be given by 
\[U_{dipole}=\frac{-1}{2\varepsilon_{0} c} \frac{\partial \alpha}{\partial \lambda} \Delta \lambda(m) \; I\]
and is evaluated with the value above to
\[\frac{U(\omega)}{\Delta \lambda}=5.9*10^{-39} \frac{J m^{2}}{nm_{TO} W}=2.8*10^{-16} \frac{K m^{2}}{nm_{TO} W}.\]

\section{Outcoupling Calculations}
\label{outcouplingmath}
To calculate the outcoupling surface integral we used analytical calculations in Mathematica. The code is below:\\
\includegraphics[height=20cm]{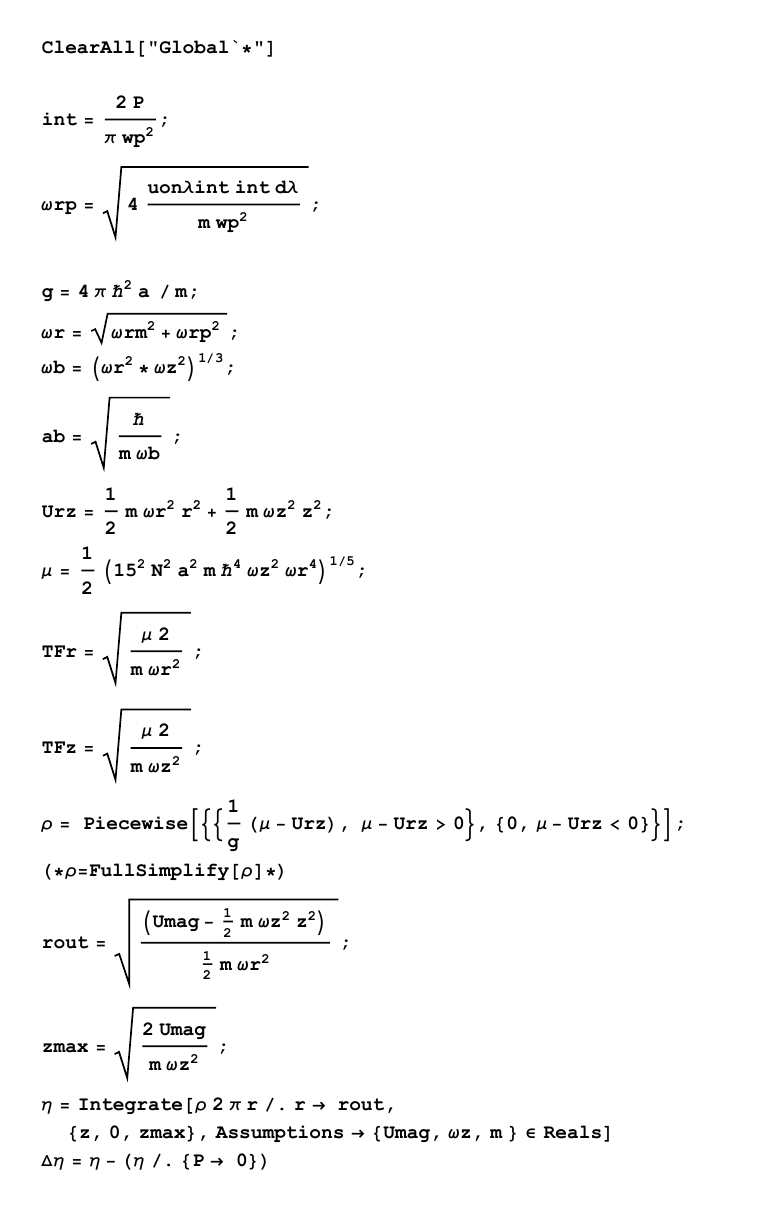}
\newpage
This gives the outcoupling density $\eta$ as\\
\includegraphics[height=4cm]{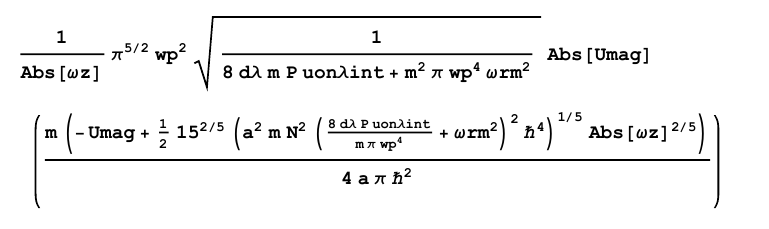}